\documentclass[11pt]{article}

\usepackage[margin=1in]{geometry}
\usepackage{setspace}
\usepackage{graphicx}
\usepackage{booktabs}
\usepackage{amsmath}
\usepackage{natbib}
\usepackage{tabularx}
\usepackage{placeins}
\usepackage{float}
\setcitestyle{aysep={}}
\usepackage[colorlinks=true, citecolor=black, linkcolor=black, urlcolor=blue]{hyperref}
\usepackage{tikz}
\usetikzlibrary{arrows.meta,positioning,calc}
\usepackage{lineno}

\usepackage{longtable}
\usepackage{pdflscape}
\usepackage{placeins}
\usepackage{titletoc}

\title{Machine Learning Models Trained on a Longitudinal Survey Dataset
Poorly Predict Hundreds of Life Outcomes}

\author{
Emily M. Cantrell,$^{1,2,*}$ 
Pranay Anchuri,$^{3}$ 
Hanzhang Ren,$^{4}$ 
Matthew J. Salganik$^{1,3}$
}

\date{October 12, 2026}

\begin{document}

\maketitle
\vspace{-2em}

\begin{center}
\small
\textit{$^{1}$Princeton University Department of Sociology} \\
\textit{$^{2}$Princeton University School of Public and International Affairs} \\
\textit{$^{3}$Princeton University Center for Information Technology Policy} \\
\textit{$^{4}$Stanford University Department of Sociology} \\
\vspace{0.5em}
$^{*}$Corresponding author: \texttt{e.cantrell@princeton.edu}
\end{center}

\begingroup
\renewcommand{\thefootnote}{}
\footnotetext{
We thank Ayomikun Gbadamosi and Zack Tseng for valuable research assistance and Abdullah Almaatouq, Mohammed Alsobay, Vineet Bansal, Josh Carmichael, Rachel Dwyer, Kathy Edin, Sayash Kapoor, Sanni Kuikka, and Brandon Stewart for helpful feedback.

Replication materials are available on our \href{https://github.com/emilycantrell/predicting-life-outcomes-ffcws}{GitHub Repository}.

This project received funding from Princeton DataX Initiative, Princeton Catalysis Initiative, and the Center for Statistics and Machine Learning (in the form of Azure computing credits). The Future of Families and Child Wellbeing Study was supported by the Eunice Kennedy Shriver National Institute of Child Health and Human Development (NICHD), the Office Of The Director, National Institutes Of Health (OD), and the National Institute On Minority Health And Health Disparities (NIMHD) of the National Institutes of Health under award numbers R01HD036916, R01HD039135, and R01HD040421, as well as a consortium of private foundations. The content of this manuscript is solely the responsibility of the authors and does not necessarily represent the official views of the National Institutes of Health.

This research was approved by the Princeton University Institutional Review Board (Protocol \#12432).
}
\setcounter{footnote}{0}
\endgroup


\begin{abstract}
Understanding the predictability of life outcomes is important for public policy and science, but existing studies focus on a limited number of outcomes and therefore struggle to describe the broad patterns of predictability. We developed a high-throughput approach that uses a single dataset to study the predictability of many future outcomes using past data. We deployed this approach with hundreds of outcomes and thousands of machine learning models per outcome in the Future of Families and Child Wellbeing Study. Three findings emerged. First, most outcomes were not very predictable, but some were much more predictable than others. Second, even for the most predictable outcomes, our complex prediction procedure typically exhibited little skill relative to a simple procedure. Third, we found modest patterns in how predictability varied across outcomes. We conclude by discussing scope conditions and how future research might address them.
\end{abstract}

\section{Introduction}

Developing a scientific understanding of the predictability of life outcomes has become increasingly important as predictive models are deployed in high-stakes fields such as child protective services, criminal justice, education, and medicine \citep{BlombergEtAl2010COMPAS, VaithianathanEtAl2013, LevyChasalowRiley2021, APPR2023PSA, Wang2024}. The predictability of life outcomes is also important for scientific reasons, such as providing a diagnostic of the completeness of scientific theories, data, and methods \citep{Breiman2001, Shmueli2010, HofmanSharmaWatts2017, FudenbergEtAl2022, Verhagen2024, vandeRijtEtAl2026}. Thus, both policymakers and scientists can benefit from facts and theories about the capabilities and limits of predictive models that use social data.

A growing body of research has examined how well life outcomes can be predicted years into the future. These studies typically find low predictability, although performance varies across outcomes and datasets \citep{Kuikka2024, SavcisensEtAl2024, ZhengCheng2025, Badolato2026, HastingsParolin2026, SivakEtAl2026}. For example, in a prior study called the Fragile Families Challenge, hundreds of researchers competed to predict six future outcomes for youth and their families using longitudinal survey data \citep{SalganikEtAl2020}. Despite substantial methodological diversity and expertise across teams, none of the submissions achieved very accurate predictions. Furthermore, the skill of the best submissions was low: they only modestly outperformed a simple four-variable linear regression model designed by a domain expert.

Although there have been a variety of studies on the predictability of life outcomes, most prior work focuses on a small number of outcomes, limiting the ability to compare predictability across different domains of social life within a common framework. As a result, it remains unclear whether some types of outcomes are systematically more predictable than others, and whether there exist highly predictable outcomes that have not yet been identified. 

We address this limitation with a high-throughput framework for studying the predictability of life outcomes. By ``high-throughput''---a term we borrow from high-throughput biology \citep{Loging2007}---we mean that one standardized modeling and evaluation procedure is applied to hundreds of outcomes, making predictive performance directly comparable across them. This comparability makes it possible to help develop ``stylized facts''~\citep{Hirschman2016} about how predictability varies across domains of social life and to identify outcomes whose predictability is atypical, both of which can be important for theory construction and testing~\citep{LundbergEtAl2024, RuefBirkhead2026}.

We applied this framework to predict nearly 500 outcomes spanning domains such as education, finances, health, and social relationships using a rich longitudinal survey dataset. For each outcome, we trained and evaluated thousands of supervised machine learning models, resulting in millions of models trained across all outcomes. By holding the data source and evaluation framework constant while varying the outcome, our approach allows us to examine patterns in predictability across domains of social life. The same framework can be applied to other datasets to assess the generality of these patterns.

The paper proceeds as follows: Section~\ref{sec:data} introduces the dataset and describes the outcomes and features. Section~\ref{sec:methods} details our methods. Section~\ref{sec:results} presents our findings: (1) most outcomes were not very predictable, (2) the winning machine learning pipelines offered little skill compared to an extremely simple benchmark, and (3) while some patterns in predictability emerged across outcome types, these patterns were weak. Section~\ref{sec:discussion} discusses scope conditions, and Section~\ref{sec:conclusion} concludes the paper. We also provide appendices that include: (1) ethical considerations, (2) additional results, (3) additional methodological information, and (4) the ``Recommendations for Machine-learning-based Science'' (REFORMS) reporting checklist~\citep{Kapoor2024}.

\section{Data}
\label{sec:data}

Our high-throughput approach to studying the predictability of life outcomes could be applied to any dataset that has individual-level data about people over time. In social research, there are three main types of datasets with this property: longitudinal surveys, administrative records held by governments, and digital trace data held by companies. For our study we chose a longitudinal survey, the Future of Families and Child Wellbeing Study (FFCWS). This dataset has three advantages for our research. First, the FFCWS is a high-quality, long-running study designed and used by domain experts; it has been used in over 1,700 publications \citep{FFCWSPublicationsArchive2026}. Therefore, it offers measurements of the types of constructs commonly studied in life course research. Second, the FFCWS is accompanied by detailed metadata about the measurement properties and substantive content of the variables, which allowed us to automate large-scale data preparation and to systematically assess patterns in what types of outcomes are most predictable \citep{KindelEtAl2019}. Third, the FFCWS was used in the Fragile Families Challenge,\footnote{The FFCWS was originally named the “Fragile Families and Child Wellbeing Study.” It was changed to the “Future of Families and Child Wellbeing Study” in 2023.} a mass collaboration where hundreds of researchers competed to predict six outcomes \citep{SalganikEtAl2020}. The Challenge results provide a credible external benchmark against which we can compare our high-throughput approach.

\subsection{Overview of the FFCWS Data}
\label{sec:ffcws_data}

The FFCWS is a probability sample of children born in large U.S. cities between 1998 and 2000, with an oversample of births to unmarried mothers \citep{ReichmanEtAl2001}. The study includes variables from a wide range of domains that experts have deemed important in a person’s life trajectory, including family and social ties, health, education, employment, finances, housing, cognitive and behavioral development, child care, parenting, demographics, and more.

\begin{figure}[htbp]
\centering
\includegraphics[width=0.6\textwidth]{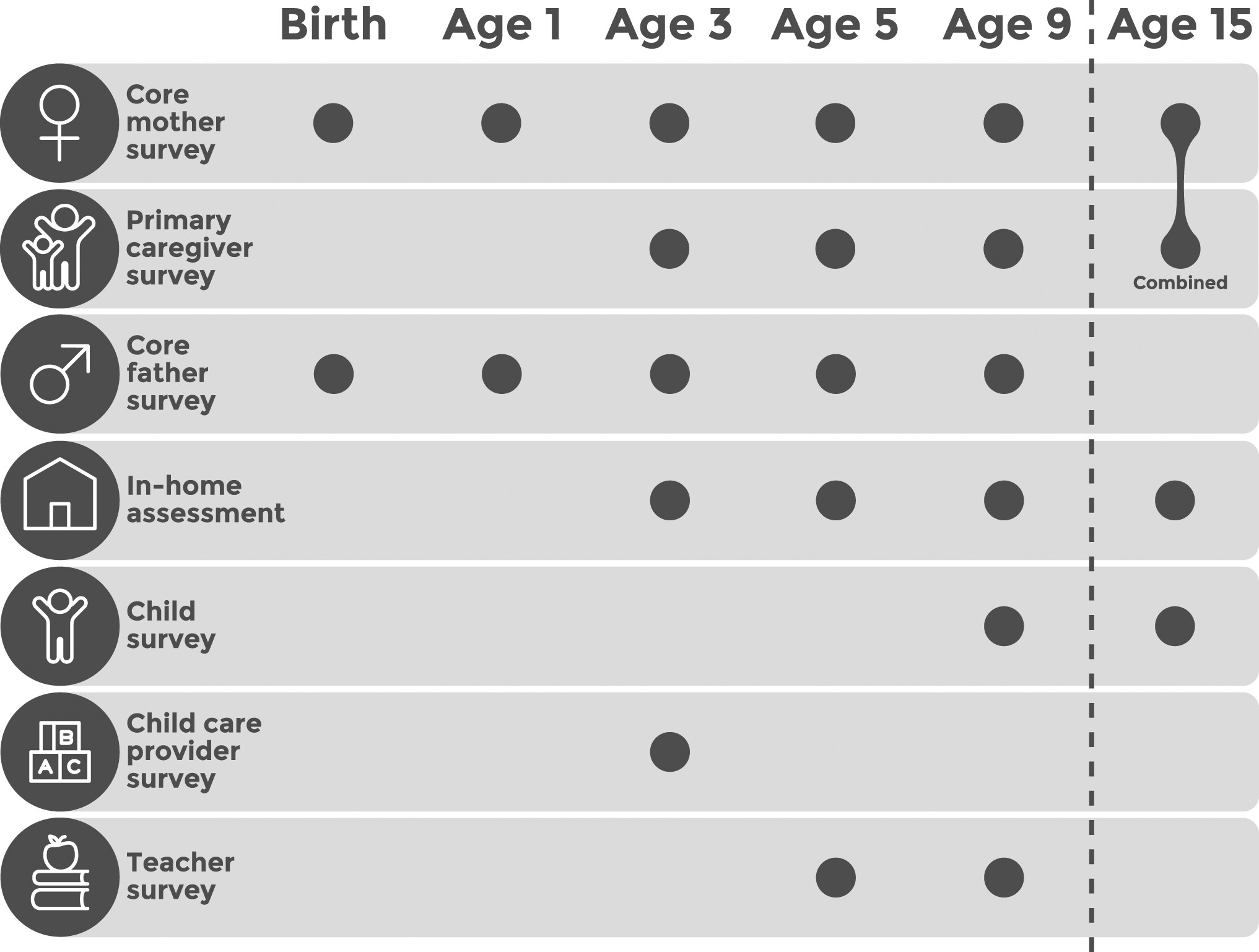}
\caption{Surveys and assessments in the FFCWS. We used thousands of variables collected from birth to year 9 to predict 484 outcomes for children and families at year 15. Figure reproduced from \citet{SalganikEtAl2020}.}
\label{fig:ffcws_waves}
\end{figure}

To enable comparison between our results and the Fragile Families Challenge results \citep{SalganikEtAl2020}, we used data from the first five waves of the FFCWS as our predictors. These waves started at the child’s birth and followed up at ages 1, 3, 5, and 9. We used data from the sixth wave, child age 15, as our outcomes. Each wave contains data from some combination of key people in the child’s life, including the mother, father, primary caregiver (who is typically, but not always, the mother), child care providers, and teachers. At ages 9 and 15, the child was also surveyed. Additionally, some waves contain in-home observations and assessments (Figure~\ref{fig:ffcws_waves}). 

\subsection{Outcomes and Features}
\label{sec:outcomes_and_features}

As described in the previous section, we used data from birth to year 9 to predict outcomes from year 15. We provide an overview of the outcomes and features below. 

\subsubsection{Outcomes: Year 15}
\label{sec:outcomes}

The FFCWS was not designed for high-throughput studies of predictability, so selecting the outcomes for our study required balancing our desire for comprehensiveness with our desire for interpretability. Ultimately, we predicted 484 outcomes. These outcomes covered a range of topics like education, finances, health, and social relationships, pertaining to the focal child, the primary caregiver, and other family members, friends, and neighbors. We reached our final set of outcomes through a process of expanding and then filtering the year-15 variables, as described below. 

Consistent with the goal of comprehensive coverage of the data, we started with all 1,616 variables measured in the main FFCWS data file from year 15. Next, we used these variables to calculate 41 scales, each of which combines information from multiple raw variables into a single composite measure \citep{FFCWSScalesDocumentation}. Finally, to facilitate comparison with the results of the Fragile Families Challenge, we calculated two Challenge outcomes that were not included among the raw year-15 variables or scales \citep{SalganikEtAl2020}.

After expanding the set of outcomes to contain scales and Challenge outcomes, we filtered the set to facilitate interpretable results. Many of the variables in the year-15 data file have characteristics that would make it difficult to interpret patterns in predictability when comparing to other outcomes. We therefore excluded 1,175 variables that did not make sense to predict in our framework. Most importantly, we excluded several hundred variables that could be skipped in the survey administration patterns depending on participants' answers to prior questions, since these variables covered different populations than the other variables. We also applied a variety of other exclusion criteria such as removing unordered categorical variables. For more information, including a flowchart that maps the total set of outcomes in the main year-15 data file to the final set of outcomes we predicted, see Appendix~\ref{app:strobe_flowcharts}. Table~\ref{tab:example_outcomes} shows some example outcomes; a full list of outcomes is available in Appendix~\ref{app:performance_for_all_outcomes}.

\begin{table}[!h]
\centering
\caption{Selected examples of outcomes.}
\label{tab:example_outcomes}
\fontsize{10}{12}\selectfont
\begin{tabularx}{.8\linewidth}{Xl}
\toprule
Outcome & Respondent\\
\midrule
Ever taken honors courses in school? & Youth\\
Frequency youth sold marijuana or other drugs & Youth\\
Number of days during school week youth eats breakfast & Youth\\
Youth's description of own health & Youth\\
Doctor diagnosed youth with seizures & Primary caregiver\\
``Parental monitoring'' scale (supervision and rule setting) & Primary caregiver\\
Total household income from all sources in past year (dollars) & Primary caregiver\\
Youth ever been suspended/expelled from school? & Primary caregiver\\
Home interior---Dark inside home & Interviewer\\
Did youth appear anxious/nervous? & Interviewer\\
\bottomrule
\end{tabularx}
\end{table}

Because of our high-throughput approach, our set of outcomes includes some variables that are not typically used as prediction targets. For example, one of the year-15 variables represents the youth’s skin color darkness, which would typically be used as an independent variable in analyses of how colorism relates to health, criminal justice system contact, or other experiences, rather than as an outcome \citep{GellerMonk2019, Monk2021}. In other words, our outcomes are defined by what data were collected, not by what kinds of variables are typically predicted. Given the way the outcome set was constructed, we treat it as a census of outcomes from the FFCWS that meet our inclusion criteria, not a random sample from some set of all possible outcomes. The potential to generalize from this set of outcomes is discussed in Section~\ref{sec:discussion}.

\subsubsection{Features: Years 0--9}
\label{sec:features}

The features we used to generate predictions included all variables in the main data file from birth to year 9, subject to a set of restrictions designed to improve predictive performance (see Section~\ref{sec:data_preparation}). In addition, we constructed 188 scales from years 0--9 to capture composite measures that may provide stronger predictors than individual variables \citep{FFCWSScalesDocumentation}. After initial data cleaning and preprocessing, the resulting data matrix contained over 15,000 features. These features reflect constructs that researchers deemed valuable to measure, and thus represent a broad set of theoretically motivated predictors of social outcomes. The subset of features used in any given model was determined during the feature selection step (see Section~\ref{sec:machine_learning_pipelines}). For more information, including a flowchart that maps the total set of variables from years 0--9 to the final set of features in our data matrix, see Appendix~\ref{app:strobe_flowcharts}.
 
\section{Methods}
\label{sec:methods}

\subsection{High-Throughput Approach}

Our study is high-throughput in two dimensions, with each dimension serving a distinct methodological purpose. First, the study is high-throughput in terms of predicting nearly 500 outcomes, which allows us to search for highly predictable outcomes and assess patterns in predictability across domains of social life. Second, the study is high-throughput in terms of training thousands of models per outcome, which helps produce robust estimates of predictive performance by ensuring that our results are not driven by one particular modeling approach. This doubly high-throughput design introduces substantial computational and engineering challenges. Due to the large number of outcomes and models, failure modes that are rare at the level of individual outcomes or models could occur regularly in our infrastructure. Thus, our approaches to data preparation and model training required care to ensure robustness against edge cases. 

\subsection{Data Preparation}
\label{sec:data_preparation}

The FFCWS data are distributed in a format that requires data preparation before applying supervised machine learning methods \citep{KindelEtAl2019}. In order to give the machine learning algorithms a stronger chance of success, we invested hundreds of researcher hours in developing an automated preprocessing infrastructure to clean the thousands of variables in the FFCWS dataset and to apply systematic feature engineering at scale. Our data preparation steps included translating variables to use consistent units across respondents, dropping variables in which values were coded in an unusual way (e.g., special negative values for unusual responses), and dropping features that did not seem plausibly useful for prediction without further data cleaning (e.g., raw calendar dates). We also iteratively refined our methods to improve predictive performance in the presence of the “curse of dimensionality,” which refers to challenges that arise when the number of features is large relative to the sample size \citep{AltmanKrzywinski2018}. For example, we removed features with extremely low base rates because they tended to degrade model performance by introducing spurious correlations between features and outcomes. Finally, we performed standard preprocessing procedures like imputation, one-hot encoding, and scaling. To the best of our knowledge, our approach to data preparation far exceeded what any participants did during the Fragile Families Challenge, in part because we had access to the metadata that were created after the Challenge \citep{KindelEtAl2019}. For additional information, see Appendix~\ref{app:strobe_flowcharts}.

\subsection{Machine Learning Pipelines}
\label{sec:machine_learning_pipelines}

During the Fragile Families Challenge, participants used many different methods to predict outcomes \citep{SalganikLundbergKindelMcLanahan2019}. Most of these approaches involved three main steps: data preparation, feature selection, and modeling. We sought to create a high-throughput approach that would mimic the kinds of pipelines that researchers might typically utilize in predictive modeling. We restricted our approach to techniques that were commonly available in mature open-source libraries such as \texttt{scikit-learn} at the time we began this research \citep{PedregosaEtAl2011}.

\begin{figure}[htbp]
\centering
\includegraphics[width=1\textwidth]{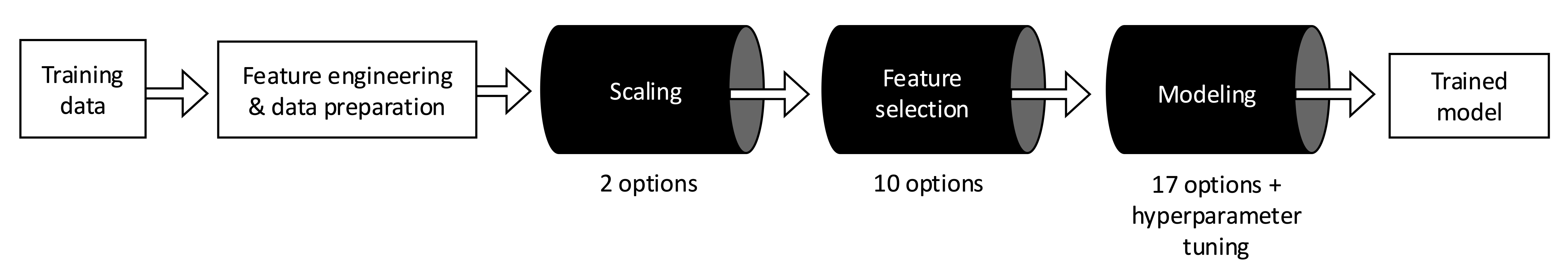}
\caption{Combinatorial structure of our machine learning pipelines. Each of the three steps could be fulfilled by one mechanism out of a set of several options appropriate for that step.}
\label{fig:pipelines}
\end{figure}

Our approach created a pipeline for every possible combination of: 2 scaling options (standard scaler or quantile transformer with uniform distribution), 10 feature selection options (selecting varied numbers of features between 1 and 50, using F-statistics for linear association or Spearman rank correlations to rank features in order of importance), and 17 model classes (e.g., LASSO regression, AdaBoost decision tree regression, multi-layer perceptron neural networks, or $k$-nearest neighbors regression). This cross-combination resulted in up to 340 possible pipelines for each outcome (Figure~\ref{fig:pipelines}). 

Consistent with best practice, we also conducted hyperparameter tuning. For each model class, we tested the default hyperparameters plus up to 63 random draws from a grid of candidate hyperparameter values \citep{BergstraBengio2012}. We iteratively refined the hyperparameter ranges and the number of draws through empirical testing. See Appendix~\ref{app:hyperparameter_tuning} for details. 

Overall, based on the total possible combinations of scaling options, feature selection options, modeling options, and hyperparameter draws, we trained up to 15,160 models per outcome. A full run for all outcomes and all pipelines took 959 compute hours---equivalent to about 40 days on a single machine---but parallelization in our computing architecture reduced the actual elapsed time to approximately four days.

As a preview of our results, we found that our pipelines achieved similar predictive performance to the top-performing models from the Challenge \citep{SalganikEtAl2020}. This comparison provides some evidence that our automated procedure produced credible estimates of predictability in this dataset; see Appendix~\ref{app:ffc} for more information. We discuss the generalizability of our approach and the implications of newer machine learning methods in Section~\ref{sec:discussion}.

\subsection{Evaluation Strategy}\label{sec:eval_strategy}

Reliably selecting and estimating the performance of the best pipeline is a difficult problem in limited data. As always, training and evaluation must be conducted on separate data to avoid overfitting. Further, additional complexity arises due to our large number of pipelines. If we selected a winning pipeline based on its performance in a test sample and used its score on that same sample as our estimate of performance, our estimates would have an optimistic bias. This phenomenon is sometimes referred to as the ``winner's curse'' \citep{Andrews2024}. To ensure an unbiased estimate of generalization performance, pipeline selection and pipeline evaluation must be conducted on different samples~\citep{CawleyTalbot2010}. 

To address these requirements, we used nested cross-validation. Conceptually, this approach separates the data into three roles: training, pipeline selection, and evaluation. Specifically, we implemented five outer folds and five inner folds. Within each outer training set, the inner folds were used to identify the best-performing pipeline. The selected pipeline was then evaluated on the corresponding held-out outer fold. Further details on our nested cross-validation procedure are provided in Appendix~\ref{app:estimand_and_estimation_procedure}.

This approach offers two advantages. First, it ensures a clear separation between pipeline selection and evaluation, avoiding bias from the winner’s curse. Second, by rotating through multiple splits of the data, it reduces sensitivity to any particular train-test split and allows all observations to contribute to evaluation. 

As an additional check on overfitting, we used just half of our data for all infrastructure development and results in this manuscript (this sample $n=2{,}121$; total sample $n=4{,}242$).\footnote{The FFCWS contains data on 4,898 families from 20 cities. Our sample size is smaller because we omit two cities where some questions were asked differently due to pilot testing. Also, this sample size refers to the total number of families, but the effective sample size for each outcome varies because of missingness patterns.} Once we receive feedback from the reviewers and editors, we will finalize all of our modeling decisions and then run the full procedure once on the other half of the data that we have not yet used in this study. If results are similar across the two samples of $n = 2{,}121$, we will then re-run the full procedure on the full sample of $n = 4{,}242$ and report the full sample results in the final publication. 

\subsection{Metric: \texorpdfstring{$R^2_{\mathrm{CV}}$}{R-squared CV}} 
\label{sec:metric}

There are many metrics for evaluating predictive performance. We use $R^2$ for three reasons. First, $R^2$ is based on mean squared error, an evaluation metric with strong theoretical foundations \citep{Murphy1987, Gneiting2007}. Second, $R^2$ facilitates comparisons across outcomes with different data types and measurement units. Because it is normalized by the performance of a baseline model (predicting the mean of the training data), $R^2$ is unitless and can be compared across continuous, ordered categorical, and binary outcomes.\footnote{For binary outcomes, we generated predicted probabilities rather than class labels, allowing $R^2$ to be computed in a manner consistent with continuous and ordered categorical outcomes. For more on evaluating predicted probabilities for binary outcomes, see \citet{Brier1950} and \citet{Greenhill2011}.} Third, $R^2$ has a clear and interpretable scale. A value of $R^2 = 1$ indicates perfect prediction, while $R^2 = 0$ indicates performance equivalent to predicting the mean of the training data. Negative values indicate performance worse than this baseline, which can occur when models are trained and tested on separate samples. For these reasons, $R^2$ is widely used in social science and data science, including in the Fragile Families Challenge, to which we compare our results.

Specifically, we use $R^2_{\mathrm{CV}}$, where the CV subscript stands for ``cross-validation'' (see Section~\ref{sec:eval_strategy}):

\begin{equation}
    R^2_{\mathrm{CV}} = 1 - \frac{\sum\limits_{k=1}^{5} \sum\limits_{i \in k}{(y_i-\hat{y}_i)^2}}{\sum\limits_{k=1}^{5}\sum\limits_{i \in k}{(y_i-\bar{y}_{\text{Training}_k})^2}}
    \label{eq:r2_cv}
\end{equation}
where $i$ is the family, $k$ is the cross-validation fold, $\hat{y}_i$ is the prediction from our procedure, and $\bar{y}_{\text{Training}_k}$ is the mean of the training data associated with cross-validation fold $k$. We calculated confidence intervals (CIs) around our estimates using standard methods, as described in Appendix~\ref{app:uncertainty_quantification}.

\section{Results}
\label{sec:results}

\subsection{Most Outcomes Are Not Very Predictable}

\begin{figure}[htbp]
\centering
\includegraphics[width=.75\textwidth]{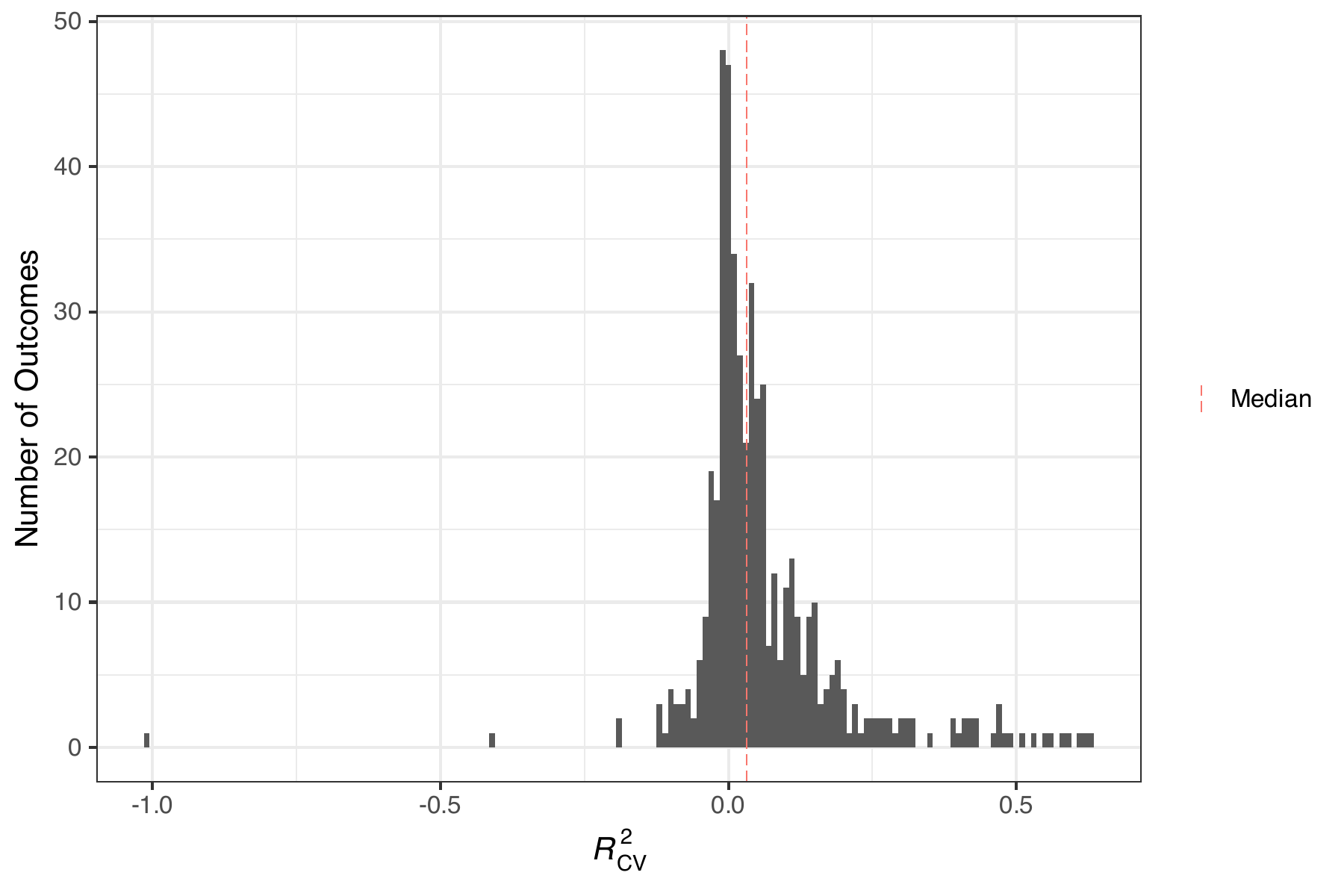}

\caption{Distribution of $\widehat{R^2}_{\mathrm{CV}}$ across outcomes.}
\label{fig:r2_histogram}
\end{figure}

Our main empirical result is that predictability was very low for most outcomes we examined (Figure~\ref{fig:r2_histogram}). The median $\widehat{R^2}_{\mathrm{CV}}$ across outcomes was 0.03. Further, $\widehat{R^2}_{\mathrm{CV}} \leq 0$ for 31 percent of the outcomes. This means that for 31 percent of the outcomes, our selected pipeline did not make predictions any better than predicting the training-set mean for every observation. 

Although predictability was low overall, Figure~\ref{fig:r2_histogram} shows substantial variation around the median, with some outcomes having much higher levels of predictability. Table~\ref{tab:top10} shows the top ten most predictable outcomes. These included several measures related to the youth's physical characteristics, a measure of the primary caregiver's marital status with the youth's other biological parent, and a measure of the primary caregiver's household income. Appendix~\ref{app:performance_for_all_outcomes} provides a table of $\widehat{R^2}_{\mathrm{CV}}$ for all outcomes. We discuss patterns in which kinds of outcomes were more or less predictable in Section~\ref{sec:patterns_predictability}.

\begin{table}[!h]
\centering
\caption{Top ten most predictable outcomes.}
\label{tab:top10}
\centering
\fontsize{10}{12}\selectfont
\begin{tabularx}{\linewidth}{lXl}
\toprule
$\widehat{R^2}_{\mathrm{CV}}$ (95\% CI) & Outcome & Respondent\\
\midrule
0.63 (0.57, 0.68) & Biological primary caregiver married to biological parent & Primary caregiver\\
0.62 (0.53, 0.70) & Youth's height (cm) from multiple measures & Youth\\
0.61 (0.57, 0.65) & Youth's self-reported height (cm) & Youth\\
0.59 (0.52, 0.67) & Youth's body mass index (BMI) (from in-home measurement) & Youth\\
0.58 (0.53, 0.64) & Youth's skin color darkness & Interviewer\\
0.56 (0.48, 0.64) & Youth's weight (lbs) from multiple measures & Youth\\
0.55 (0.47, 0.63) & Youth's weight-for-age percentile (from in-home measurement) & Youth\\
0.53 (0.45, 0.61) & Total household income from all sources in past year (dollars) & Primary caregiver\\
0.51 (0.43, 0.58) & Youth's BMI percentile (from in-home measurement) & Youth\\
0.49 (0.42, 0.57) & Youth's self-reported weight (lbs) & Youth\\
\bottomrule
\end{tabularx}
\end{table}

The results that most outcomes were not very predictable and that some outcomes were more predictable than others are broadly consistent with the Fragile Families Challenge, the previously mentioned mass collaboration where hundreds of researchers competed to predict six year-15 outcomes \citep{SalganikEtAl2020}. Further, we found that for the outcomes predicted in both studies, the performance of our selected pipelines was similar to the performance of the winning models in the Challenge (see Appendix~\ref{app:ffc}). Although this direct comparison is limited to a small number of outcomes, it provides some evidence that our data preparation approach, machine learning pipelines, and evaluation strategy produced reasonable estimates of predictive performance in the FFCWS dataset.

\subsection{Skill Is Low} 
\label{sec:skill_is_low}

Although most outcomes were difficult to predict, some outcomes were more predictable than others (Figure~\ref{fig:r2_histogram}). However, high predictive accuracy does not necessarily indicate that our modeling procedure uncovered complex structure in the data. Therefore, we compared our winning pipelines' performance to a benchmark to assess the incremental gains from our complex modeling procedure. We refer to this comparison as \emph{skill}. The concept of skill originates in meteorological forecasting, where the performance of complex models is evaluated relative to simple reference forecasts such as persistence (assuming tomorrow’s weather will be the same as today’s) or climatology (assuming tomorrow’s weather will match historical averages for that date) \citep{Murphy1992}.  We operationalized skill as:

\begin{equation}
\text{Skill} = R^2_{\mathrm{CV}, \text{ winning pipeline}} - R^2_{\mathrm{CV}, \text{ benchmark}}.
\label{skill_equation}
\end{equation}

When constructing the benchmark procedure, we opted for extreme simplicity for both substantive and practical reasons. Substantively, the simple benchmark was motivated by prior research showing that the winning submissions from the Fragile Families Challenge only modestly outperformed a four-variable model created by a domain expert~\citep{SalganikEtAl2020}. Practically, a simple procedure was necessary because it was not possible to have domain experts handcraft sophisticated benchmarks for hundreds of outcomes. Therefore, for each outcome, the benchmark consisted of a pipeline with standard scaling, selection of the top feature based on its linear association with the outcome, and a linear or logistic regression model (see Appendix~\ref{app:simple_benchmark_models}). In other words, the benchmark models were one-feature regressions that required no domain expertise to create.

\begin{figure}[H]
\centering
\includegraphics[width=.6\textwidth]{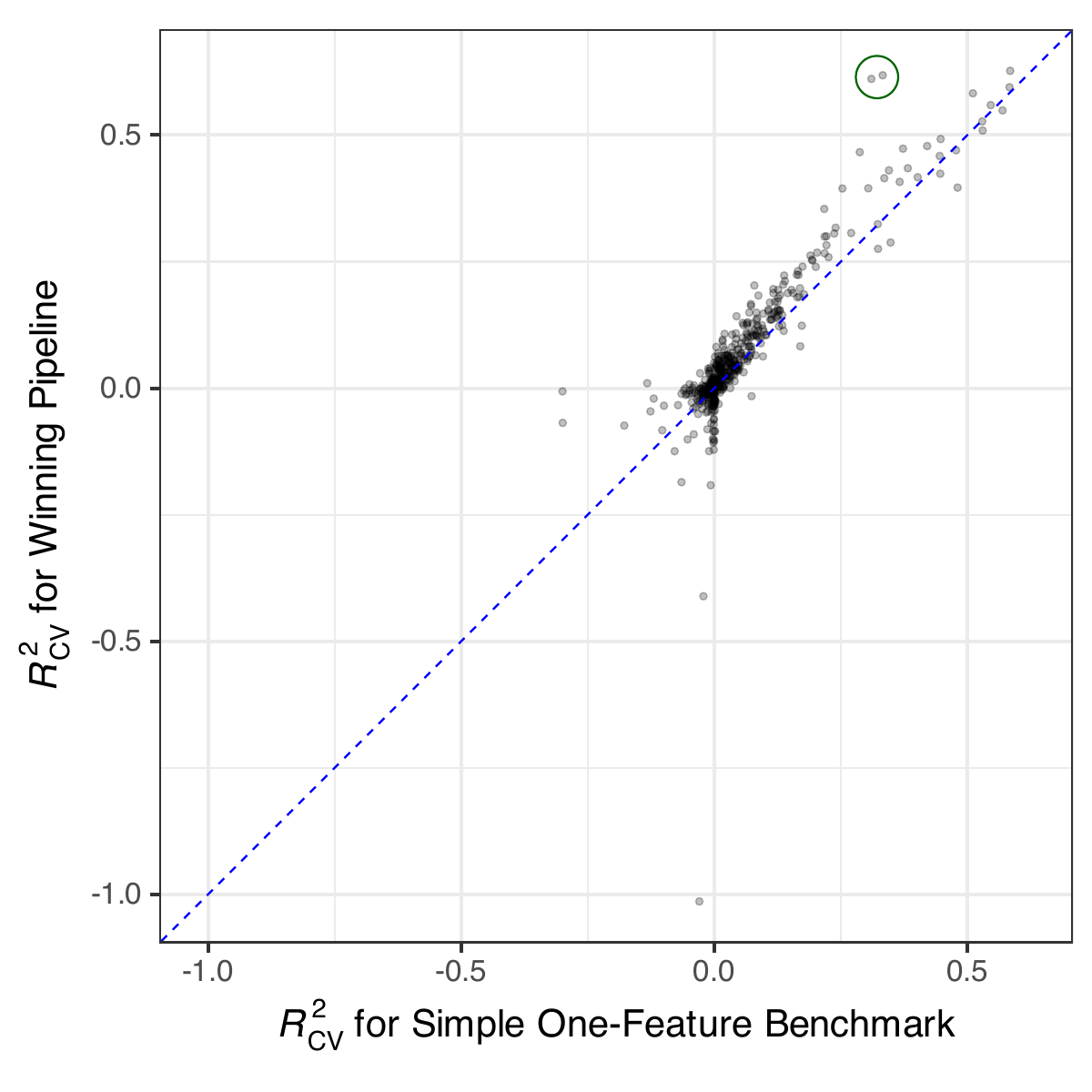}
\caption{$\widehat{R^2}_{\mathrm{CV}}$ for the simple one-feature benchmark versus the winning pipeline for each outcome. The two circled points are discussed in the main text because predictive accuracy and skill were both unusually high.}
\label{fig:benchmark_vs_winners}
\end{figure}

Figure~\ref{fig:benchmark_vs_winners} compares the predictive accuracy of each outcome’s winning pipeline to that of the simple benchmark. Across outcomes, the median skill was 0.01. In other words, for the median outcome in our setting, the benefits of our complex modeling procedure were very small relative to the simple procedure. The low level of median skill might be expected given the low level of median accuracy. However, even among the winning pipelines in the highest decile of accuracy, where the median accuracy was $\widehat{R^2}_{\mathrm{CV}} = 0.35$, the median skill was 0.06. Moreover, for 31 percent of all outcomes we predicted, the estimated skill was negative, meaning the benchmark performed better than the winning pipeline. We attribute this negative skill to the difficulty of pipeline selection in this setting.\footnote{Negative skill occurred even though the simple benchmark was included among the candidate pipelines during model selection, because sampling variability meant that the pipeline selected as best on the inner folds was not always the one that performed best on the outer folds.} Overall, these results are consistent with prior research showing that simple models often perform similarly to more complex ones \citep{SalganikEtAl2020, Semenova2022, MorucciSpirling2024}.

Even though accuracy and skill were low for most outcomes, there were some outcomes for which both were unusually high. However, these apparent outliers need to be interpreted in light of our extremely simple benchmark. As an exploration, we examined two such outcomes, circled in Figure~\ref{fig:benchmark_vs_winners}. These outcomes turned out to be two measures of the same construct: the youth's height at age 15 as measured by the interviewer ($\widehat{R^2}_{\mathrm{CV}} = 0.62$ [95\% CI: 0.53, 0.70]; $\text{skill} = 0.29$) and as self-reported by the youth ($\widehat{R^2}_{\mathrm{CV}} = 0.61$ [95\% CI: 0.57, 0.65]; $\text{skill} = 0.30$). For both outcomes, the top feature was a binary indicator of the youth’s sex assigned at birth. We then added one additional handpicked feature to the simple pipeline: the youth’s height percentile at age 9. The resulting two-feature model nearly matched the performance of the winning pipeline.\footnote{For the youth’s height as measured by the interviewer at age 15, the one-feature benchmark had $\widehat{R^2}_{\mathrm{CV}} = 0.33$ [95\% CI: 0.27, 0.39], the two-feature model had $\widehat{R^2}_{\mathrm{CV}} = 0.62$ [95\% CI: 0.55, 0.68], and the winning pipeline had $\widehat{R^2}_{\mathrm{CV}} = 0.62$ [95\% CI: 0.53, 0.70]. For the youth’s self-reported height at age 15, the one-feature benchmark had $\widehat{R^2}_{\mathrm{CV}} = 0.31$ [95\% CI: 0.27, 0.35], the two-feature model had $\widehat{R^2}_{\mathrm{CV}} = 0.58$ [95\% CI: 0.54, 0.62], and the winning pipeline had $\widehat{R^2}_{\mathrm{CV}} = 0.61$ [95\% CI: 0.57, 0.65].} This example illustrates that high skill relative to an extremely simple benchmark does not necessarily indicate that the complex procedure uncovered interesting or important structure in the data. 

\subsection{Predictability Shows Modest Patterns Across Outcomes}
\label{sec:patterns_predictability}

Figure~\ref{fig:r2_histogram} shows that there is variability in the predictability of outcomes. An important question is how this variability relates to characteristics of the outcomes. For example, if outcomes related to health are more predictable than outcomes related to education, then this kind of ``stylized fact''~\citep{Hirschman2016} would be useful both for practitioners and for researchers developing theory.

We are not aware of any general-purpose categorization scheme that could be used to organize the 484 outcomes in this study along multiple dimensions. However, the FFCWS is accompanied by metadata that categorize outcomes along dimensions that the study designers considered meaningful for their study~\citep{KindelEtAl2019}, including: (1) the topical domain of the question (e.g., religion, mental health), (2) who the question was about (e.g., youth, parent), and (3) who answered the question (e.g., youth, parent). In addition to using these existing metadata categories, we also categorized outcomes by whether they were composite measures (scales) constructed from multiple survey items. We focus on these simple categories with the goal of identifying broad empirical regularities in the predictability of life outcomes.

\subsubsection{Physical Characteristics Are the Most Predictable, on Average}
\label{sec:physical_characteristics}

\begin{figure}[htbp]
\centering
\includegraphics[width=.8\textwidth]{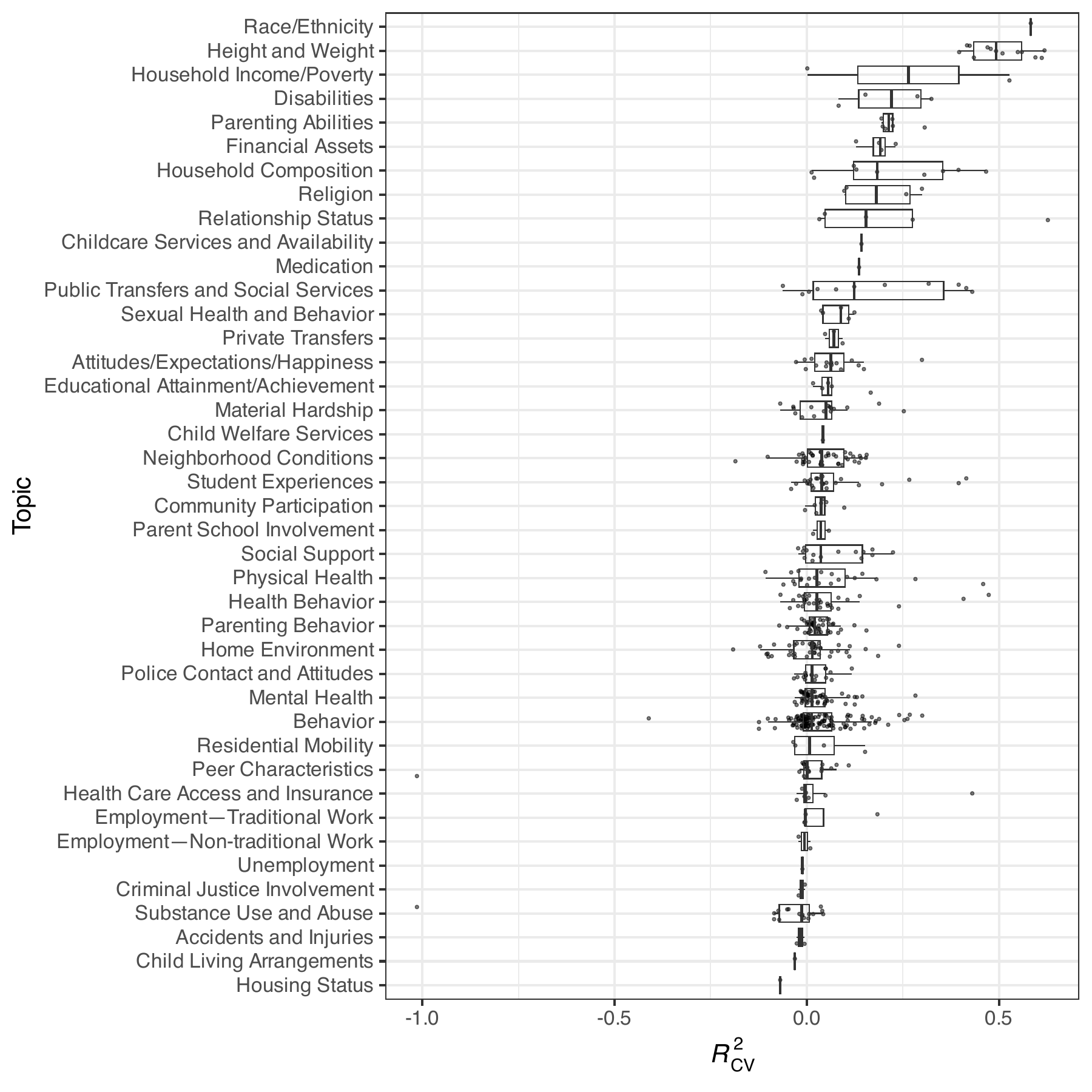}
\caption{$\widehat{R^2}_{\mathrm{CV}}$ by topic. Each outcome may have up to two topics, so an outcome can appear twice on the plot. Points are vertically jittered to reduce overplotting.}
\label{fig:subtopics}
\end{figure}

The FFCWS metadata categorize our outcomes into 41 topic areas.\footnote{The “topic areas” we analyze here are called “subtopics” in the FFCWS metadata.} Figure~\ref{fig:subtopics} displays box-and-whisker plots for the predictability of outcomes grouped by topic. Two topic areas stand out as having much higher predictability, on average, than the others: \textit{race/ethnicity} (which contains a single outcome estimating the youth’s skin color darkness; $\widehat{R^2}_{\mathrm{CV}} = 0.58$), and \textit{height and weight} (which contains multiple representations of the youth’s height and weight, such as direct measurements, self-reported measurements, percentiles, and BMI; median $\widehat{R^2}_{\mathrm{CV}} = 0.49$). Notably, both of these topic areas describe physical characteristics. 

\subsubsection{Parents Are More Predictable Than the Youth, on Average}
\label{sec:parent_vs_youth}

One potential source of variation in predictability is whether outcomes refer to the youth or their parent(s). On one hand, we might expect the youth-focused outcomes to be easier to predict since the youth is the primary focus of the FFCWS study. On the other hand, we might expect the parent-focused outcomes to be easier to predict because adults may have more stable characteristics than children. 

\begin{figure}[htbp]
\centering
\includegraphics[width=0.75\textwidth]{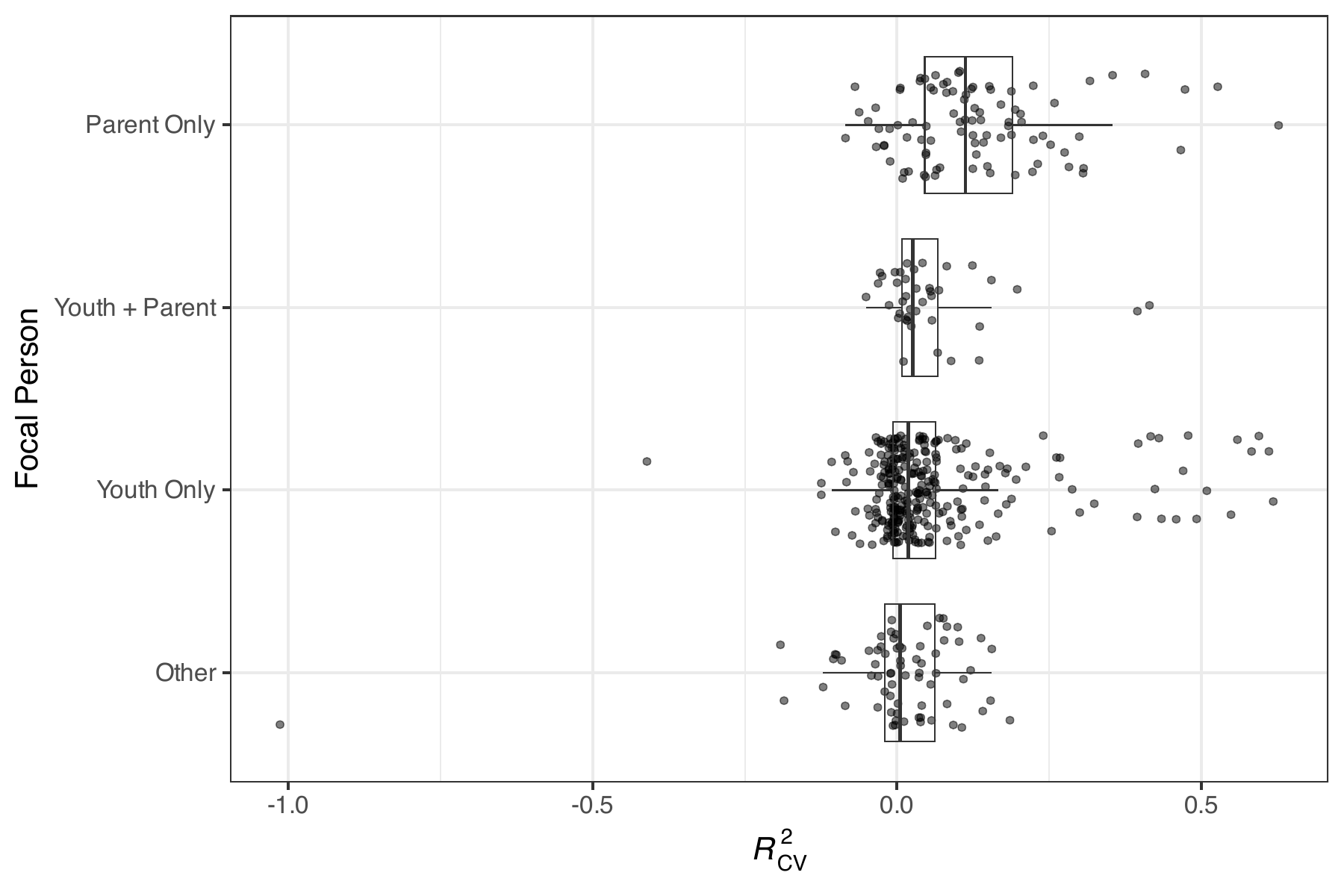}
\caption{$\widehat{R^2}_{\mathrm{CV}}$ by focal person. Points are vertically jittered to reduce overplotting.}
\label{fig:focal_person}
\end{figure}

The FFCWS metadata assign each outcome one or more values for “focal person,” where the focal person is the person the question is about, such as the youth, the youth’s parent(s),\footnote{We combined three focal person codes as “parent” in this analysis: “primary caregiver” (who is typically, but not always, the mother), “mother,” and “father.”} or other people like romantic partners, friends, or neighbors. Some outcomes are assigned multiple focal person values; for example, the outcome ``Nights during week family eats dinner together'' was labeled as focusing on the parent and youth. As Figure~\ref{fig:focal_person} shows, outcomes that focused exclusively on a parent tended to be the most predictable, with median $\widehat{R^2}_{\mathrm{CV}} = 0.11$, compared to median $\widehat{R^2}_{\mathrm{CV}} = 0.03$ for outcomes that focused jointly on a parent and the youth, 0.02 for outcomes that focused on the youth alone, and 0.00 for outcomes that focused on other people. One way to quantify the magnitude of this pattern non-parametrically is to note that if we randomly select one outcome that focuses exclusively on a parent and one that focuses exclusively on the youth, the parent-focused outcome has a higher $\widehat{R^2}_{\mathrm{CV}}$ 72 percent of the time. However, Figure~\ref{fig:focal_person} also shows that there is substantial variation in predictability among outcomes with the same focal person classification, indicating that who the outcome is about captures only part of the variation in predictability.

\begin{figure}[htbp]
\centering
\includegraphics[width=0.75\textwidth]{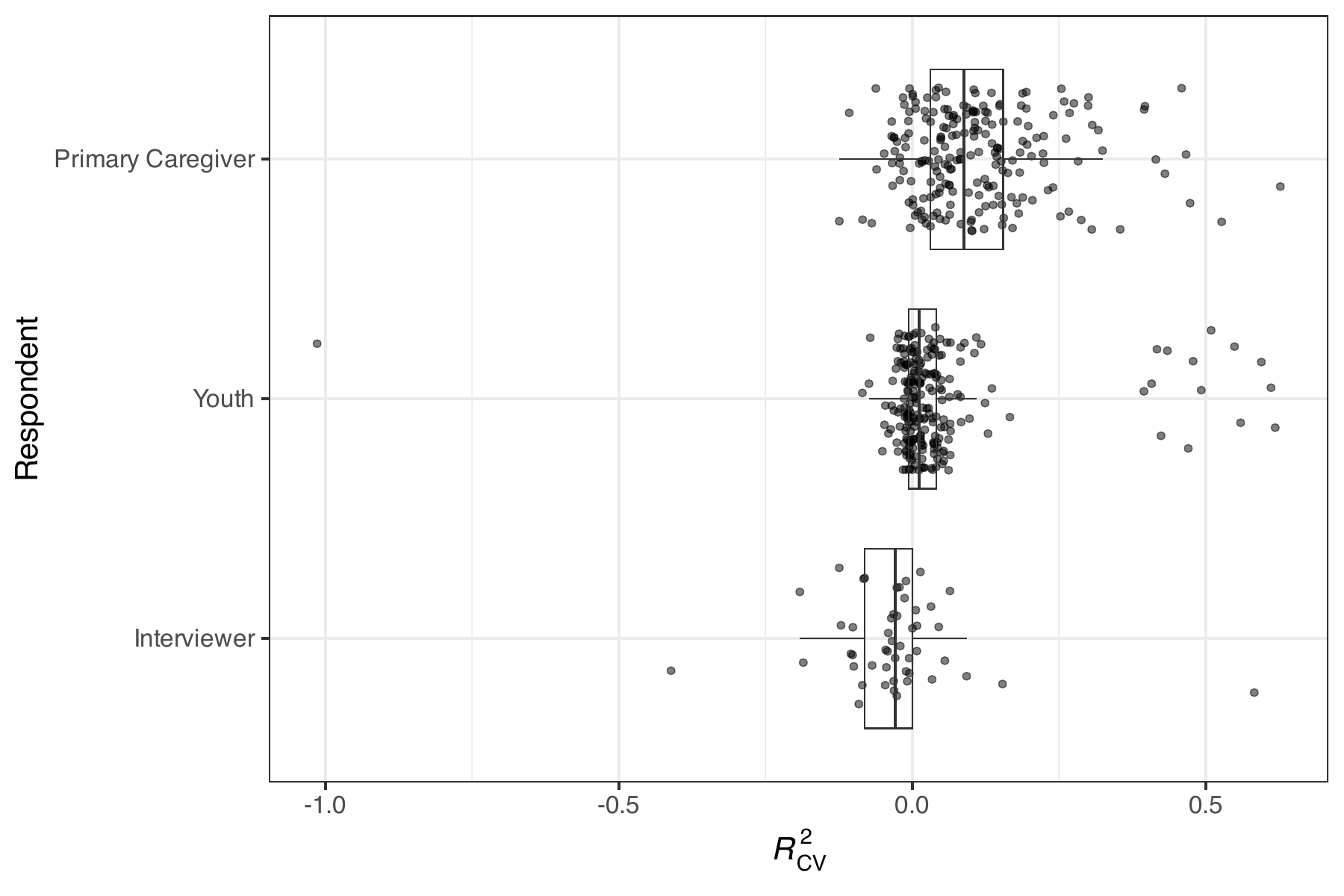}
\caption{$\widehat{R^2}_{\mathrm{CV}}$ by respondent. Points are vertically jittered to reduce overplotting.}
\label{fig:respondent}
\end{figure}

In addition to categorizing the outcomes by focal person, the FFCWS metadata categorize the outcomes by who reported them. There were three respondent types in the year-15 survey wave: the youth, the primary caregiver, and the interviewer who conducted in-home observations. The youth could report outcomes where the primary caregiver was a focal person, and vice versa. Figure~\ref{fig:respondent} shows the predictability of outcomes grouped by respondent. We found that the outcomes reported by the primary caregiver had the highest predictability on average, with median $\widehat{R^2}_{\mathrm{CV}} = 0.09$, compared to median $\widehat{R^2}_{\mathrm{CV}} = 0.01$ for youth-reported outcomes and median $\widehat{R^2}_{\mathrm{CV}} = -0.03$ for interviewer-reported outcomes.\footnote{In-home observations were conducted for only a subset of families due to budget constraints, so the sample size for interviewer-reported outcomes is smaller ($n = 500$ or less depending on missingness in the outcome), which might lead to lower $\widehat{R^2}_{\mathrm{CV}}$.} If we randomly select an outcome where the primary caregiver is the respondent, and randomly select an outcome where the youth is the respondent, the primary-caregiver-reported outcome has a higher $\widehat{R^2}_{\mathrm{CV}}$ 74 percent of the time. However, once again, there is a high degree of heterogeneity within these groups of outcomes.

One reason parent-focused and primary-caregiver-reported outcomes might be more predictable than youth-focused and youth-reported outcomes is that many parent measures were also collected in earlier survey waves, while many youth measures were not collected at earlier waves because they were not developmentally appropriate. For example, questions about the youth’s grade point average and whether the youth had ever dated anyone were not asked at ages 9 and younger. As shown in Appendix~\ref{app:accounting_for_prior_measurement}, the pattern of parent-focused and primary-caregiver-reported outcomes being more predictable than youth-focused and youth-reported outcomes holds for both outcomes that were previously measured and outcomes that were not previously measured. Thus, the greater predictability of parent outcomes cannot be explained solely by whether the outcome was measured in earlier waves. The pattern is consistent with the possibility that differences across life course stages contribute to the greater predictability of parent outcomes, although our design cannot distinguish this explanation from other differences between parent and youth outcomes.

\subsubsection{Scales Are More Predictable Than Other Outcomes, on Average}
\label{sec:scales}

\begin{figure}[htbp]
\centering
\includegraphics[width=.75\textwidth]{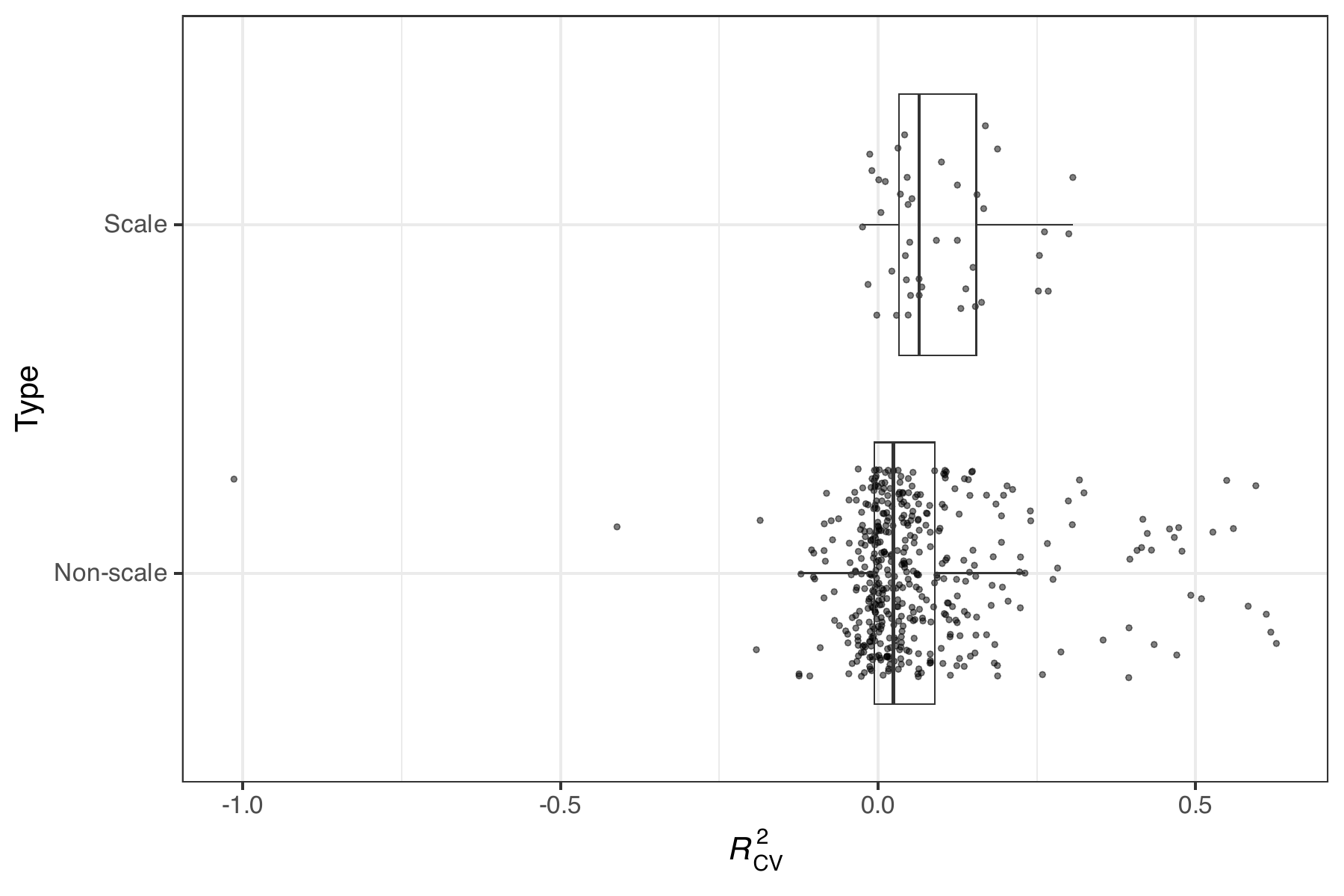}
\caption{$\widehat{R^2}_{\mathrm{CV}}$ for scales versus other outcomes. Points are vertically jittered to reduce overplotting.}
\label{fig:scales}
\end{figure}

We expect that it should be easier to predict a general experience, like a teen getting good grades in most of her classes, than a specific experience, like a teen getting an ‘A’ in math class or a ‘B’ in English, because idiosyncratic noise in individual measures should cancel out when outcomes are aggregated. The FFCWS scales offer an opportunity to test this theory. As stated above, a scale is a composite measure that combines information from multiple variables. For example, the “material hardship” scale takes the average of 11 variables indicating experiences with hunger, eviction, difficulty paying bills, and other related topics \citep{MayerJencks1989, BendheimThoman2024Year15Guide}. As expected, we found that the median $\widehat{R^2}_{\mathrm{CV}}$ for scales (0.06) was higher than the median $\widehat{R^2}_{\mathrm{CV}}$ for other outcomes in our set (0.02); see Figure~\ref{fig:scales}. If we randomly select an outcome that is a scale, and randomly select an outcome that is not a scale, the scale has a higher $\widehat{R^2}_{\mathrm{CV}}$ 66 percent of the time. However, as with the analyses above, the distributions overlap substantially, and many non-scale outcomes are more predictable than the median scale outcome. Thus, whether an outcome is a scale is associated with predictability on average, but there remains substantial variation in predictability within both groups.

Overall, this pattern and the patterns in Sections~\ref{sec:physical_characteristics} and~\ref{sec:parent_vs_youth} show that the categories we used do not cleanly separate our outcomes into high and low predictability groups. While there are some differences between the group medians, there is substantial within-group variability.

\section{Discussion}
\label{sec:discussion}

Developing an understanding of the predictability of life outcomes informs the responsible use of predictive models in applied contexts and contributes to scientific theory about the life course. Our study advances this research agenda in three ways. First, we introduce a high-throughput methodology for systematically characterizing patterns in the predictability of life outcomes at scale. Second, we show that even with rich longitudinal data about individuals over many years, the predictability of life outcomes can be low across a broad range of domains. Third, we find that our high-throughput machine learning procedure offers little skill relative to a simple procedure in this dataset. To our knowledge, this is the first high-throughput study of the predictability of life outcomes, so important questions remain. In this section, we discuss generalizability in terms of methods, data, and prediction tasks, and examine how our research setting compares with predictive modeling in U.S. social service settings.

\subsection{Generalizability Across Methods}
\label{sec:generalizability_across_ML}

Our standardized high-throughput approach may perform differently from an approach developed specifically for a single prediction task. A researcher focused on one outcome could use substantive knowledge and experimentation to tailor feature engineering and modeling choices to that outcome. Such customization could yield higher predictive accuracy and skill than our automated procedure, although the similarity between our results and the Fragile Families Challenge results for several outcomes provides some evidence that our procedure can perform comparably to models tailored for individual prediction tasks.

Our findings are also specific to the types of machine learning methods included in our framework. We used supervised machine learning pipelines constructed from off-the-shelf procedures similar to those used by participants in the Fragile Families Challenge. We focused on this class of methods because they are widely used and are likely to be deployed in applied settings. However, newer methods are rapidly emerging for predicting life outcomes \citep{Athey2024, SavcisensEtAl2024, VafaEtAl2024CAREER, HollmannEtAl2025, Shmatko2025, VerhagenEtAl2025BookOfLife, SatishEtAl2026}. These methods are not yet widely used to make predictions about individuals, and most are not sufficiently standardized to be incorporated into a high-throughput framework like ours. However, as these approaches become more widely adopted and standardized, evaluating their accuracy and skill at scale will become both more feasible and more important for science and policy.

Some of these emerging methods could not only improve predictive performance, but also make high-throughput studies of predictability substantially easier to conduct. One particularly relevant approach is tabular foundation models, such as TabPFN \citep{HollmannEtAl2025, GrinsztajnEtAl2026}. Because TabPFN is pretrained to perform well across heterogeneous tabular prediction tasks, it might improve predictive performance relative to the methods considered here. Furthermore, and we think more importantly, tabular foundation models could dramatically streamline our high-throughput approach by automating some data preprocessing steps and removing the need for outcome-specific pipeline selection. How these new methods would perform in our setting is an open empirical question.

Additionally, conclusions about skill depend heavily on the design of the benchmark. If researchers can develop principled methods for producing sophisticated benchmarks in high-throughput contexts, then outcomes where models demonstrate high skill would be a promising place to search for new scientific insights~\citep{Verhagen2024}.

\subsection{Generalizability Across Data}

The FFCWS is just one of many longitudinal survey datasets. It is a useful dataset choice for the goals of this project as discussed in Section~\ref{sec:data}. Future work should extend the high-throughput approach for assessing the predictability of life outcomes to longitudinal survey datasets at different stages of the life course and with different populations.

Beyond differences within longitudinal surveys, our results may differ in other types of data sources, such as administrative data or digital trace data. Large-scale administrative datasets such as Dutch or Nordic registries provide particularly rich settings for studying predictability across outcomes, populations, and time horizons \citep{Kuikka2024, SavcisensEtAl2024, SivakEtAl2026}. These datasets typically have much larger sample sizes than surveys, which can increase predictive accuracy, although with diminishing returns \citep{Viering_Loog_2023, CantrellEtAl2026MoreData}. At the same time, they contain different types of variables than surveys: rather than being designed to measure constructs of interest to social scientists, registry and digital trace data are generated for administrative purposes~\citep{Salganik2018}. Comparisons across types of datasets can help clarify how predictive performance relates to both sample size and the types of constructs measured. For example, \citet{SivakEtAl2026} found that small-sample survey data outperformed large-sample registry data in predicting births and adoptions, largely due to the inclusion of powerful features about fertility intentions in the survey data.

The skill of complex modeling procedures may also vary across datasets. For example, in registry datasets, the gains from more flexible models may be larger due to higher sample sizes and less feature engineering pre-embedded in the data \citep{Perlich_et_al_2003, Viering_Loog_2023, MorucciSpirling2024, CantrellEtAl2026MoreData}.

\subsection{Generalizability Across Prediction Tasks}
\label{sec:generalizability_prediction_tasks}

Within any given dataset, there are many possible prediction tasks that could be defined. In this study, we used data from years 0--9 to predict outcomes at year 15, but alternative choices of background period and time horizon could yield different levels of predictability. Future research could systematically vary task definitions to assess how predictability varies with these design choices \citep{YanRahal2025}. Researchers may also vary how outcomes are selected. For example, future research could focus on a targeted set of outcomes to test a specific hypothesis, or could take an exploratory approach like the one used here by examining a broad range of outcomes. Such efforts can be guided by theoretical perspectives on the origins of unpredictability in the life course \citep{LiouEtAl2023, LundbergEtAl2024}.

\subsection{Comparison to Prediction in Social Services Settings}
\label{comparison_to_social_services}

A key question for generalizability is how our study relates to predictive models used in U.S. social services and criminal justice contexts. The FFCWS has some important similarities to the populations in these settings. In particular, its oversample of families with nonmarital births results in a sample with substantial representation of low-income families and people of color, populations that are disproportionately represented in social services and criminal justice contexts where predictive models are deployed~\citep{Eubanks2018}. However, predictive modeling applications are heterogeneous in their goals, data inputs, and methodologies, and our research setting differs from deployed applications in several important ways. Here we compare our data and methods to common characteristics of predictive modeling tasks in U.S. social services agencies and the criminal justice system.

Compared to the types of administrative data commonly used in social services, the FFCWS dataset has both advantages and disadvantages for prediction. On one hand, our training sample has substantially fewer people than many administrative datasets. On the other hand, we observe more information about each person, with rich longitudinal data on individuals and their families over many years. While it is possible that predictive models with administrative data in applied settings could be more accurate than ours due to larger sample sizes, it is also possible that such models could be less accurate than ours due to weaker predictive features. For example, as stated above, \citet{SivakEtAl2026} found higher predictive accuracy in survey data than administrative registry data. Furthermore, the types of outcomes predicted in social services settings likely differ from the outcomes measured in the FFCWS.

Applied prediction tasks may also allow for more outcome-specific model development than our high-throughput approach. As discussed in Section~\ref{sec:generalizability_across_ML}, a researcher focused on predicting a single outcome could tailor the feature engineering and modeling approach to the specific prediction task. Such customization might produce higher predictive accuracy than our standardized high-throughput approach.

Another important difference between our study and many applied prediction tasks is the time horizon. We predicted outcomes six years into the future, whereas models deployed in social services often make predictions for much shorter time frames, such as weeks or months. Predictive accuracy likely declines as the time horizon increases, since there is more opportunity for small differences between individuals to widen and for “consequential intervening events” to occur \citep{LundbergEtAl2024}. Further research is needed to assess the rate at which predictive performance changes as the time horizon increases, and how this interacts with data richness and sample size.

Finally, one aspect of our evaluation framework was more favorable to prediction than deployed settings: our training, selection, and evaluation data all came from a single cohort with outcomes measured at the same time. In a real-world, deployed setting, models need to predict outcomes in a later time period, which introduces the possibility of “temporal drift,” in which patterns the model learned from earlier data no longer hold \citep{LogeswaranOliver2025}. Thus, the temporal structure of our evaluation setting may produce more optimistic estimates of predictive accuracy than would be achieved in deployment.

\section{Conclusion}
\label{sec:conclusion}

This study introduces a high-throughput framework for systematically evaluating the predictability of life outcomes by comparing hundreds of outcomes under a common dataset, modeling approach, and evaluation procedure. This framework makes it possible to assess patterns in predictability across outcomes, thereby informing the responsible use of predictive models in high-stakes social settings and advancing scientific understanding.

Across a wide range of domains, we found that most outcomes were difficult to predict and that our high-throughput machine learning procedure had limited skill compared to a simple procedure. These results demonstrate that extensive application of contemporary machine learning methods to rich longitudinal data does not necessarily yield high predictive accuracy or substantial skill. 

For policymakers using predictive models in high-stakes settings, our findings underscore the importance of evaluating---not assuming---both accuracy and skill. For scientists, this study encourages a shift in focus from studying predictive performance for a small number of outcomes to mapping patterns in predictability across many outcomes. By identifying systematic variation in predictability, high-throughput approaches can contribute to a broader theoretical understanding of the predictability of the life course. In this view, low predictability is not merely a limitation to overcome but a phenomenon to explain.

\clearpage
\nolinenumbers
\bibliography{references}

\clearpage

\appendix

\startcontents[appendices]

\begin{center}
    {\Large\bfseries Appendices}
\end{center}

\section*{Contents}
\printcontents[appendices]{}{1}{}
\clearpage

\renewcommand{\thefigure}{\thesection\arabic{figure}}
\setcounter{figure}{0}

\renewcommand{\thetable}{\thesection\arabic{table}}
\setcounter{table}{0}

\numberwithin{equation}{section}
\renewcommand{\theequation}{\thesection\arabic{equation}}

\renewcommand{\theHfigure}{\thesection.\arabic{figure}}
\renewcommand{\theHtable}{\thesection.\arabic{table}}



\clearpage
\section{Ethics}
\label{app:ethics}

Our procedures were approved by the Princeton Institutional Review Board (Protocol \#12432). However, predicting individuals’ life outcomes raises ethical questions that extend beyond regulatory compliance. This appendix situates our study in relation to established ethical principles.

Given the extraordinarily low levels of predictability in our results, a reader might wonder why extensive ethical consideration is necessary. Two points are important. First, we did not know these results in advance; we made ethical decisions under uncertainty about what the study would reveal. Second, the predictability of life outcomes may change as data sources and modeling approaches improve. We therefore view this appendix not only as documentation for this specific study, but also as a starting point for future research in settings where predictive performance may be higher.

\subsection{Ethical Principles}

This study is informed by four widely used ethical principles for scientific research: (1) respect for persons, (2) beneficence, (3) justice, and (4) respect for law and public interest \citep{BelmontReport1979, MenloReport2012}. We assess our research design in terms of each principle.

Respect for persons emphasizes the autonomy of research participants. The data used in our study were collected under protocols that included informed consent from adults and assent from children when appropriate \citep{LundbergEtAl2019Ethics}. 

Beneficence requires balancing risks and benefits for study participants. In our study, the risks and benefits are similar to those associated with other secondary analyses of longitudinal survey data. 

Justice concerns the distribution of risks and benefits across social groups. The Future of Families and Child Wellbeing Study (FFCWS) sample includes a large number of Black, Hispanic, and low-income families \citep{ReichmanEtAl2001}, and these groups are disproportionately affected by predictive systems in domains such as social services and criminal justice \citep{Eubanks2018}. Therefore, any improvements in practice would benefit the groups in the study.

Respect for law and public interest extends considerations beyond research participants to broader societal impacts. Research on predictive modeling may influence how such models are perceived and used in practice. Our findings should not be interpreted as supporting or opposing the use of predictive models in specific policy domains. The effects of such systems depend on how they are designed, implemented, and governed. Our contribution is descriptive: to characterize what is currently predictable using widely available methods with a high-quality data source. Respect for law and public interest also motivates careful consideration of which outcomes are appropriate to predict in a given research context, as discussed in the next section.

\subsection{Framework for Excluding Outcomes for Ethical Concerns}

None of the predictive models in this study will be deployed in real-life settings, but scientific knowledge about the predictability of specific life outcomes could be used in ways that are either beneficial or harmful. We therefore adopted a decision framework that may also be useful for future research. We chose to focus on the U.S. in this framework since the sampling frame for the FFCWS study is within the U.S. and we are based at U.S. institutions. 

For each outcome, we asked ourselves two questions: Given current cultural and political circumstances in the U.S. (as of 2022 when we made these decisions), can we reasonably foresee a real-world circumstance where prediction of the outcome using the FFCWS or another dataset: (1) could be used to harm many people, or (2) could be used to help many people? Based on our answers to these questions, we made decisions as outlined in Table \ref{table:ethics}.

\begin{table}[htbp]
\centering
\caption{Ethical decision-making framework for whether to predict an outcome in this study.}
\label{table:ethics}
\begin{tabularx}{\textwidth}{l >{\centering\arraybackslash}X >{\centering\arraybackslash}X}
\toprule
 & \textbf{Helpful} & \textbf{Not helpful} \\
\midrule
\textbf{Not harmful} & Predict & Predict \\
\midrule
\textbf{Harmful} 
& \parbox[c]{\linewidth}{\centering Default is to predict;\\ discuss as a team}
& \parbox[c]{\linewidth}{\centering Default is to \textbf{not} predict;\\ discuss as a team} \\
\bottomrule
\end{tabularx}
\end{table}

In our team discussions about outcomes for which predictive models in deployed contexts could plausibly be harmful, we took into consideration existing governance structures that might mitigate these risks.

Following this framework, we excluded two outcomes because they are closely related to sexual orientation, and models predicting similar outcomes could present potential harms in deployed contexts without corresponding foreseeable benefits:
\begin{itemize}
    \item \texttt{k6f2}, which asked the youth, “Have you ever liked a girl as more than just a friend?”
    \item \texttt{k6f3}, which asked the youth, “Have you ever liked a boy as more than just a friend?”
\end{itemize}

These judgments necessarily involve uncertainty and may change over time as social contexts and technological capabilities evolve. We therefore do not present them as definitive rules, but as an example of how ethical considerations can be incorporated into high-throughput research on prediction.


\clearpage

\setcounter{figure}{0}
\setcounter{table}{0}

\section{Additional Results}
\label{app:results}

\subsection{Our Pipelines Performed Similarly to the Winners of the Fragile Families Challenge}
\label{app:ffc}

To assess the performance of our pipelines, we compared our results to the results of the Fragile Families Challenge \citep{SalganikEtAl2020}. This comparison is informative because dozens of teams competed in the Challenge to produce the most accurate predictive models, thus providing a useful empirical benchmark for the performance achievable using these data. However, the comparison is not exact because the data splits and evaluation procedures differed slightly.

\begin{figure}[htbp]
\centering
\includegraphics[width=.6\textwidth]{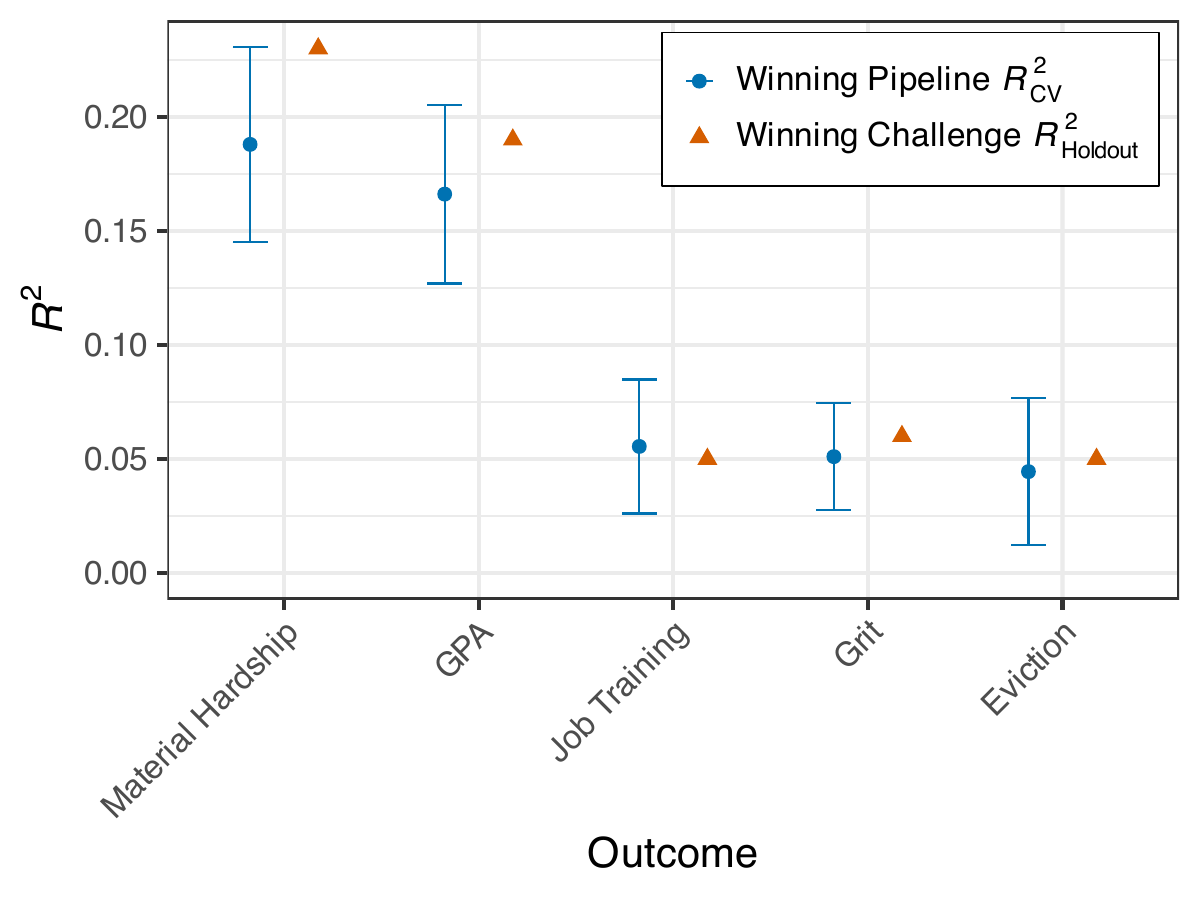}
\caption{Winning pipeline versus winning Fragile Families Challenge submission for five outcomes. As described in the text, the training sample sizes and the evaluation procedures differed between studies, so these comparisons are approximate.}
\label{fig:ffc_comparison}
\end{figure}

For the five outcomes that were analyzed in both studies,\footnote{The sixth outcome in the Challenge was job layoff for the primary caregiver. We did not predict this outcome because it could be skipped in the survey logic; for more information about our exclusion criteria, see Section~\ref{app:strobe_flowcharts}.} the $\widehat{R^2}_{\mathrm{CV}}$ estimates produced by our high-throughput framework are comparable in magnitude to the best $\widehat{R^2}_{\mathrm{Holdout}}$ scores achieved by Challenge participants. As shown in Figure~\ref{fig:ffc_comparison}, the Challenge $\widehat{R^2}_{\mathrm{Holdout}}$ estimate lies within the 95 percent confidence interval of our $\widehat{R^2}_{\mathrm{CV}}$ estimate for all five outcomes, although the Challenge point estimate is slightly higher for four of the five. 

Two aspects of the Challenge evaluation procedure may contribute to this slightly higher performance. First, the training sample size in the Challenge was slightly larger: while our full sample corresponded to the Challenge training set ($n = 2{,}121$),\footnote{This sample size will increase when we test our infrastructure on a sample of $n = 4{,}242$ after the journal review process, as discussed in Section~\ref{sec:eval_strategy}, ``Evaluation Strategy,'' in the main text.} we used only 80 percent of these cases for training within each outer fold of cross-validation. Second, the Challenge selected winning submissions based on performance on the same holdout data used for final evaluation, which may introduce an optimistic bias to the estimates. In contrast, our procedure used separate data splits for pipeline selection and evaluation to mitigate this risk of bias.

\FloatBarrier


\subsection{Greater Predictability of Parent Outcomes Is Not Limited to Previously Measured Constructs}\label{app:accounting_for_prior_measurement}

\begin{figure}[htbp]
\centering
\includegraphics[width=.8\textwidth]{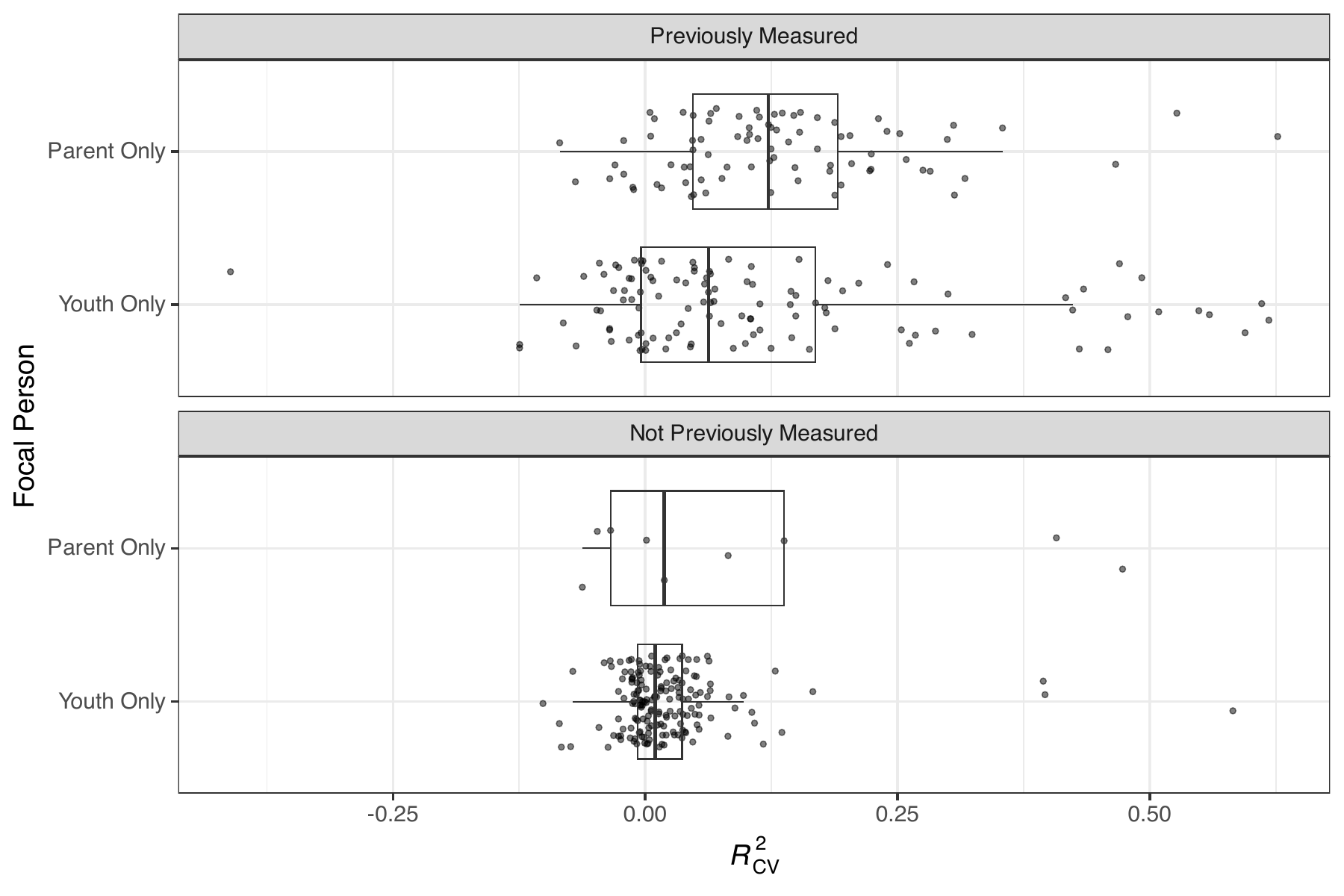}
\caption{$\widehat{R^2}_{\mathrm{CV}}$ by focal person and by whether the same construct was measured at an earlier survey wave. Points are vertically jittered to reduce overplotting.}
\label{fig:focal_person_by_measured_earlier}
\end{figure}

\begin{figure}[htbp]
\centering
\includegraphics[width=.8\textwidth]{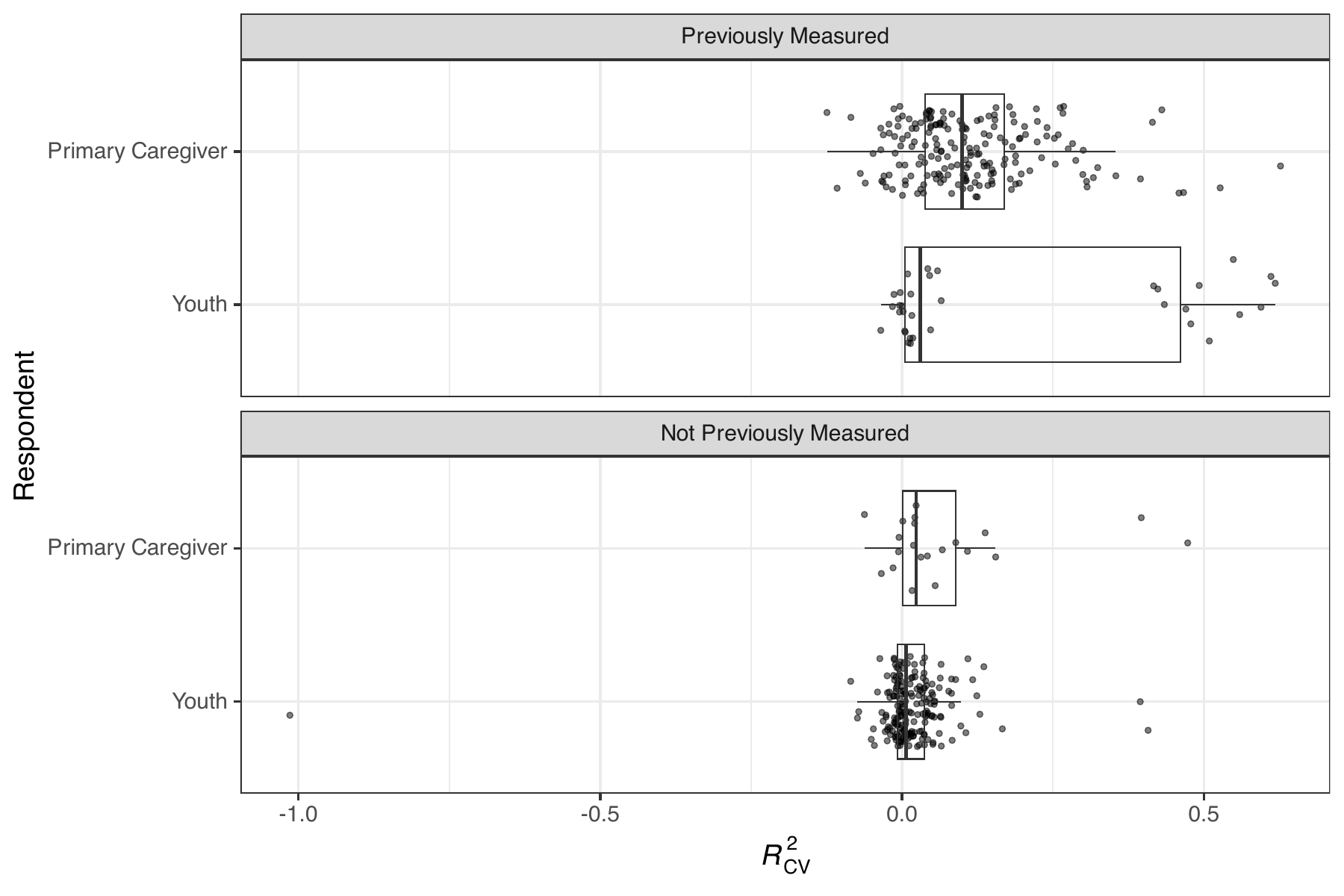}
\caption{$\widehat{R^2}_{\mathrm{CV}}$ by respondent and by whether the same construct was measured at an earlier survey wave. Points are vertically jittered to reduce overplotting.}
\label{fig:respondent_by_measured_earlier}
\end{figure}

In the main text, we show that on average, parent-focused and primary-caregiver-reported outcomes are more predictable than youth-focused and youth-reported outcomes (Section~\ref{sec:parent_vs_youth}). This finding might be because parent-focused and primary-caregiver-reported outcomes are more likely to represent constructs that were also measured at earlier survey waves. However, further analysis shows that when outcomes are divided by whether they were measured in previous waves, the pattern of parent-focused and primary-caregiver-reported outcomes being more predictable, on average, still holds, as shown in Figure~\ref{fig:focal_person_by_measured_earlier} (focal person) and Figure~\ref{fig:respondent_by_measured_earlier} (respondent). 

Our methodology for this analysis was as follows. To determine whether the construct represented by each outcome had also been measured at any previous FFCWS survey wave, we manually reviewed all 484 outcomes. We defined a construct as “previously measured” if the same construct was collected from the same respondent at any prior wave and was included in our feature data prior to feature selection. We considered variables that were similar to represent the same construct even if there were minor wording changes between waves, changes to the set of response options, or changes to the variable calculation process (for scales and constructed variables). Additionally, if multiple variables from a prior wave could be combined to represent the same construct as the year-15 variable, we considered that construct to be previously measured. For example, there was a year-15 question that asked the primary caregiver whether they had used any of a list of illicit drugs, whereas in year 9 there was a series of questions that asked about each of those drugs individually; we considered this construct to be previously measured. For outcomes where the primary caregiver was the respondent, we considered survey items from either the primary caregiver survey or the mother survey at earlier waves to be from the same respondent, since the mother was typically the primary caregiver. Two authors independently reviewed 20 percent of the outcomes to assess whether each construct had been previously measured, and compared our classifications. Inter-rater reliability was high, so the remaining outcomes were classified by a single author. A \href{https://github.com/emilycantrell/predicting-life-outcomes-ffcws/blob/main/analysis/metadata_for_analysis/same_construct_previously_measured_FINAL.csv}{CSV file} with our classifications of whether constructs were previously measured and notes on the reasons for our decisions is available on our GitHub repository.

\FloatBarrier

\subsection{Winning Pipeline and Benchmark Performance for All Outcomes}
\label{app:performance_for_all_outcomes}

Table \ref{tab:all_outcomes_results} provides results for all outcomes in our analysis. The table is ordered by the predictive performance of the winning pipeline, from highest to lowest $\widehat{R^2}_{\mathrm{CV}}$. For each outcome, we report the variable label, respondent, and $\widehat{R^2}_{\mathrm{CV}}$ with 95 percent confidence intervals (CIs) for the winning pipeline and the benchmark pipeline. A machine-readable \href{https://github.com/emilycantrell/predicting-life-outcomes-ffcws/blob/main/analysis/figures_and_analyses_for_paper/precomputed_results/results_and_metadata_for_all_outcomes_manuscript_submission.csv}{CSV file} containing these results and additional metadata for all outcomes is available on our GitHub repository.

In the outcome names, ``scale'' indicates the outcome is either a scale we constructed following the FFCWS instructions \citep{FFCWSScalesDocumentation}, or a variable we constructed to match an outcome from the Fragile Families Challenge \citep{SalganikEtAl2020} using the same code infrastructure we used for scales; see Section~\ref{app:strobe_flowcharts} for more details. In the variable labels, ``constructed'' indicates the variable was created by FFCWS staff (e.g., by deriving it from other variables), rather than collected directly as a survey item \citep{BendheimThoman2024Year15Guide}. Abbreviations in the table are as follows: ADD/ADHD: attention-deficit or attention-deficit/hyperactivity disorder; BMI: body mass index; BSI: Brief Symptom Inventory; CBCL: Child Behavior Checklist; CES-D: Center for Epidemiologic Studies Depression Scale; CHAOS: Confusion, Hubbub, and Order Scale; CIDI: Composite International Diagnostic Interview; CIDI-SF: Composite International Diagnostic Interview--Short Form; CPS: Child Protective Services; EPOCH: Engagement, Perseverance, Optimism, Connectedness, and Happiness; GPA: grade point average; SNAP: Supplemental Nutrition Assistance Program; SSDI: Social Security Disability Insurance; SSI: Supplemental Security Income; TANF: Temporary Assistance for Needy Families; and UI/WC: unemployment insurance/workers' compensation.

\begin{landscape}
\begingroup
\singlespacing
\fontsize{8}{9}\selectfont
\setlength{\tabcolsep}{3pt}

\begin{longtable}[t]{>{\raggedright\arraybackslash}p{3.8cm}>{\raggedright\arraybackslash}p{9.1cm}>{\raggedright\arraybackslash}p{2.6cm}>{\raggedright\arraybackslash}p{3.0cm}>{\raggedright\arraybackslash}p{3.0cm}}
\caption{Predictive performance for all outcomes.\label{tab:all_outcomes_results}}\\
\toprule
Outcome & Variable label & Respondent & Winning pipeline $\widehat{R^2}_{\mathrm{CV}}$ (95\% CI) & Benchmark pipeline $\widehat{R^2}_{\mathrm{CV}}$ (95\% CI)\\
\midrule
\endfirsthead
\caption[]{Predictive performance for all outcomes. \textit{(continued)}}\\
\toprule
Outcome & Variable label & Respondent & Winning pipeline $\widehat{R^2}_{\mathrm{CV}}$ (95\% CI) & Benchmark pipeline $\widehat{R^2}_{\mathrm{CV}}$ (95\% CI)\\
\midrule
\endhead

\endfoot
\bottomrule
\endlastfoot
cp6pmarb & Constructed—Biological primary caregiver married to biological parent & Primary caregiver & 0.63 (0.57, 0.68) & 0.58 (0.53, 0.64)\\
\addlinespace[6pt]
ch6chtcm & Constructed—Youth's height (cm) from multiple measures & Youth & 0.62 (0.53, 0.70) & 0.33 (0.27, 0.39)\\
\addlinespace[6pt]
ck6chtcm & Constructed—Youth's self-reported height (cm) & Youth & 0.61 (0.57, 0.65) & 0.31 (0.27, 0.35)\\
\addlinespace[6pt]
ch6cbmi & Constructed—Youth's BMI (from in-home measurement) & Youth & 0.59 (0.52, 0.67) & 0.58 (0.50, 0.67)\\
\addlinespace[6pt]
h6a8 & Youth's skin color darkness & Interviewer & 0.58 (0.53, 0.64) & 0.51 (0.45, 0.57)\\
\addlinespace[6pt]
ch6cwtlb & Constructed—Youth's weight (lbs) from multiple measures & Youth & 0.56 (0.48, 0.64) & 0.55 (0.45, 0.64)\\
\addlinespace[6pt]
ch6wap & Constructed—Youth's weight-for-age percentile (from in-home measurement) & Youth & 0.55 (0.47, 0.63) & 0.57 (0.49, 0.64)\\
\addlinespace[6pt]
p6k57 & Total household income from all sources in past year (dollars) & Primary caregiver & 0.53 (0.45, 0.61) & 0.53 (0.44, 0.61)\\
\addlinespace[6pt]
ch6bmip & Constructed—Youth's BMI percentile (from in-home measurement) & Youth & 0.51 (0.43, 0.58) & 0.53 (0.45, 0.61)\\
\addlinespace[6pt]
ck6cwtlb & Constructed—Youth's self-reported weight (lbs) & Youth & 0.49 (0.42, 0.57) & 0.45 (0.37, 0.52)\\
\addlinespace[6pt]
ch6hap & Constructed—Youth's height-for-age percentile (from in-home measurement) & Youth & 0.48 (0.40, 0.56) & 0.42 (0.35, 0.50)\\
\addlinespace[6pt]
p6h74 & Ever regularly smoked cigarettes? & Primary caregiver & 0.47 (0.43, 0.52) & 0.37 (0.32, 0.42)\\
\addlinespace[6pt]
ck6wap & Constructed—Youth's weight-for-age percentile (from youth's self-report) & Youth & 0.47 (0.42, 0.52) & 0.48 (0.43, 0.53)\\
\addlinespace[6pt]
p6e30 & Number of biological children in household (youth excluded) & Primary caregiver & 0.47 (0.41, 0.52) & 0.29 (0.24, 0.34)\\
\addlinespace[6pt]
p6b2 & Doctor diagnosed youth with asthma & Primary caregiver & 0.46 (0.40, 0.52) & 0.45 (0.39, 0.51)\\
\addlinespace[6pt]
ck6hap & Constructed—Youth's height-for-age percentile (from youth's self-report) & Youth & 0.43 (0.39, 0.48) & 0.38 (0.34, 0.43)\\
\addlinespace[6pt]
p6b31 & Youth covered by Medicaid/public insurance plan? & Primary caregiver & 0.43 (0.39, 0.47) & 0.35 (0.30, 0.39)\\
\addlinespace[6pt]
ck6cbmi & Constructed—Youth's BMI (from youth's self-report) & Youth & 0.42 (0.36, 0.49) & 0.45 (0.37, 0.52)\\
\addlinespace[6pt]
ck6bmip & Constructed—Youth's BMI percentile (from youth's self-report) & Youth & 0.42 (0.37, 0.46) & 0.40 (0.35, 0.45)\\
\addlinespace[6pt]
p6j35 & Youth gets free/reduced price school lunch & Primary caregiver & 0.41 (0.37, 0.46) & 0.34 (0.28, 0.39)\\
\addlinespace[6pt]
k6d46 & How often primary caregiver smokes? & Youth & 0.41 (0.36, 0.46) & 0.37 (0.31, 0.42)\\
\addlinespace[6pt]
ch6waist & Constructed—Teen waist from multiple measures (cm) & Primary caregiver & 0.40 (0.29, 0.50) & 0.48 (0.38, 0.58)\\
\addlinespace[6pt]
p6j34 & Youth gets free/reduced price school breakfast & Primary caregiver & 0.39 (0.35, 0.44) & 0.30 (0.25, 0.35)\\
\addlinespace[6pt]
k6c42 & Number of full siblings living with youth & Youth & 0.39 (0.34, 0.45) & 0.25 (0.20, 0.31)\\
\addlinespace[6pt]
cp6kids & Constructed—Primary caregiver household number of children (\textless{}18) & Primary caregiver & 0.35 (0.31, 0.40) & 0.22 (0.17, 0.26)\\
\addlinespace[6pt]
p6b11 & Doctor diagnosed youth with autism & Primary caregiver & 0.32 (0.16, 0.48) & 0.32 (0.16, 0.49)\\
\addlinespace[6pt]
p6j24 & Received household income from food stamps/SNAP in past year? & Primary caregiver & 0.32 (0.28, 0.36) & 0.24 (0.20, 0.28)\\
\addlinespace[6pt]
p6\_scale\_parentingAggrava\newline{}tion & Theme—Aggravation in Parenting & Primary caregiver & 0.31 (0.26, 0.35) & 0.27 (0.23, 0.32)\\
\addlinespace[6pt]
cp6hhsize & Number of people in household—Includes primary caregiver and focal child & Primary caregiver & 0.31 (0.26, 0.35) & 0.24 (0.19, 0.29)\\
\addlinespace[6pt]
p6\_scale\_CBCL\_attention & Theme—Child Behavior Problems (CBCL): Scale—Attention Problems & Primary caregiver & 0.30 (0.25, 0.35) & 0.22 (0.18, 0.27)\\
\addlinespace[6pt]
p6i31 & My religious faith is an important guide for my daily life & Primary caregiver & 0.30 (0.24, 0.35) & 0.22 (0.16, 0.27)\\
\addlinespace[6pt]
p6b10 & Doctor diagnosed youth with ADD/ADHD & Primary caregiver & 0.29 (0.22, 0.36) & 0.35 (0.28, 0.42)\\
\addlinespace[6pt]
p6h2 & Primary caregiver's description of own health in past year & Primary caregiver & 0.28 (0.24, 0.32) & 0.22 (0.18, 0.26)\\
\addlinespace[6pt]
cp6pcohb & Constructed—Biological primary caregiver cohabiting with biological parent (unmarried) & Primary caregiver & 0.28 (0.17, 0.38) & 0.32 (0.23, 0.42)\\
\addlinespace[6pt]
p6\_scale\_CBCL\_aggressive & Theme—Child Behavior Problems (CBCL): Scale—Aggressive Behavior & Primary caregiver & 0.27 (0.22, 0.31) & 0.20 (0.15, 0.25)\\
\addlinespace[6pt]
p6c4 & Number of schools youth attended since first grade & Primary caregiver & 0.27 (0.22, 0.31) & 0.22 (0.17, 0.26)\\
\addlinespace[6pt]
p6\_scale\_CBCL & Theme—Child Behavior Problems (CBCL) & Primary caregiver & 0.26 (0.22, 0.31) & 0.19 (0.14, 0.24)\\
\addlinespace[6pt]
p6i32 & How often you attend religious services? & Primary caregiver & 0.26 (0.22, 0.30) & 0.23 (0.18, 0.27)\\
\addlinespace[6pt]
p6\_scale\_CBCL\_externalizi\newline{}ng & Theme—Child Behavior Problems (CBCL): Scale—Externalizing Behaviors & Primary caregiver & 0.25 (0.21, 0.30) & 0.19 (0.14, 0.24)\\
\addlinespace[6pt]
p6\_scale\_materialHardship\newline{}\_sincePastWave & Theme—Material Hardship: Scale—Since Past Wave & Primary caregiver & 0.25 (0.21, 0.30) & 0.19 (0.15, 0.24)\\
\addlinespace[6pt]
p6b46 & Youth can't concentrate, can't pay attention for long & Primary caregiver & 0.24 (0.20, 0.28) & 0.17 (0.13, 0.22)\\
\addlinespace[6pt]
p6h78 & Number of other people in household who smoke & Primary caregiver & 0.24 (0.20, 0.28) & 0.20 (0.16, 0.24)\\
\addlinespace[6pt]
p6k63 & Own a vehicle, such as car, truck, or van? & Primary caregiver & 0.23 (0.19, 0.28) & 0.17 (0.12, 0.21)\\
\addlinespace[6pt]
p6i27 & Could count on someone to co-sign a bank loan for \$1000? & Primary caregiver & 0.22 (0.18, 0.27) & 0.16 (0.12, 0.20)\\
\addlinespace[6pt]
p6d34 & Taking care of children is more work than pleasure & Primary caregiver & 0.22 (0.18, 0.27) & 0.17 (0.12, 0.21)\\
\addlinespace[6pt]
p6d32 & Being a parent is harder than I thought it would be & Primary caregiver & 0.22 (0.18, 0.26) & 0.14 (0.10, 0.18)\\
\addlinespace[6pt]
p6b39 & Youth is disobedient at school & Primary caregiver & 0.21 (0.16, 0.26) & 0.14 (0.10, 0.18)\\
\addlinespace[6pt]
p6d33 & I feel trapped by my responsibilities as a parent & Primary caregiver & 0.20 (0.16, 0.25) & 0.14 (0.09, 0.18)\\
\addlinespace[6pt]
p6j33 & Filed a federal tax return for previous tax year? & Primary caregiver & 0.20 (0.16, 0.25) & 0.08 (0.04, 0.12)\\
\addlinespace[6pt]
p6d36 & Rating of self as parent/guardian & Primary caregiver & 0.20 (0.16, 0.24) & 0.17 (0.13, 0.21)\\
\addlinespace[6pt]
p6c21 & Youth ever been suspended/expelled? & Primary caregiver & 0.20 (0.16, 0.23) & 0.12 (0.08, 0.15)\\
\addlinespace[6pt]
p6k67 & Have a credit card? & Primary caregiver & 0.19 (0.16, 0.23) & 0.13 (0.09, 0.16)\\
\addlinespace[6pt]
p6d35 & I feel tired/worn out/exhausted from raising a family & Primary caregiver & 0.19 (0.16, 0.23) & 0.15 (0.12, 0.19)\\
\addlinespace[6pt]
p6b48 & Youth is impulsive or acts without thinking & Primary caregiver & 0.19 (0.14, 0.23) & 0.16 (0.11, 0.20)\\
\addlinespace[6pt]
p6\_scale\_materialHardship\newline{}\_pastYear & Theme—Material Hardship: Scale—Past Year & Primary caregiver & 0.19 (0.15, 0.23) & 0.15 (0.10, 0.19)\\
\addlinespace[6pt]
p6k73 & Have any savings? & Primary caregiver & 0.19 (0.15, 0.23) & 0.12 (0.08, 0.15)\\
\addlinespace[6pt]
p6d6 & Number of televisions in home & Primary caregiver & 0.19 (0.14, 0.23) & 0.18 (0.13, 0.22)\\
\addlinespace[6pt]
p6k6 & Did any regular work for pay in last week? & Primary caregiver & 0.18 (0.14, 0.22) & 0.13 (0.09, 0.17)\\
\addlinespace[6pt]
cp6adult & Constructed—Primary caregiver household number of adults ($\geq$18) & Primary caregiver & 0.18 (0.14, 0.23) & 0.09 (0.06, 0.12)\\
\addlinespace[6pt]
p6b14 & Youth had eczema/skin allergy in past year & Primary caregiver & 0.18 (0.13, 0.23) & 0.17 (0.12, 0.22)\\
\addlinespace[6pt]
p6b47 & Youth can't sit still, is restless or hyperactive & Primary caregiver & 0.18 (0.13, 0.23) & 0.16 (0.12, 0.21)\\
\addlinespace[6pt]
p6b44 & Youth has temper tantrums or a hot temper & Primary caregiver & 0.18 (0.14, 0.22) & 0.13 (0.09, 0.17)\\
\addlinespace[6pt]
p6i23 & Could count on someone to loan \$200 during next year? & Primary caregiver & 0.17 (0.12, 0.22) & 0.12 (0.08, 0.16)\\
\addlinespace[6pt]
p6i25 & Could count on someone to provide place to live? & Primary caregiver & 0.17 (0.12, 0.22) & 0.12 (0.08, 0.17)\\
\addlinespace[6pt]
p6\_scale\_CBCL\_ruleBreakin\newline{}g & Theme—Child Behavior Problems (CBCL): Scale—Rule-Breaking Behavior & Primary caregiver & 0.17 (0.13, 0.21) & 0.11 (0.07, 0.15)\\
\addlinespace[6pt]
k6\_scale\_GPA & Theme—GPA & Youth & 0.17 (0.13, 0.21) & 0.07 (0.04, 0.11)\\
\addlinespace[6pt]
p6\_scale\_CBCL\_internalizi\newline{}ng & Theme—Child Behavior Problems (CBCL): Scale—Internalizing Behaviors & Primary caregiver & 0.16 (0.12, 0.21) & 0.07 (0.03, 0.11)\\
\addlinespace[6pt]
p6\_scale\_neighborhoodColl\newline{}ectiveEfficacy\_PCG & Theme—Neighborhood Collective Efficacy: Scale—Primary Caregiver Perception & Primary caregiver & 0.16 (0.12, 0.19) & 0.13 (0.09, 0.16)\\
\addlinespace[6pt]
p6d28 & Shouted or swore at youth in past year & Primary caregiver & 0.16 (0.12, 0.19) & 0.11 (0.08, 0.14)\\
\addlinespace[6pt]
cp6pmarp & Constructed—Biological primary caregiver married to new partner & Primary caregiver & 0.15 (0.10, 0.21) & 0.13 (0.08, 0.18)\\
\addlinespace[6pt]
o6a2 & General condition of buildings on block & Interviewer & 0.15 (0.09, 0.22) & 0.13 (0.07, 0.19)\\
\addlinespace[6pt]
p6\_scale\_CHAOS\_PCG & Theme—Environmental Confusion (CHAOS): Scale—Primary Caregiver Perception & Primary caregiver & 0.15 (0.12, 0.19) & 0.13 (0.09, 0.16)\\
\addlinespace[6pt]
p6b12 & Doctor diagnosed youth with other learning disability & Primary caregiver & 0.15 (0.09, 0.22) & 0.11 (0.05, 0.16)\\
\addlinespace[6pt]
p6j1 & Moved since last interview? & Primary caregiver & 0.15 (0.11, 0.19) & 0.12 (0.09, 0.15)\\
\addlinespace[6pt]
p6\_scale\_CBCL\_anxiousDepr\newline{}essed & Theme—Child Behavior Problems (CBCL): Scale—Anxious/Depressed Behavior & Primary caregiver & 0.15 (0.11, 0.19) & 0.07 (0.03, 0.11)\\
\addlinespace[6pt]
p6b59 & Youth argues a lot & Primary caregiver & 0.15 (0.11, 0.19) & 0.08 (0.04, 0.12)\\
\addlinespace[6pt]
p6h1 & Primary caregiver's description of satisfaction with own life & Primary caregiver & 0.15 (0.11, 0.19) & 0.12 (0.08, 0.16)\\
\addlinespace[6pt]
p6i29 & Someone who would pay for youth's activities if unable to afford? & Primary caregiver & 0.15 (0.11, 0.19) & 0.10 (0.06, 0.13)\\
\addlinespace[6pt]
p6b38 & Youth is disobedient at home & Primary caregiver & 0.15 (0.11, 0.18) & 0.11 (0.08, 0.15)\\
\addlinespace[6pt]
p6b1 & Primary caregiver's description of youth's health & Primary caregiver & 0.14 (0.10, 0.18) & 0.13 (0.10, 0.17)\\
\addlinespace[6pt]
p6b58 & Youth talks too much & Primary caregiver & 0.14 (0.11, 0.18) & 0.13 (0.09, 0.16)\\
\addlinespace[6pt]
p6i26 & Could count on someone to help with emergency child care? & Primary caregiver & 0.14 (0.10, 0.18) & 0.04 (0.01, 0.08)\\
\addlinespace[6pt]
p6i11 & Gangs are a problem in this neighborhood & Primary caregiver & 0.14 (0.10, 0.18) & 0.12 (0.08, 0.16)\\
\addlinespace[6pt]
p6\_scale\_neighborhoodColl\newline{}ectiveEfficacy\_cohesionAn\newline{}dTrustPCG & Theme—Neighborhood Collective Efficacy: Scale—Cohesion and Trust, Primary Caregiver Perception & Primary caregiver & 0.14 (0.10, 0.18) & 0.13 (0.09, 0.16)\\
\addlinespace[6pt]
p6h79a & Ever drank an alcoholic beverage & Primary caregiver & 0.14 (0.09, 0.19) & 0.08 (0.06, 0.11)\\
\addlinespace[6pt]
p6b26 & Youth takes doctor-prescribed medication? & Primary caregiver & 0.14 (0.10, 0.17) & 0.11 (0.08, 0.15)\\
\addlinespace[6pt]
p6i12 & Afraid to let youth outside because of neighborhood violence? & Primary caregiver & 0.14 (0.10, 0.17) & 0.11 (0.07, 0.15)\\
\addlinespace[6pt]
k6b18 & Ever taken honors courses in school? & Youth & 0.14 (0.10, 0.17) & 0.11 (0.08, 0.14)\\
\addlinespace[6pt]
p6d37 & How close do you feel to youth? & Primary caregiver & 0.13 (0.10, 0.17) & 0.08 (0.05, 0.12)\\
\addlinespace[6pt]
p6\_scale\_CIDISFDepression\newline{}\_liberalSincePastWaveBina\newline{}ry & Theme—Mental Health Depression (CIDI-SF): Scale—Liberal, Since Past Wave, Binary & Primary caregiver & 0.13 (0.09, 0.17) & 0.07 (0.04, 0.09)\\
\addlinespace[6pt]
k6c43 & Number of half siblings living with youth & Youth & 0.13 (0.08, 0.17) & 0.06 (0.02, 0.09)\\
\addlinespace[6pt]
p6k70 & Have other loans (not house, vehicle, credit card)? & Primary caregiver & 0.13 (0.09, 0.17) & 0.06 (0.04, 0.09)\\
\addlinespace[6pt]
p6i30 & Special person you feel close with and can depend on? & Primary caregiver & 0.13 (0.08, 0.17) & 0.08 (0.04, 0.12)\\
\addlinespace[6pt]
p6\_scale\_CBCL\_withdrawn & Theme—Child Behavior Problems (CBCL): Scale—Withdrawn & Primary caregiver & 0.12 (0.09, 0.16) & 0.06 (0.03, 0.10)\\
\addlinespace[6pt]
p6h3 & Have serious health problem that limits work? & Primary caregiver & 0.12 (0.07, 0.18) & 0.13 (0.09, 0.18)\\
\addlinespace[6pt]
p6\_scale\_CIDISFDepression\newline{}\_liberalSincePastWave & Theme—Mental Health Depression (CIDI-SF): Scale—Liberal, Since Past Wave & Primary caregiver & 0.12 (0.08, 0.16) & 0.06 (0.03, 0.09)\\
\addlinespace[6pt]
p6i1 & Number of families on block known well & Primary caregiver & 0.12 (0.09, 0.16) & 0.09 (0.06, 0.13)\\
\addlinespace[6pt]
k6f1b & How parents would feel about you having sex? & Youth & 0.12 (0.08, 0.16) & 0.08 (0.06, 0.11)\\
\addlinespace[6pt]
p6j27 & Received household income from SSI/SSDI in past year? & Primary caregiver & 0.12 (0.07, 0.18) & 0.17 (0.13, 0.22)\\
\addlinespace[6pt]
p6e31 & Number of non-biological children in household (partners' excluded) & Primary caregiver & 0.12 (0.01, 0.24) & 0.13 (0.02, 0.24)\\
\addlinespace[6pt]
p6i7 & People around here are willing to help their neighbors & Primary caregiver & 0.12 (0.09, 0.15) & 0.09 (0.05, 0.12)\\
\addlinespace[6pt]
k6e7f & The police create more problems than they solve & Youth & 0.12 (0.09, 0.15) & 0.10 (0.07, 0.13)\\
\addlinespace[6pt]
p6b51 & Youth lies or cheats & Primary caregiver & 0.11 (0.08, 0.15) & 0.08 (0.05, 0.11)\\
\addlinespace[6pt]
p6b64 & Youth swears or uses obscene language & Primary caregiver & 0.11 (0.08, 0.15) & 0.09 (0.05, 0.12)\\
\addlinespace[6pt]
p6i13 & Frequency saw person get hit, slapped, punched in past year & Primary caregiver & 0.11 (0.06, 0.17) & 0.14 (0.08, 0.19)\\
\addlinespace[6pt]
p6d14 & Home is a real zoo & Primary caregiver & 0.11 (0.07, 0.15) & 0.10 (0.06, 0.13)\\
\addlinespace[6pt]
p6d17 & Atmosphere in house is calm & Primary caregiver & 0.11 (0.08, 0.14) & 0.07 (0.04, 0.10)\\
\addlinespace[6pt]
k6d63 & How many of your friends do you think have had sex? & Youth & 0.11 (0.08, 0.14) & 0.06 (0.04, 0.09)\\
\addlinespace[6pt]
p6b5 & Doctor diagnosed youth with depression/anxiety & Primary caregiver & 0.11 (0.06, 0.16) & 0.04 (0.02, 0.07)\\
\addlinespace[6pt]
p6b66 & Youth is unhappy, sad, or depressed & Primary caregiver & 0.11 (0.07, 0.14) & 0.02 (-0.01, 0.05)\\
\addlinespace[6pt]
p6i10 & People in this neighborhood do not share the same values & Primary caregiver & 0.11 (0.07, 0.14) & 0.07 (0.04, 0.09)\\
\addlinespace[6pt]
p6b57 & Youth is unusually loud & Primary caregiver & 0.11 (0.07, 0.14) & 0.08 (0.05, 0.11)\\
\addlinespace[6pt]
k6d38 & Days engage in physical activity for 30+ minutes in typical week & Youth & 0.11 (0.08, 0.14) & 0.04 (0.01, 0.06)\\
\addlinespace[6pt]
p6j41 & Did not pay utility bills in full in past year & Primary caregiver & 0.11 (0.07, 0.14) & 0.10 (0.07, 0.14)\\
\addlinespace[6pt]
p6b68 & Youth worries & Primary caregiver & 0.11 (0.07, 0.14) & 0.08 (0.05, 0.10)\\
\addlinespace[6pt]
p6b19 & Youth had trouble breathing/chest problem in past year & Primary caregiver & 0.10 (0.05, 0.16) & 0.10 (0.06, 0.14)\\
\addlinespace[6pt]
p6b37 & Youth destroys things belonging to family or others & Primary caregiver & 0.10 (0.05, 0.15) & 0.08 (0.04, 0.12)\\
\addlinespace[6pt]
p6d13 & Can't hear yourself think at home & Primary caregiver & 0.10 (0.07, 0.14) & 0.08 (0.06, 0.11)\\
\addlinespace[6pt]
p6i33 & Had transforming religious experience since last interview? & Primary caregiver & 0.10 (0.07, 0.13) & 0.08 (0.05, 0.11)\\
\addlinespace[6pt]
p6i8 & This is a close-knit neighborhood & Primary caregiver & 0.10 (0.07, 0.14) & 0.09 (0.06, 0.12)\\
\addlinespace[6pt]
p6b35 & Youth is cruel, bullies, or shows meanness to others & Primary caregiver & 0.10 (0.06, 0.14) & 0.07 (0.04, 0.11)\\
\addlinespace[6pt]
cp6md\_case\_lib & Constructed—Primary caregiver meets liberal depression criteria in past year & Primary caregiver & 0.10 (0.06, 0.14) & 0.07 (0.04, 0.10)\\
\addlinespace[6pt]
p6\_scale\_neighborhoodColl\newline{}ectiveEfficacy\_socCtrlPCG & Theme—Neighborhood Collective Efficacy: Scale—Social Control, Primary Caregiver Perception & Primary caregiver & 0.10 (0.07, 0.13) & 0.09 (0.05, 0.12)\\
\addlinespace[6pt]
p6b43 & Youth is stubborn, sullen, or irritable & Primary caregiver & 0.10 (0.06, 0.14) & 0.06 (0.02, 0.09)\\
\addlinespace[6pt]
k6b22e & Spend time on religious services & Youth & 0.10 (0.06, 0.14) & 0.04 (0.02, 0.07)\\
\addlinespace[6pt]
p6b53 & Youth is too fearful or anxious & Primary caregiver & 0.10 (0.07, 0.13) & 0.02 (-0.01, 0.04)\\
\addlinespace[6pt]
p6k60 & Given/loaned money to friends/relatives in past year? & Primary caregiver & 0.09 (0.06, 0.13) & 0.06 (0.04, 0.09)\\
\addlinespace[6pt]
o6a9 & Rate condition of street in front of home & Interviewer & 0.09 (0.03, 0.15) & 0.02 (-0.04, 0.07)\\
\addlinespace[6pt]
p6\_scale\_CIDISFDepression\newline{}\_liberalPastYear & Theme—Mental Health Depression (CIDI-SF): Scale—Liberal, Past Year & Primary caregiver & 0.09 (0.06, 0.13) & 0.07 (0.03, 0.10)\\
\addlinespace[6pt]
p6c27 & How likely youth will graduate from college? & Primary caregiver & 0.09 (0.06, 0.12) & 0.05 (0.02, 0.08)\\
\addlinespace[6pt]
k6f1a & How parents would feel about you dating or going out? & Youth & 0.09 (0.06, 0.12) & 0.09 (0.06, 0.12)\\
\addlinespace[6pt]
p6b65 & Youth is underactive, slow moving, or lacks energy & Primary caregiver & 0.09 (0.05, 0.12) & 0.04 (0.01, 0.07)\\
\addlinespace[6pt]
k6d27 & Share a bed or bedroom with someone else? & Youth & 0.08 (0.05, 0.11) & 0.07 (0.04, 0.10)\\
\addlinespace[6pt]
p6b18 & Youth had stuttering or stammering problem in past year & Primary caregiver & 0.08 (-0.06, 0.23) & 0.17 (0.07, 0.27)\\
\addlinespace[6pt]
k6d57 & How often primary caregiver drinks alcohol? & Youth & 0.08 (0.05, 0.12) & 0.07 (0.04, 0.10)\\
\addlinespace[6pt]
p6i3 & Neighbors would get involved if children spray paint buildings with graffiti & Primary caregiver & 0.08 (0.05, 0.11) & 0.06 (0.03, 0.09)\\
\addlinespace[6pt]
k6d35f & Friends/family encourage diet to lose/control weight & Youth & 0.08 (0.05, 0.11) & 0.07 (0.04, 0.09)\\
\addlinespace[6pt]
k6e2e & Gangs are a problem in this neighborhood & Youth & 0.08 (0.05, 0.11) & 0.02 (-0.01, 0.05)\\
\addlinespace[6pt]
p6d2 & Talked about youth's day with youth in past month & Primary caregiver & 0.08 (0.05, 0.11) & 0.00 (-0.02, 0.03)\\
\addlinespace[6pt]
p6d16 & Usually able to stay on top of things & Primary caregiver & 0.08 (0.05, 0.11) & 0.03 (0.01, 0.06)\\
\addlinespace[6pt]
k6b14 & How many friends will likely graduate from high school? & Youth & 0.08 (0.05, 0.10) & 0.05 (0.03, 0.07)\\
\addlinespace[6pt]
p6j18 & Received household income from welfare/TANF in past year? & Primary caregiver & 0.08 (0.03, 0.12) & 0.07 (0.04, 0.11)\\
\addlinespace[6pt]
p6i9 & People in this neighborhood generally don't get along with each other & Primary caregiver & 0.08 (0.04, 0.11) & 0.05 (0.02, 0.07)\\
\addlinespace[6pt]
p6c23 & Youth required to attend summer school since last interview? & Primary caregiver & 0.08 (0.05, 0.10) & 0.02 (0.00, 0.04)\\
\addlinespace[6pt]
p6j43 & Borrowed money from friends/family to help pay bills in past year & Primary caregiver & 0.07 (0.04, 0.10) & 0.04 (0.01, 0.06)\\
\addlinespace[6pt]
p6i2 & Neighbors would get involved if children skip school and hang out on street & Primary caregiver & 0.07 (0.04, 0.10) & 0.07 (0.04, 0.10)\\
\addlinespace[6pt]
p6b45 & Youth threatens people & Primary caregiver & 0.07 (0.03, 0.11) & 0.01 (-0.03, 0.05)\\
\addlinespace[6pt]
p6\_scale\_conflictTactics\_\newline{}nonviolentDisciplinePCGPe\newline{}rception & Theme—Conflict Tactics: Scale—Non-violent Discipline, Primary Caregiver Perception & Primary caregiver & 0.07 (0.04, 0.10) & 0.05 (0.03, 0.07)\\
\addlinespace[6pt]
p6b52 & Youth is nervous, high-strung, or tense & Primary caregiver & 0.07 (0.04, 0.10) & 0.05 (0.02, 0.08)\\
\addlinespace[6pt]
p6d27 & Took away privileges or grounded youth in past year & Primary caregiver & 0.07 (0.03, 0.10) & 0.03 (0.01, 0.05)\\
\addlinespace[6pt]
k6d39 & Days participate in vigorous physical activity in typical week & Youth & 0.07 (0.03, 0.10) & 0.08 (0.05, 0.11)\\
\addlinespace[6pt]
p6k2 & Completed any training or schooling since last interview? & Primary caregiver & 0.06 (0.03, 0.10) & 0.06 (0.04, 0.08)\\
\addlinespace[6pt]
k6d3 & Youth's description of own health & Youth & 0.06 (0.04, 0.09) & 0.02 (0.00, 0.04)\\
\addlinespace[6pt]
p6b41 & Youth gets in many fights & Primary caregiver & 0.06 (0.02, 0.11) & 0.03 (-0.01, 0.06)\\
\addlinespace[6pt]
k6\_scale\_legalCynicism & Theme—Legal Cynicism & Youth & 0.06 (0.04, 0.09) & 0.02 (0.00, 0.04)\\
\addlinespace[6pt]
k6\_scale\_impulsivity & Theme—Impulsivity & Youth & 0.06 (0.03, 0.10) & 0.05 (0.03, 0.07)\\
\addlinespace[6pt]
o6c6 & Interior—Dark inside home & Interviewer & 0.06 (0.01, 0.11) & 0.03 (-0.01, 0.06)\\
\addlinespace[6pt]
p6b40 & Youth feels worthless or inferior & Primary caregiver & 0.06 (0.03, 0.09) & 0.01 (-0.01, 0.04)\\
\addlinespace[6pt]
p6b50 & Youth hangs around with others who get in trouble & Primary caregiver & 0.06 (0.03, 0.10) & 0.01 (-0.02, 0.03)\\
\addlinespace[6pt]
k6b15 & How many friends will likely graduate from college? & Youth & 0.06 (0.03, 0.09) & 0.03 (0.01, 0.05)\\
\addlinespace[6pt]
k6d2aj & I often get into trouble because I don't think before I act & Youth & 0.06 (0.04, 0.09) & 0.04 (0.02, 0.06)\\
\addlinespace[6pt]
p6j39 & Did not pay rent/mortgage in full in past year & Primary caregiver & 0.06 (0.03, 0.09) & 0.07 (0.04, 0.10)\\
\addlinespace[6pt]
p6b13 & Youth had food/digestive allergy in past year & Primary caregiver & 0.06 (-0.03, 0.15) & 0.10 (0.03, 0.16)\\
\addlinespace[6pt]
p6j37 & Received free food or meals in past year & Primary caregiver & 0.06 (0.03, 0.09) & 0.05 (0.02, 0.09)\\
\addlinespace[6pt]
k6e7a & I have a great deal of respect for the police & Youth & 0.06 (0.04, 0.09) & 0.00 (-0.02, 0.02)\\
\addlinespace[6pt]
k6d44 & Cigarettes easily available to you in home? & Youth & 0.06 (0.03, 0.10) & 0.07 (0.04, 0.10)\\
\addlinespace[6pt]
p6b49 & Youth doesn't seem to feel guilty after misbehaving & Primary caregiver & 0.06 (0.03, 0.09) & 0.03 (0.00, 0.05)\\
\addlinespace[6pt]
p6j47 & Telephone disconnected because not enough money in past year & Primary caregiver & 0.06 (0.02, 0.10) & 0.03 (0.00, 0.06)\\
\addlinespace[6pt]
k6d61d & Gotten into a serious physical fight & Youth & 0.06 (0.03, 0.08) & 0.02 (0.00, 0.04)\\
\addlinespace[6pt]
p6b55 & Youth has trouble sleeping & Primary caregiver & 0.06 (0.02, 0.09) & 0.07 (0.03, 0.10)\\
\addlinespace[6pt]
p6d4 & Helped youth with homework in past month & Primary caregiver & 0.06 (0.03, 0.08) & 0.02 (0.00, 0.04)\\
\addlinespace[6pt]
p6d1 & Talked about current events with youth in past month & Primary caregiver & 0.06 (0.03, 0.09) & 0.06 (0.03, 0.09)\\
\addlinespace[6pt]
p6i6 & Neighbors would get involved if fire station was threatened and budget cut & Primary caregiver & 0.06 (0.03, 0.09) & 0.04 (0.01, 0.07)\\
\addlinespace[6pt]
p6k4 & Taken any classes to improve job skills since last interview? & Primary caregiver & 0.06 (0.03, 0.08) & 0.01 (-0.01, 0.02)\\
\addlinespace[6pt]
p6j38 & Hungry because could not afford food in past year & Primary caregiver & 0.06 (0.02, 0.09) & 0.01 (-0.01, 0.03)\\
\addlinespace[6pt]
o6a1 & Garbage, litter, broken glass in street & Interviewer & 0.06 (0.00, 0.11) & 0.01 (-0.04, 0.06)\\
\addlinespace[6pt]
p6d29 & Hit or slapped youth in past year & Primary caregiver & 0.05 (0.02, 0.09) & 0.04 (0.01, 0.07)\\
\addlinespace[6pt]
k6d31 & Days in typical week youth eats vegetables or fruit & Youth & 0.05 (0.03, 0.08) & 0.04 (0.02, 0.06)\\
\addlinespace[6pt]
k6d2z & I often make up my mind without taking the time to consider & Youth & 0.05 (0.03, 0.08) & 0.04 (0.02, 0.06)\\
\addlinespace[6pt]
k6d37 & Days physically active for 60+ minutes in past week & Youth & 0.05 (0.02, 0.08) & 0.07 (0.04, 0.09)\\
\addlinespace[6pt]
p6d5 & Did activities with youth in past month & Primary caregiver & 0.05 (0.03, 0.08) & 0.02 (0.00, 0.03)\\
\addlinespace[6pt]
k6\_scale\_BSI18Anxiety & Theme—BSI 18 for Anxiety & Youth & 0.05 (0.02, 0.09) & 0.02 (0.00, 0.04)\\
\addlinespace[6pt]
k6b11 & How likely are you to graduate from college? & Youth & 0.05 (0.03, 0.08) & 0.03 (0.01, 0.06)\\
\addlinespace[6pt]
k6\_scale\_EPOCH\_perseveran\newline{}ce & Theme—Positive Adolescent Functioning (EPOCH): Scale—Perseverance & Youth & 0.05 (0.03, 0.07) & 0.01 (-0.01, 0.03)\\
\addlinespace[6pt]
k6b22a & Spend time on athletic or sports teams & Youth & 0.05 (0.02, 0.08) & 0.03 (0.01, 0.05)\\
\addlinespace[6pt]
k6e7d & Laws are made to be broken & Youth & 0.05 (0.02, 0.08) & 0.00 (-0.02, 0.02)\\
\addlinespace[6pt]
k6\_scale\_neighborhoodColl\newline{}ectiveEfficacy\_cohesionAn\newline{}dTrustKid & Theme—Neighborhood Collective Efficacy: Scale—Cohesion and Trust, Child Perception & Youth & 0.05 (0.03, 0.07) & 0.03 (0.01, 0.05)\\
\addlinespace[6pt]
k6e14 & Ever seen someone stopped by police in your neighborhood? & Youth & 0.05 (0.02, 0.08) & 0.02 (0.01, 0.04)\\
\addlinespace[6pt]
p6b56 & Youth clings to adults or too dependent & Primary caregiver & 0.05 (0.02, 0.08) & 0.02 (-0.01, 0.04)\\
\addlinespace[6pt]
k6e10 & Ever been stopped by the police? & Youth & 0.05 (0.02, 0.08) & 0.05 (0.03, 0.07)\\
\addlinespace[6pt]
p6b23 & Youth saw doctor for an illness in past year? & Primary caregiver & 0.05 (0.02, 0.08) & 0.01 (0.00, 0.03)\\
\addlinespace[6pt]
p6h37 & Felt worried/anxious for one month or longer in past year? & Primary caregiver & 0.05 (0.01, 0.09) & 0.04 (0.02, 0.06)\\
\addlinespace[6pt]
p6i19 & Received financial help or money from anyone in past year? & Primary caregiver & 0.05 (0.02, 0.08) & 0.03 (0.01, 0.05)\\
\addlinespace[6pt]
cp6md\_case\_con & Constructed—Primary caregiver meets conservative depression criteria in past year & Primary caregiver & 0.05 (0.01, 0.09) & 0.03 (0.00, 0.06)\\
\addlinespace[6pt]
k6\_scale\_delinquentBehavi\newline{}or\_self & Theme—Delinquent Behavior: Scale—Self & Youth & 0.05 (0.02, 0.07) & 0.00 (-0.02, 0.03)\\
\addlinespace[6pt]
k6\_scale\_CESD & Theme—CES-D for Depression & Youth & 0.05 (0.02, 0.07) & 0.01 (-0.01, 0.03)\\
\addlinespace[6pt]
cp6pcohp & Constructed—Biological primary caregiver cohabiting with new partner (unmarried) & Primary caregiver & 0.05 (0.02, 0.08) & 0.04 (0.01, 0.06)\\
\addlinespace[6pt]
k6d1b & I try to comfort others when they are upset & Youth & 0.05 (0.02, 0.07) & 0.00 (-0.01, 0.02)\\
\addlinespace[6pt]
p6\_scale\_CIDISFGAD\_sinceP\newline{}astWave & Theme—Mental Health for Generalized Anxiety Disorder (CIDI-SF): Scale—Since Past Wave & Primary caregiver & 0.05 (0.01, 0.08) & 0.05 (0.02, 0.08)\\
\addlinespace[6pt]
k6d36e & Hours per day plays games on computer/TV/device on weekdays & Youth & 0.05 (0.02, 0.07) & 0.04 (0.02, 0.06)\\
\addlinespace[6pt]
o6g2 & Youth's understanding of the questions & Interviewer & 0.04 (-0.01, 0.10) & 0.05 (0.00, 0.10)\\
\addlinespace[6pt]
p6b54 & Youth feels too guilty & Primary caregiver & 0.04 (0.02, 0.07) & 0.01 (-0.02, 0.03)\\
\addlinespace[6pt]
p6\_scale\_eviction & Theme—Eviction & Primary caregiver & 0.04 (0.01, 0.08) & 0.05 (0.01, 0.09)\\
\addlinespace[6pt]
k6\_scale\_extracurricularI\newline{}nvolvement & Theme—Adolescent Extracurricular and Community Involvement & Youth & 0.04 (0.02, 0.07) & 0.04 (0.02, 0.06)\\
\addlinespace[6pt]
k6c3 & Days per week you eat with family & Youth & 0.04 (0.02, 0.06) & 0.02 (0.01, 0.04)\\
\addlinespace[6pt]
k6f63 & Ever tried marijuana? & Youth & 0.04 (0.02, 0.07) & 0.04 (0.02, 0.05)\\
\addlinespace[6pt]
p6d38 & Talked to youth about sex or birth control? & Primary caregiver & 0.04 (0.02, 0.06) & 0.01 (-0.01, 0.02)\\
\addlinespace[6pt]
p6j59 & CPS contacted you with concerns since last interview? & Primary caregiver & 0.04 (0.01, 0.07) & 0.01 (-0.01, 0.03)\\
\addlinespace[6pt]
k6\_scale\_EPOCH\_optimism & Theme—Positive Adolescent Functioning (EPOCH): Scale—Optimism & Youth & 0.04 (0.02, 0.07) & 0.01 (0.00, 0.03)\\
\addlinespace[6pt]
k6b31c & Friends have trouble getting along with other students & Youth & 0.04 (0.02, 0.06) & 0.02 (0.00, 0.04)\\
\addlinespace[6pt]
k6d2r & The plans I make don't work out because I haven't gone over them & Youth & 0.04 (0.02, 0.06) & 0.03 (0.01, 0.05)\\
\addlinespace[6pt]
k6d2ak & I feel so restless I can't sit still & Youth & 0.04 (0.02, 0.06) & 0.01 (-0.01, 0.03)\\
\addlinespace[6pt]
p6c25 & Youth repeated any grades since last interview? & Primary caregiver & 0.04 (0.02, 0.06) & -0.01 (-0.03, 0.01)\\
\addlinespace[6pt]
p6i4 & Neighbors would get involved if children show disrespect to an adult & Primary caregiver & 0.04 (0.02, 0.07) & 0.04 (0.02, 0.07)\\
\addlinespace[6pt]
p6h86 & Ever used illegal drugs in past year? & Primary caregiver & 0.04 (-0.02, 0.10) & 0.03 (0.00, 0.07)\\
\addlinespace[6pt]
k6d1a & I understand others' feelings like when they are happy, sad, or mad & Youth & 0.04 (0.01, 0.06) & 0.05 (0.02, 0.07)\\
\addlinespace[6pt]
k6b31a & Friends work hard in school & Youth & 0.04 (0.02, 0.06) & 0.01 (-0.01, 0.03)\\
\addlinespace[6pt]
p6i14 & Frequency saw person attacked with weapon in past year & Primary caregiver & 0.04 (-0.02, 0.10) & -0.01 (-0.07, 0.04)\\
\addlinespace[6pt]
k6e2d & People in this neighborhood do not share the same values & Youth & 0.04 (0.01, 0.06) & 0.00 (-0.02, 0.02)\\
\addlinespace[6pt]
k6b31b & Friends have trouble getting along with their teachers & Youth & 0.04 (0.02, 0.06) & 0.01 (0.00, 0.03)\\
\addlinespace[6pt]
p6i15 & Frequency saw person shot at in past year & Primary caregiver & 0.04 (0.01, 0.07) & 0.04 (-0.01, 0.09)\\
\addlinespace[6pt]
k6d8 & Physical development earlier or later than most? & Youth & 0.04 (0.01, 0.06) & 0.05 (0.02, 0.07)\\
\addlinespace[6pt]
k6b22b & Spend time on group performance activities & Youth & 0.04 (0.02, 0.06) & 0.03 (0.01, 0.05)\\
\addlinespace[6pt]
k6e4c & I feel unsafe walking around my neighborhood at night & Youth & 0.04 (0.01, 0.06) & 0.01 (0.00, 0.03)\\
\addlinespace[6pt]
k6d2t & I get suddenly scared for no reason & Youth & 0.04 (0.02, 0.06) & 0.01 (-0.01, 0.03)\\
\addlinespace[6pt]
k6d59 & Alcohol easily available to you in home? & Youth & 0.04 (0.00, 0.07) & 0.01 (-0.02, 0.04)\\
\addlinespace[6pt]
k6e2c & People in this neighborhood generally don't get along with each other & Youth & 0.04 (0.01, 0.06) & 0.01 (-0.01, 0.03)\\
\addlinespace[6pt]
k6d35c & Friends/family encourage exercise to lose/control weight & Youth & 0.04 (0.01, 0.06) & 0.04 (0.02, 0.06)\\
\addlinespace[6pt]
p6i5 & Neighbors would get involved if fight broke out in front of house/building & Primary caregiver & 0.04 (0.00, 0.07) & 0.05 (0.02, 0.08)\\
\addlinespace[6pt]
k6e4a & There is so much traffic on streets that walking is difficult & Youth & 0.04 (0.01, 0.06) & -0.01 (-0.03, 0.01)\\
\addlinespace[6pt]
p6d8 & Youth has regular access to internet at home? & Primary caregiver & 0.04 (0.00, 0.07) & 0.02 (0.00, 0.03)\\
\addlinespace[6pt]
k6\_scale\_CHAOS\_kid & Theme—Environmental Confusion (CHAOS): Scale—Child Perception & Youth & 0.04 (0.02, 0.05) & 0.01 (-0.01, 0.03)\\
\addlinespace[6pt]
k6d32 & Days in typical week youth eats fast food & Youth & 0.03 (0.01, 0.06) & 0.04 (0.02, 0.06)\\
\addlinespace[6pt]
k6d2n & I feel sad & Youth & 0.03 (0.01, 0.06) & -0.01 (-0.03, 0.01)\\
\addlinespace[6pt]
k6d1l & I report accidents to appropriate persons & Youth & 0.03 (0.01, 0.06) & 0.02 (0.00, 0.03)\\
\addlinespace[6pt]
o6g3 & Youth's ability to articulate answers & Interviewer & 0.03 (-0.03, 0.10) & 0.01 (-0.04, 0.06)\\
\addlinespace[6pt]
k6d1h & I show interest in a variety of things & Youth & 0.03 (0.01, 0.06) & 0.05 (0.02, 0.07)\\
\addlinespace[6pt]
k6f4 & Ever dated someone? & Youth & 0.03 (0.00, 0.06) & 0.03 (0.01, 0.05)\\
\addlinespace[6pt]
k6d29 & Days during school week youth eats breakfast & Youth & 0.03 (0.01, 0.06) & 0.03 (0.01, 0.05)\\
\addlinespace[6pt]
o6c7 & Interior—Crowded inside home & Interviewer & 0.03 (-0.03, 0.09) & 0.02 (0.00, 0.04)\\
\addlinespace[6pt]
p6d31 & Nights during week family eats dinner together & Primary caregiver & 0.03 (0.00, 0.06) & 0.06 (0.03, 0.08)\\
\addlinespace[6pt]
p6\_scale\_parentalMonitori\newline{}ng\_PCG & Theme—Parental Monitoring: Scale—Primary Caregiver-Perceived Parental Monitoring & Primary caregiver & 0.03 (0.01, 0.05) & 0.02 (0.00, 0.04)\\
\addlinespace[6pt]
p6b16 & Youth had frequent headaches/migraines in past year & Primary caregiver & 0.03 (0.00, 0.06) & 0.03 (0.00, 0.06)\\
\addlinespace[6pt]
p6b36 & Youth cries a lot & Primary caregiver & 0.03 (0.00, 0.07) & 0.02 (0.00, 0.04)\\
\addlinespace[6pt]
k6d1j & I am liked by others & Youth & 0.03 (0.01, 0.05) & -0.01 (-0.03, 0.01)\\
\addlinespace[6pt]
k6d33 & Number of non-diet sweetened drinks in typical day & Youth & 0.03 (0.01, 0.05) & -0.03 (-0.05, -0.01)\\
\addlinespace[6pt]
k6\_scale\_EPOCH\_happiness & Theme—Positive Adolescent Functioning (EPOCH): Scale—Happiness & Youth & 0.03 (0.01, 0.05) & -0.01 (-0.03, 0.01)\\
\addlinespace[6pt]
k6d2c & I feel I cannot shake off the blues, even with help & Youth & 0.03 (0.01, 0.05) & 0.04 (0.02, 0.06)\\
\addlinespace[6pt]
p6j36 & Youth gets free/reduced price snacks/dinner in after school program & Primary caregiver & 0.03 (0.00, 0.06) & 0.01 (0.00, 0.02)\\
\addlinespace[6pt]
k6d1k & I invite others to my home & Youth & 0.03 (0.01, 0.05) & 0.01 (-0.01, 0.02)\\
\addlinespace[6pt]
k6d21 & How often snore during sleep? & Youth & 0.03 (0.00, 0.05) & 0.04 (0.02, 0.06)\\
\addlinespace[6pt]
cp6gad & Constructed—Primary caregiver meets anxious criteria in last year (CIDI) & Primary caregiver & 0.03 (0.00, 0.05) & 0.02 (0.00, 0.04)\\
\addlinespace[6pt]
k6c4e & The atmosphere in your house is calm & Youth & 0.03 (0.01, 0.04) & 0.03 (0.01, 0.04)\\
\addlinespace[6pt]
k6b22d & Spend time on school activities & Youth & 0.02 (0.01, 0.04) & 0.01 (-0.01, 0.02)\\
\addlinespace[6pt]
k6e7c & There are no right or wrong ways to make money & Youth & 0.02 (0.00, 0.05) & 0.02 (0.00, 0.04)\\
\addlinespace[6pt]
p6d12 & Youth has desk or table to do homework? & Primary caregiver & 0.02 (0.00, 0.05) & 0.03 (0.01, 0.05)\\
\addlinespace[6pt]
p6d26 & Explained why something youth did was wrong in past year & Primary caregiver & 0.02 (0.00, 0.05) & 0.03 (0.01, 0.05)\\
\addlinespace[6pt]
p6d25 & How often knows what youth spends money on? & Primary caregiver & 0.02 (0.00, 0.04) & 0.00 (-0.02, 0.02)\\
\addlinespace[6pt]
k6\_scale\_socialSkills & Theme—“Adolescent Social Skills” and “Adaptive Social Behavior Inventory” & Youth & 0.02 (0.00, 0.05) & 0.01 (-0.01, 0.03)\\
\addlinespace[6pt]
k6d2ac & I feel depressed & Youth & 0.02 (0.00, 0.04) & 0.00 (-0.02, 0.02)\\
\addlinespace[6pt]
p6b20 & Health problems limit youth's usual activities? & Primary caregiver & 0.02 (-0.02, 0.06) & 0.01 (0.00, 0.03)\\
\addlinespace[6pt]
p6b69 & Youth ever been stopped by police? & Primary caregiver & 0.02 (-0.02, 0.06) & 0.02 (0.00, 0.03)\\
\addlinespace[6pt]
k6d2d & I have spells of terror or panic & Youth & 0.02 (0.00, 0.04) & 0.00 (-0.02, 0.01)\\
\addlinespace[6pt]
k6b22c & Spend time on scouts or hobby clubs & Youth & 0.02 (0.00, 0.04) & 0.01 (-0.01, 0.02)\\
\addlinespace[6pt]
p6d15 & Children have regular bedtime routine & Primary caregiver & 0.02 (0.00, 0.04) & -0.01 (-0.02, 0.01)\\
\addlinespace[6pt]
k6c4c & The children have a regular bedtime routine & Youth & 0.02 (0.00, 0.04) & 0.01 (-0.01, 0.03)\\
\addlinespace[6pt]
p6j16 & Others moved in because of financial problems in past year? & Primary caregiver & 0.02 (-0.01, 0.04) & 0.01 (0.00, 0.03)\\
\addlinespace[6pt]
k6c4d & You are usually able to stay on top of things & Youth & 0.02 (0.00, 0.04) & 0.00 (-0.02, 0.01)\\
\addlinespace[6pt]
k6d25 & Time looking at screen two hours before bed (minutes) & Youth & 0.02 (0.00, 0.04) & 0.03 (0.01, 0.04)\\
\addlinespace[6pt]
k6c6a & Primary caregiver knows what you do during your free time & Youth & 0.02 (0.00, 0.03) & 0.00 (-0.02, 0.02)\\
\addlinespace[6pt]
k6b12 & Have you talked to an adult about applying to college? & Youth & 0.02 (0.00, 0.04) & 0.01 (0.00, 0.02)\\
\addlinespace[6pt]
k6d2v & I am a hard worker & Youth & 0.02 (0.00, 0.03) & 0.02 (0.01, 0.04)\\
\addlinespace[6pt]
p6d39 & Talked to youth about drugs or alcohol? & Primary caregiver & 0.02 (-0.01, 0.04) & 0.03 (0.01, 0.05)\\
\addlinespace[6pt]
p6d3 & Made sure youth completed homework in past month & Primary caregiver & 0.02 (0.00, 0.04) & -0.02 (-0.03, 0.00)\\
\addlinespace[6pt]
p6k1 & Currently attending school or training programs? & Primary caregiver & 0.02 (-0.01, 0.05) & 0.01 (-0.01, 0.02)\\
\addlinespace[6pt]
k6d36g & Hours per day watches TV/videos/movies on weekdays & Youth & 0.02 (0.00, 0.03) & 0.00 (-0.02, 0.01)\\
\addlinespace[6pt]
k6c4a & You can't hear yourself think in your home & Youth & 0.02 (0.00, 0.03) & 0.01 (-0.01, 0.03)\\
\addlinespace[6pt]
k6d2p & I often say and do things without considering the consequences & Youth & 0.02 (-0.01, 0.04) & 0.02 (0.00, 0.03)\\
\addlinespace[6pt]
k6e8 & Ever been victim of a crime? & Youth & 0.02 (-0.01, 0.04) & 0.00 (0.00, 0.00)\\
\addlinespace[6pt]
k6e4b & I feel unsafe walking around my neighborhood during the day & Youth & 0.02 (0.00, 0.03) & -0.01 (-0.02, 0.01)\\
\addlinespace[6pt]
k6c40 & Another adult in life with whom you share ideas/talk? & Youth & 0.01 (0.00, 0.03) & 0.00 (0.00, 0.01)\\
\addlinespace[6pt]
p6d24 & How often knows what youth does during free time? & Primary caregiver & 0.01 (-0.01, 0.04) & 0.03 (0.01, 0.05)\\
\addlinespace[6pt]
k6c9f & Primary caregiver missed events or activities that were important to you & Youth & 0.01 (-0.01, 0.03) & 0.02 (0.01, 0.04)\\
\addlinespace[6pt]
k6c4b & It's a real zoo in your home & Youth & 0.01 (0.00, 0.03) & 0.00 (-0.02, 0.02)\\
\addlinespace[6pt]
k6c9a & Primary caregiver explained to you why something you did was wrong & Youth & 0.01 (0.00, 0.03) & 0.00 (-0.03, 0.02)\\
\addlinespace[6pt]
o6a7a & Exterior—Peeling paint & Interviewer & 0.01 (-0.05, 0.07) & 0.04 (0.01, 0.07)\\
\addlinespace[6pt]
k6d36d & Hours per day electronically communicates with friends on weekdays & Youth & 0.01 (-0.02, 0.05) & 0.00 (-0.02, 0.02)\\
\addlinespace[6pt]
k6e15 & Ever seen someone stopped by police in your school? & Youth & 0.01 (-0.01, 0.03) & 0.00 (-0.01, 0.01)\\
\addlinespace[6pt]
k6d2j & I feel tense or keyed up & Youth & 0.01 (0.00, 0.03) & 0.01 (0.00, 0.03)\\
\addlinespace[6pt]
k6d36a & Hours per day hangs out with friends on weekdays & Youth & 0.01 (0.00, 0.03) & 0.00 (-0.02, 0.01)\\
\addlinespace[6pt]
k6c44 & Number of step siblings living with youth & Youth & 0.01 (-0.01, 0.03) & 0.01 (-0.03, 0.04)\\
\addlinespace[6pt]
k6e7e & If I fight with someone, it's nobody else's business & Youth & 0.01 (-0.01, 0.03) & 0.00 (-0.01, 0.02)\\
\addlinespace[6pt]
p6j42 & Utilities turned off because not enough money in past year & Primary caregiver & 0.01 (-0.01, 0.04) & 0.01 (-0.01, 0.03)\\
\addlinespace[6pt]
k6d2a & I don't spend enough time thinking over a situation before I act & Youth & 0.01 (0.00, 0.02) & -0.01 (-0.02, 0.01)\\
\addlinespace[6pt]
k6\_scale\_neighborhoodColl\newline{}ectiveEfficacy\_kid & Theme—Neighborhood Collective Efficacy: Scale—Child Perception & Youth & 0.01 (-0.01, 0.03) & 0.01 (0.00, 0.03)\\
\addlinespace[6pt]
k6c9b & Primary caregiver took away privileges or grounded you & Youth & 0.01 (-0.02, 0.04) & 0.00 (-0.02, 0.02)\\
\addlinespace[6pt]
k6b7 & Hours spent doing homework each weekend day & Youth & 0.01 (-0.01, 0.03) & -0.13 (-0.25, -0.01)\\
\addlinespace[6pt]
k6c9c & Primary caregiver shouted, yelled, screamed, swore or cursed at you & Youth & 0.01 (-0.01, 0.03) & 0.01 (0.00, 0.03)\\
\addlinespace[6pt]
p6k36 & Worked in own business in past year? & Primary caregiver & 0.01 (-0.02, 0.04) & 0.01 (0.00, 0.02)\\
\addlinespace[6pt]
k6d2f & I am a cheerful person & Youth & 0.01 (-0.01, 0.03) & -0.01 (-0.03, 0.01)\\
\addlinespace[6pt]
k6c5 & Firearms kept in or around your home? & Youth & 0.01 (-0.01, 0.03) & -0.01 (-0.02, 0.01)\\
\addlinespace[6pt]
o6c11 & Interior—Youth's artwork or photographs displayed & Interviewer & 0.01 (-0.02, 0.03) & 0.00 (-0.02, 0.01)\\
\addlinespace[6pt]
o6f3 & How attentive was the youth when completing the interview? & Interviewer & 0.01 (-0.04, 0.06) & -0.05 (-0.11, 0.01)\\
\addlinespace[6pt]
k6e1 & How many families on block you know well? & Youth & 0.01 (0.00, 0.02) & 0.00 (-0.02, 0.01)\\
\addlinespace[6pt]
k6d2k & Once I make a plan to get something done, I stick to it & Youth & 0.01 (-0.01, 0.02) & -0.01 (-0.02, 0.01)\\
\addlinespace[6pt]
k6d2b & I love life & Youth & 0.01 (-0.01, 0.03) & -0.01 (-0.03, 0.01)\\
\addlinespace[6pt]
k6c8 & How often spend time alone in home without adult present? & Youth & 0.01 (-0.01, 0.03) & 0.02 (0.00, 0.04)\\
\addlinespace[6pt]
k6d48 & Ever drank alcohol more than two times without parents? & Youth & 0.01 (-0.01, 0.03) & 0.00 (-0.01, 0.01)\\
\addlinespace[6pt]
k6d22 & Nights during week has problems falling asleep & Youth & 0.01 (-0.01, 0.02) & 0.00 (-0.01, 0.02)\\
\addlinespace[6pt]
o6a4 & Vacant, abandoned, boarded up buildings & Interviewer & 0.01 (-0.07, 0.08) & 0.00 (-0.09, 0.09)\\
\addlinespace[6pt]
p6b21 & Days youth absent from school due to health in past year & Primary caregiver & 0.01 (-0.01, 0.02) & -0.04 (-0.08, 0.00)\\
\addlinespace[6pt]
k6d62g & Friends deliberately damaged property that did not belong to them & Youth & 0.01 (-0.01, 0.02) & -0.02 (-0.04, -0.01)\\
\addlinespace[6pt]
p6k5 & Received employment counseling since last interview? & Primary caregiver & 0.01 (-0.01, 0.02) & 0.00 (-0.01, 0.01)\\
\addlinespace[6pt]
k6c9e & Primary caregiver listened to your side of an argument & Youth & 0.01 (-0.01, 0.02) & -0.01 (-0.03, 0.00)\\
\addlinespace[6pt]
k6d1g & I easily change from one activity to another & Youth & 0.00 (-0.01, 0.02) & 0.02 (0.00, 0.03)\\
\addlinespace[6pt]
p6h5 & Primary caregiver covered by any health insurance? & Primary caregiver & 0.00 (-0.02, 0.03) & 0.01 (-0.01, 0.02)\\
\addlinespace[6pt]
k6\_scale\_conflictTactics\_\newline{}nonviolentDisciplineKidPe\newline{}rception & Theme—Conflict Tactics: Scale—Non-violent Discipline, Child Perception & Youth & 0.00 (-0.01, 0.02) & 0.01 (0.00, 0.03)\\
\addlinespace[6pt]
k6d62b & Friends drank alcohol more than two times without their parents & Youth & 0.00 (-0.02, 0.03) & 0.01 (-0.01, 0.02)\\
\addlinespace[6pt]
k6d36h & Hours per day reads a book or magazine for fun on weekdays & Youth & 0.00 (-0.01, 0.02) & -0.01 (-0.02, 0.00)\\
\addlinespace[6pt]
k6d2ag & I feel nervous or shaky inside & Youth & 0.00 (-0.02, 0.03) & 0.02 (0.00, 0.04)\\
\addlinespace[6pt]
k6d1f & I am self-confident in social situations & Youth & 0.00 (-0.02, 0.02) & -0.02 (-0.03, 0.00)\\
\addlinespace[6pt]
k6b6 & Hours spent doing homework each weekday & Youth & 0.00 (-0.02, 0.02) & -0.02 (-0.04, 0.00)\\
\addlinespace[6pt]
k6d2o & I think good things are going to happen to me & Youth & 0.00 (-0.02, 0.02) & 0.00 (-0.02, 0.01)\\
\addlinespace[6pt]
k6d2ae & In uncertain times, I expect the best & Youth & 0.00 (-0.01, 0.02) & -0.02 (-0.03, 0.00)\\
\addlinespace[6pt]
k6c6b & Primary caregiver knows what you spend money on & Youth & 0.00 (-0.01, 0.02) & 0.01 (-0.01, 0.03)\\
\addlinespace[6pt]
k6d2m & I finish whatever I begin & Youth & 0.00 (-0.02, 0.02) & -0.01 (-0.02, 0.01)\\
\addlinespace[6pt]
k6b31d & Friends skip school for a full day without an excuse & Youth & 0.00 (-0.01, 0.01) & 0.02 (0.00, 0.03)\\
\addlinespace[6pt]
k6d61m & Were you loud, rowdy, or unruly in a public place & Youth & 0.00 (-0.01, 0.02) & -0.01 (-0.03, 0.00)\\
\addlinespace[6pt]
p6k59 & Experienced large drop in household income since last interview? & Primary caregiver & 0.00 (-0.02, 0.02) & 0.02 (0.00, 0.03)\\
\addlinespace[6pt]
k6d2s & I feel happy & Youth & 0.00 (-0.02, 0.02) & -0.01 (-0.02, 0.01)\\
\addlinespace[6pt]
k6\_scale\_EPOCH & Theme—Positive Adolescent Functioning (EPOCH) & Youth & 0.00 (-0.02, 0.02) & -0.01 (-0.02, 0.00)\\
\addlinespace[6pt]
p6b7 & Doctor diagnosed youth with limb problems & Primary caregiver & 0.00 (-0.05, 0.05) & -0.01 (-0.02, 0.00)\\
\addlinespace[6pt]
p6b42 & Youth physically attacks people & Primary caregiver & 0.00 (-0.03, 0.03) & -0.04 (-0.07, -0.01)\\
\addlinespace[6pt]
k6d2w & I believe that things will work out, no matter how difficult & Youth & 0.00 (-0.02, 0.02) & -0.01 (-0.03, 0.00)\\
\addlinespace[6pt]
p6b63 & Youth steals outside the home & Primary caregiver & 0.00 (-0.02, 0.02) & 0.00 (-0.02, 0.01)\\
\addlinespace[6pt]
o6a6f & Garbage/litter outside home & Interviewer & 0.00 (-0.08, 0.08) & 0.03 (0.00, 0.06)\\
\addlinespace[6pt]
k6c9d & Primary caregiver hit or slapped you & Youth & 0.00 (-0.02, 0.02) & -0.03 (-0.05, -0.01)\\
\addlinespace[6pt]
k6d62i & Friends used or threatened to use a weapon to get something & Youth & 0.00 (-0.02, 0.02) & -0.06 (-0.08, -0.03)\\
\addlinespace[6pt]
k6d1i & I start conversations rather than waiting for others to talk first & Youth & 0.00 (-0.01, 0.01) & -0.01 (-0.02, 0.01)\\
\addlinespace[6pt]
k6d35e & Friends/family encourage not eating high-fat food & Youth & 0.00 (-0.02, 0.02) & 0.00 (-0.01, 0.02)\\
\addlinespace[6pt]
k6d2y & When I have a problem, I have someone who will be there for me & Youth & 0.00 (-0.01, 0.01) & -0.03 (-0.04, -0.01)\\
\addlinespace[6pt]
k6d62k & Friends stole something worth less than \$50 & Youth & 0.00 (-0.02, 0.01) & -0.02 (-0.04, 0.00)\\
\addlinespace[6pt]
k6\_scale\_delinquentBehavi\newline{}or\_peers & Theme—Delinquent Behavior: Scale—Peer & Youth & 0.00 (-0.02, 0.01) & -0.03 (-0.05, -0.01)\\
\addlinespace[6pt]
k6d2x & I feel life is not worth living & Youth & 0.00 (-0.03, 0.02) & 0.02 (0.00, 0.04)\\
\addlinespace[6pt]
p6b28 & Youth has usual place for routine health care? & Primary caregiver & 0.00 (-0.02, 0.01) & 0.00 (0.00, 0.00)\\
\addlinespace[6pt]
k6e16 & Know anyone who has been stopped by police? & Youth & 0.00 (-0.02, 0.01) & 0.00 (0.00, 0.00)\\
\addlinespace[6pt]
k6e7b & It's okay to do anything you want & Youth & 0.00 (-0.02, 0.01) & -0.01 (-0.03, 0.01)\\
\addlinespace[6pt]
k6d36b & Hours per day hangs out with family members on weekdays & Youth & 0.00 (-0.02, 0.01) & -0.01 (-0.03, 0.02)\\
\addlinespace[6pt]
k6b25 & Hours per week worked for pay during school year & Youth & 0.00 (-0.01, 0.01) & -0.04 (-0.11, 0.02)\\
\addlinespace[6pt]
k6e2a & People around here are willing to help their neighbors & Youth & 0.00 (-0.03, 0.02) & 0.00 (-0.02, 0.02)\\
\addlinespace[6pt]
k6e17\_1 & Parent was stopped by police & Youth & 0.00 (-0.02, 0.01) & 0.00 (0.00, 0.00)\\
\addlinespace[6pt]
p6b24 & Youth saw doctor for regular check-up in past year? & Primary caregiver & 0.00 (-0.01, 0.01) & 0.00 (0.00, 0.00)\\
\addlinespace[6pt]
k6d61l & Taken part in a group fight & Youth & 0.00 (-0.04, 0.04) & 0.01 (-0.02, 0.05)\\
\addlinespace[6pt]
k6d1d & I join group activities without being told to & Youth & 0.00 (-0.02, 0.01) & 0.00 (-0.02, 0.01)\\
\addlinespace[6pt]
k6b26 & Hours per week worked for pay during summer & Youth & 0.00 (-0.02, 0.01) & -0.01 (-0.02, 0.00)\\
\addlinespace[6pt]
k6d36c & Hours per day does chores or helps out at home on weekdays & Youth & 0.00 (-0.03, 0.02) & -0.01 (-0.02, 0.00)\\
\addlinespace[6pt]
k6d61a & Painted graffiti or signs on private property/public spaces & Youth & 0.00 (-0.01, 0.00) & -0.06 (-0.11, -0.02)\\
\addlinespace[6pt]
k6e17\_5 & Other person was stopped by police & Youth & 0.00 (-0.03, 0.02) & 0.00 (-0.01, 0.01)\\
\addlinespace[6pt]
k6b22f & Spend time on volunteer service activities & Youth & 0.00 (-0.01, 0.00) & -0.01 (-0.02, 0.01)\\
\addlinespace[6pt]
k6d61g & Stolen something worth more than \$50 & Youth & 0.00 (-0.03, 0.02) & -0.05 (-0.09, -0.01)\\
\addlinespace[6pt]
p6d30 & Mornings during school week youth eats breakfast & Primary caregiver & 0.00 (-0.02, 0.01) & 0.00 (-0.02, 0.02)\\
\addlinespace[6pt]
p6b70 & Youth ever been arrested? & Primary caregiver & -0.01 (-0.04, 0.03) & 0.01 (-0.01, 0.02)\\
\addlinespace[6pt]
k6d2ah & I am optimistic about my future & Youth & -0.01 (-0.01, 0.00) & 0.00 (-0.02, 0.01)\\
\addlinespace[6pt]
o6d3 & Cleanliness of youth's hair & Interviewer & -0.01 (-0.05, 0.04) & -0.01 (-0.07, 0.05)\\
\addlinespace[6pt]
o6e16 & Youth appeared to understand questions & Interviewer & -0.01 (-0.08, 0.07) & -0.02 (-0.04, 0.00)\\
\addlinespace[6pt]
k6b9 & How important to you that you graduate from college? & Youth & -0.01 (-0.02, 0.01) & -0.03 (-0.05, -0.01)\\
\addlinespace[6pt]
k6e3a & Neighbors would get involved if children skip school & Youth & -0.01 (-0.02, 0.00) & -0.01 (-0.02, 0.01)\\
\addlinespace[6pt]
p6b25 & Times youth has been injured in past year & Primary caregiver & -0.01 (-0.02, 0.01) & 0.00 (-0.02, 0.01)\\
\addlinespace[6pt]
k6d35d & Friends/family talk about eating more health foods & Youth & -0.01 (-0.02, 0.01) & -0.02 (-0.03, 0.00)\\
\addlinespace[6pt]
k6d2ai & I feel fearful & Youth & -0.01 (-0.03, 0.01) & 0.00 (-0.02, 0.01)\\
\addlinespace[6pt]
p6b30 & Times youth stayed overnight at hospital past year & Primary caregiver & -0.01 (-0.01, 0.00) & -0.30 (-0.78, 0.18)\\
\addlinespace[6pt]
k6b24 & Worked for pay in past four weeks? & Youth & -0.01 (-0.02, 0.01) & 0.00 (-0.01, 0.00)\\
\addlinespace[6pt]
k6d62h & Friends stole something worth more than \$50 & Youth & -0.01 (-0.02, 0.00) & -0.02 (-0.05, 0.01)\\
\addlinespace[6pt]
p6d21a & Youth has friends & Primary caregiver & -0.01 (-0.04, 0.03) & 0.00 (0.00, 0.00)\\
\addlinespace[6pt]
k6d34 & Number of low-calorie drinks in typical day & Youth & -0.01 (-0.03, 0.01) & -0.02 (-0.04, -0.01)\\
\addlinespace[6pt]
k6d2e & I get so involved in activities that I forget about everything else & Youth & -0.01 (-0.02, 0.01) & -0.01 (-0.03, 0.00)\\
\addlinespace[6pt]
k6d62e & Friends asked to go drinking with them & Youth & -0.01 (-0.02, 0.01) & -0.01 (-0.02, 0.00)\\
\addlinespace[6pt]
k6e17\_2 & Sibling was stopped by police & Youth & -0.01 (-0.04, 0.02) & -0.01 (-0.02, 0.01)\\
\addlinespace[6pt]
k6d62j & Friends sold marijuana or other drugs & Youth & -0.01 (-0.03, 0.01) & -0.04 (-0.06, -0.02)\\
\addlinespace[6pt]
k6d2i & I keep at my schoolwork until I am done with it & Youth & -0.01 (-0.03, 0.01) & -0.03 (-0.04, -0.01)\\
\addlinespace[6pt]
o6a3 & Graffiti on buildings & Interviewer & -0.01 (-0.09, 0.07) & -0.01 (-0.06, 0.05)\\
\addlinespace[6pt]
k6d2ad & When I am learning something new, I lose track of time & Youth & -0.01 (-0.03, 0.01) & 0.01 (0.00, 0.03)\\
\addlinespace[6pt]
k6d2ab & I often say whatever comes into my head without thinking first & Youth & -0.01 (-0.03, 0.01) & -0.02 (-0.04, 0.00)\\
\addlinespace[6pt]
k6e2b & This is a close-knit neighborhood & Youth & -0.01 (-0.03, 0.01) & -0.01 (-0.03, 0.00)\\
\addlinespace[6pt]
k6\_scale\_neighborhoodColl\newline{}ectiveEfficacy\_socCtrlKid & Theme—Neighborhood Collective Efficacy: Scale—Social Control, Child Perception & Youth & -0.01 (-0.02, 0.00) & -0.02 (-0.03, 0.00)\\
\addlinespace[6pt]
k6d1e & I make friends easily & Youth & -0.01 (-0.03, 0.01) & -0.01 (-0.02, 0.01)\\
\addlinespace[6pt]
k6e3d & Neighbors would get involved if fight broke out in front of the house & Youth & -0.01 (-0.02, 0.00) & -0.02 (-0.03, -0.01)\\
\addlinespace[6pt]
k6d35a & Friends/family encourage youth to be physically active & Youth & -0.01 (-0.02, 0.00) & -0.02 (-0.03, 0.00)\\
\addlinespace[6pt]
k6d36f & Hours per day visits websites/shop on internet on weekdays & Youth & -0.01 (-0.02, 0.00) & -0.04 (-0.08, -0.01)\\
\addlinespace[6pt]
o6d4 & Youth's odor & Interviewer & -0.01 (-0.02, 0.00) & -0.02 (-0.15, 0.10)\\
\addlinespace[6pt]
o6c8 & Interior—Visible rooms noticeably cluttered & Interviewer & -0.01 (-0.06, 0.04) & -0.05 (-0.10, -0.01)\\
\addlinespace[6pt]
k6d62a & Friends smoked an entire cigarette & Youth & -0.01 (-0.03, 0.00) & -0.02 (-0.04, -0.01)\\
\addlinespace[6pt]
k6e5 & Ever witnessed a crime or known about a crime? & Youth & -0.01 (-0.02, 0.00) & 0.00 (-0.01, 0.00)\\
\addlinespace[6pt]
k6d62f & Friends given or sold marijuana to you & Youth & -0.01 (-0.02, 0.00) & -0.06 (-0.14, 0.01)\\
\addlinespace[6pt]
k6d2g & I have friends that I really care about & Youth & -0.01 (-0.03, 0.01) & -0.03 (-0.05, -0.02)\\
\addlinespace[6pt]
p6j21 & Received household income from UI/WC in past year? & Primary caregiver & -0.01 (-0.04, 0.02) & 0.00 (0.00, 0.00)\\
\addlinespace[6pt]
p6j46 & Someone in household did not see doctor/go to hospital because of cost & Primary caregiver & -0.01 (-0.04, 0.01) & 0.00 (-0.01, 0.01)\\
\addlinespace[6pt]
k6d61i & Used or threatened to use a weapon to get something & Youth & -0.01 (-0.02, 0.00) & -0.05 (-0.22, 0.12)\\
\addlinespace[6pt]
k6d2aa & I have a lot of fun & Youth & -0.01 (-0.04, 0.02) & -0.01 (-0.03, 0.00)\\
\addlinespace[6pt]
k6d1c & I am open and direct about what I want & Youth & -0.01 (-0.03, 0.00) & -0.01 (-0.03, 0.00)\\
\addlinespace[6pt]
k6\_scale\_EPOCH\_connectedn\newline{}ess & Theme—Positive Adolescent Functioning (EPOCH): Scale—Connectedness & Youth & -0.01 (-0.03, 0.01) & -0.02 (-0.03, 0.00)\\
\addlinespace[6pt]
k6d60 & Parents ever given you alcohol to drink? & Youth & -0.01 (-0.02, 0.00) & 0.00 (0.00, 0.00)\\
\addlinespace[6pt]
o6e14 & Youth initiated verbal exchanges & Interviewer & -0.01 (-0.04, 0.01) & 0.00 (-0.01, 0.01)\\
\addlinespace[6pt]
k6d61c & Taken something from a store without paying for it & Youth & -0.01 (-0.03, 0.01) & -0.01 (-0.02, 0.00)\\
\addlinespace[6pt]
p6b60 & Youth runs away from home & Primary caregiver & -0.01 (-0.03, 0.01) & -0.06 (-0.10, -0.02)\\
\addlinespace[6pt]
k6d2af & When something good happens to me, I have people to share news with & Youth & -0.01 (-0.02, 0.00) & -0.02 (-0.06, 0.02)\\
\addlinespace[6pt]
k6d61e & Hurt someone badly enough to need bandages or medical care & Youth & -0.01 (-0.05, 0.02) & -0.01 (-0.03, 0.01)\\
\addlinespace[6pt]
k6d2l & There are people in my life who really care about me & Youth & -0.01 (-0.02, 0.00) & -0.02 (-0.03, -0.01)\\
\addlinespace[6pt]
p6b9 & Doctor diagnosed youth with other condition & Primary caregiver & -0.01 (-0.04, 0.01) & 0.00 (-0.01, 0.00)\\
\addlinespace[6pt]
p6b4 & Doctor diagnosed youth with heart disease/condition & Primary caregiver & -0.02 (-0.20, 0.17) & 0.07 (-0.03, 0.18)\\
\addlinespace[6pt]
k6d61b & Deliberately damaged property that didn't belong to you & Youth & -0.02 (-0.04, 0.01) & -0.03 (-0.06, 0.00)\\
\addlinespace[6pt]
k6\_scale\_EPOCH\_engagement & Theme—Positive Adolescent Functioning (EPOCH): Scale—Engagement & Youth & -0.02 (-0.03, 0.00) & -0.02 (-0.03, 0.00)\\
\addlinespace[6pt]
k6d62c & Friends tried marijuana & Youth & -0.02 (-0.04, 0.00) & -0.01 (-0.03, 0.01)\\
\addlinespace[6pt]
k6d23 & Nights during week has problems staying asleep & Youth & -0.02 (-0.04, 0.00) & -0.01 (-0.02, 0.01)\\
\addlinespace[6pt]
k6e3b & Neighbors would get involved if children spray paint buildings & Youth & -0.02 (-0.03, -0.01) & -0.02 (-0.03, -0.01)\\
\addlinespace[6pt]
o6g1 & Youth's attention to interviewer & Interviewer & -0.02 (-0.06, 0.02) & -0.12 (-0.22, -0.02)\\
\addlinespace[6pt]
k6d61j & Sold marijuana or other drugs & Youth & -0.02 (-0.05, 0.01) & -0.03 (-0.07, 0.00)\\
\addlinespace[6pt]
p6h102 & Spent time in jail since last interview? & Primary caregiver & -0.02 (-0.04, 0.00) & 0.00 (0.00, 0.00)\\
\addlinespace[6pt]
p6k50 & Did anything else to earn money in past year? & Primary caregiver & -0.02 (-0.05, 0.01) & 0.00 (0.00, 0.00)\\
\addlinespace[6pt]
p6b15 & Youth had frequent diarrhea/colitis in past year & Primary caregiver & -0.02 (-0.04, 0.00) & 0.00 (0.00, 0.00)\\
\addlinespace[6pt]
o6e13 & Youth's speech was distinct, clear and audible & Interviewer & -0.02 (-0.07, 0.03) & 0.00 (-0.02, 0.01)\\
\addlinespace[6pt]
k6d35b & Friends/family engage in physical activities with youth & Youth & -0.02 (-0.04, -0.01) & 0.00 (-0.01, 0.02)\\
\addlinespace[6pt]
k6d28 & Spent night away from home without permission in past year & Youth & -0.02 (-0.05, 0.00) & 0.00 (0.00, 0.00)\\
\addlinespace[6pt]
k6e17\_4 & Neighbor was stopped by police & Youth & -0.02 (-0.05, 0.00) & 0.00 (0.00, 0.00)\\
\addlinespace[6pt]
k6\_scale\_parentalMonitori\newline{}ng\_kid & Theme—Parental Monitoring: Scale—Child-Perceived Parental Monitoring & Youth & -0.02 (-0.04, -0.01) & -0.02 (-0.04, 0.00)\\
\addlinespace[6pt]
k6d61k & Stolen something worth less than \$50 & Youth & -0.02 (-0.05, 0.00) & -0.03 (-0.05, 0.00)\\
\addlinespace[6pt]
o6c5 & Interior—Broken plaster/peeling paint & Interviewer & -0.03 (-0.16, 0.11) & 0.00 (0.00, 0.00)\\
\addlinespace[6pt]
p6b22 & Youth saw doctor for accident or injury in past year? & Primary caregiver & -0.03 (-0.04, -0.01) & 0.00 (0.00, 0.00)\\
\addlinespace[6pt]
o6e15 & Youth expressed ideas freely and easily & Interviewer & -0.03 (-0.07, 0.02) & 0.00 (-0.01, 0.01)\\
\addlinespace[6pt]
o6g5b & Did youth appear uncommunicative? & Interviewer & -0.03 (-0.08, 0.03) & -0.01 (-0.05, 0.04)\\
\addlinespace[6pt]
k6e3c & Neighbors would get involved if children show disrespect to adults & Youth & -0.03 (-0.04, -0.01) & -0.02 (-0.04, -0.01)\\
\addlinespace[6pt]
k6d2u & When I do an activity, I enjoy it so much that I lose track of time & Youth & -0.03 (-0.04, -0.01) & -0.03 (-0.05, -0.02)\\
\addlinespace[6pt]
k6b10 & How important to primary caregiver that you graduate from college? & Youth & -0.03 (-0.05, -0.01) & -0.04 (-0.11, 0.02)\\
\addlinespace[6pt]
o6f1 & Did the youth display positive emotions during the visit? & Interviewer & -0.03 (-0.07, 0.01) & 0.00 (-0.04, 0.04)\\
\addlinespace[6pt]
p6j44 & Moved in with other people because of financial problems in past year & Primary caregiver & -0.03 (-0.09, 0.03) & -0.01 (-0.03, 0.01)\\
\addlinespace[6pt]
cp6conf2 & How much time youth lives with primary caregiver & Primary caregiver & -0.03 (-0.06, 0.00) & 0.00 (0.00, 0.00)\\
\addlinespace[6pt]
o6c1 & Interior—Broken/cracked windows & Interviewer & -0.03 (-0.07, 0.00) & -0.03 (-0.06, 0.00)\\
\addlinespace[6pt]
k6d2h & I get completely absorbed in what I am doing & Youth & -0.03 (-0.05, -0.01) & -0.01 (-0.03, 0.01)\\
\addlinespace[6pt]
p6b3 & Doctor diagnosed youth with anemia & Primary caregiver & -0.03 (-0.10, 0.04) & 0.01 (-0.04, 0.05)\\
\addlinespace[6pt]
o6a6a & Unlit entrance or stairway & Interviewer & -0.03 (-0.07, 0.01) & 0.00 (0.00, 0.00)\\
\addlinespace[6pt]
o6a7b & Exterior—Crumbling or damaged walls & Interviewer & -0.03 (-0.10, 0.03) & 0.00 (0.00, 0.00)\\
\addlinespace[6pt]
k6e17\_3 & Friend was stopped by police & Youth & -0.03 (-0.07, 0.00) & 0.00 (-0.01, 0.00)\\
\addlinespace[6pt]
p6b61 & Youth sets fires & Primary caregiver & -0.03 (-0.09, 0.02) & -0.07 (-0.14, -0.01)\\
\addlinespace[6pt]
p6k72 & Filed for bankruptcy since last interview? & Primary caregiver & -0.03 (-0.06, -0.01) & -0.01 (-0.03, 0.01)\\
\addlinespace[6pt]
o6g5c & Did youth appear anxious/nervous? & Interviewer & -0.03 (-0.08, 0.01) & -0.10 (-0.22, 0.02)\\
\addlinespace[6pt]
p6j40 & Evicted for not paying rent/mortgage in past year & Primary caregiver & -0.04 (-0.11, 0.04) & -0.01 (-0.01, 0.00)\\
\addlinespace[6pt]
k6d40 & Ever smoked an entire cigarette? & Youth & -0.04 (-0.08, 0.01) & 0.00 (0.00, 0.00)\\
\addlinespace[6pt]
p6b6 & Doctor diagnosed youth with diabetes & Primary caregiver & -0.04 (-0.06, -0.02) & 0.00 (0.00, 0.00)\\
\addlinespace[6pt]
o6a7c & Exterior—Broken or cracked windows & Interviewer & -0.04 (-0.07, 0.00) & 0.00 (-0.01, 0.00)\\
\addlinespace[6pt]
k6d61f & Driven a car without its owner's permission & Youth & -0.04 (-0.07, 0.00) & -0.04 (-0.10, 0.03)\\
\addlinespace[6pt]
k6d4 & Days absent from school due to health in past year & Youth & -0.04 (-0.07, -0.01) & -0.04 (-0.12, 0.03)\\
\addlinespace[6pt]
o6g4 & Youth's cooperation throughout most of the interview & Interviewer & -0.04 (-0.11, 0.02) & -0.02 (-0.10, 0.05)\\
\addlinespace[6pt]
o6a6b & Broken steps outside house & Interviewer & -0.04 (-0.09, 0.01) & 0.00 (-0.01, 0.00)\\
\addlinespace[6pt]
o6d1\_10 & Youth's clothing—None of the above apply & Interviewer & -0.04 (-0.11, 0.02) & 0.00 (0.00, 0.01)\\
\addlinespace[6pt]
o6d1\_3 & Youth's clothing—Worn, but mended or not ripped or torn & Interviewer & -0.05 (-0.07, -0.02) & -0.01 (-0.02, 0.00)\\
\addlinespace[6pt]
k6d61h & Gone into a house or building to steal something & Youth & -0.05 (-0.09, 0.00) & -0.13 (-0.29, 0.04)\\
\addlinespace[6pt]
o6c12 & Interior—Home noisy from inside noise & Interviewer & -0.05 (-0.10, 0.01) & 0.00 (-0.03, 0.02)\\
\addlinespace[6pt]
k6f80 & Primary caregiver uses drugs? & Youth & -0.05 (-0.08, -0.02) & -0.01 (-0.02, 0.00)\\
\addlinespace[6pt]
p6b67 & Youth vandalizes & Primary caregiver & -0.05 (-0.08, -0.01) & -0.02 (-0.05, 0.02)\\
\addlinespace[6pt]
k6f83 & Parents ever given you drugs? & Youth & -0.05 (-0.10, 0.00) & -0.03 (-0.10, 0.03)\\
\addlinespace[6pt]
p6b17 & Youth had ear infection in past year & Primary caregiver & -0.06 (-0.09, -0.03) & 0.00 (0.00, 0.00)\\
\addlinespace[6pt]
p6j30 & Received household income from other cash aid in past year? & Primary caregiver & -0.06 (-0.11, -0.01) & 0.00 (0.00, 0.00)\\
\addlinespace[6pt]
o6d2 & Youth had washed or bathed & Interviewer & -0.07 (-0.17, 0.03) & -0.30 (-0.78, 0.18)\\
\addlinespace[6pt]
p6j45 & Stayed at shelter or place not meant for housing in past year & Primary caregiver & -0.07 (-0.19, 0.05) & -0.01 (-0.02, 0.00)\\
\addlinespace[6pt]
k6f82 & Drugs easily available to you in home? & Youth & -0.07 (-0.13, -0.01) & 0.00 (0.00, 0.00)\\
\addlinespace[6pt]
k6f68 & Ever tried other illegal drugs besides marijuana? & Youth & -0.07 (-0.21, 0.06) & -0.18 (-0.30, -0.06)\\
\addlinespace[6pt]
o6f4 & How cooperative was the youth during the interview? & Interviewer & -0.08 (-0.13, -0.03) & -0.01 (-0.12, 0.09)\\
\addlinespace[6pt]
o6g5a & Did youth appear suspicious? & Interviewer & -0.08 (-0.20, 0.04) & -0.10 (-0.21, 0.01)\\
\addlinespace[6pt]
p6h94 & Used prescription drugs (not prescribed) in past year? & Primary caregiver & -0.08 (-0.17, 0.00) & 0.00 (-0.01, 0.01)\\
\addlinespace[6pt]
k6f74 & Ever used prescription drugs (not prescribed)? & Youth & -0.09 (-0.20, 0.03) & 0.00 (0.00, 0.00)\\
\addlinespace[6pt]
o6e17 & At least 10 books in household & Interviewer & -0.09 (-0.16, -0.01) & 0.00 (-0.02, 0.03)\\
\addlinespace[6pt]
o6c9 & Interior—All visible rooms dirty & Interviewer & -0.09 (-0.16, -0.02) & -0.04 (-0.11, 0.03)\\
\addlinespace[6pt]
o6a6d & Large ditches outside home & Interviewer & -0.10 (-0.26, 0.06) & 0.00 (-0.01, 0.01)\\
\addlinespace[6pt]
o6g5d & Did youth appear hostile? & Interviewer & -0.10 (-0.28, 0.08) & -0.05 (-0.11, 0.00)\\
\addlinespace[6pt]
o6c13 & Interior—Home noisy from outside noise & Interviewer & -0.10 (-0.19, -0.01) & 0.00 (-0.01, 0.01)\\
\addlinespace[6pt]
o6c3 & Interior—Cracks/holes in walls/ceiling & Interviewer & -0.10 (-0.16, -0.05) & 0.00 (0.00, 0.00)\\
\addlinespace[6pt]
p6b8 & Doctor diagnosed youth with seizures & Primary caregiver & -0.11 (-0.24, 0.03) & 0.00 (-0.01, 0.00)\\
\addlinespace[6pt]
o6a6c & Broken glass/toys outside home & Interviewer & -0.12 (-0.21, -0.03) & 0.00 (0.00, 0.00)\\
\addlinespace[6pt]
p6b62 & Youth steals at home & Primary caregiver & -0.12 (-0.19, -0.06) & -0.01 (-0.03, 0.01)\\
\addlinespace[6pt]
o6f2 & Did the youth display negative emotions during the visit? & Interviewer & -0.12 (-0.28, 0.03) & -0.08 (-0.20, 0.04)\\
\addlinespace[6pt]
o6a5 & Abandoned vehicles on the block & Interviewer & -0.19 (-0.30, -0.07) & -0.06 (-0.12, -0.01)\\
\addlinespace[6pt]
o6c10 & Interior—Hazardous conditions & Interviewer & -0.19 (-0.44, 0.06) & -0.01 (-0.03, 0.01)\\
\addlinespace[6pt]
o6f5 & How cooperative was the youth while being measured? & Interviewer & -0.41 (-0.87, 0.04) & -0.02 (-0.05, 0.01)\\
\addlinespace[6pt]
k6d62d & Friends tried other drugs to get high & Youth & -1.01 (-1.28, -0.75) & -0.03 (-0.07, 0.01)\\*
\end{longtable}
\endgroup
\end{landscape}


\clearpage

\setcounter{figure}{0}
\setcounter{table}{0}

\section{Methods}

\subsection{Workflows for Constructing and Excluding Outcomes and Features}
\label{app:strobe_flowcharts}

Expanding on the overviews of outcomes, features, and data preparation in Sections~\ref{sec:data} and~\ref{sec:methods} of the main text, here we provide detailed workflows for the construction and exclusion of the outcomes and features used in our analyses. Figure~\ref{fig:outcome_strobe} shows how the raw FFCWS year 15 data are mapped to our final set of 484 outcomes, while Figure~\ref{fig:feature_strobe} shows how the raw data from years 0--9 are mapped to the feature matrices in our outer cross-validation folds.

The construction and exclusion criteria were applied sequentially, so the number of variables added or removed at each step is conditional on all preceding steps (in the figures, the order of steps corresponds to the order of the boxes and bullet points). Thus, the counts in the figures should not be interpreted as the total number of variables in the original data that met each criterion. For example, Figure~\ref{fig:outcome_strobe} reports that zero outcomes were removed for having ``unstandardized units.'' Some outcomes did meet this criterion, but they had already been removed in earlier steps.

\subsubsection{Outcome Construction and Exclusion}
\label{app:outcome_construction_and_exclusion}

\begin{figure}[htbp]
\centering
\includegraphics[width=.9\textwidth]{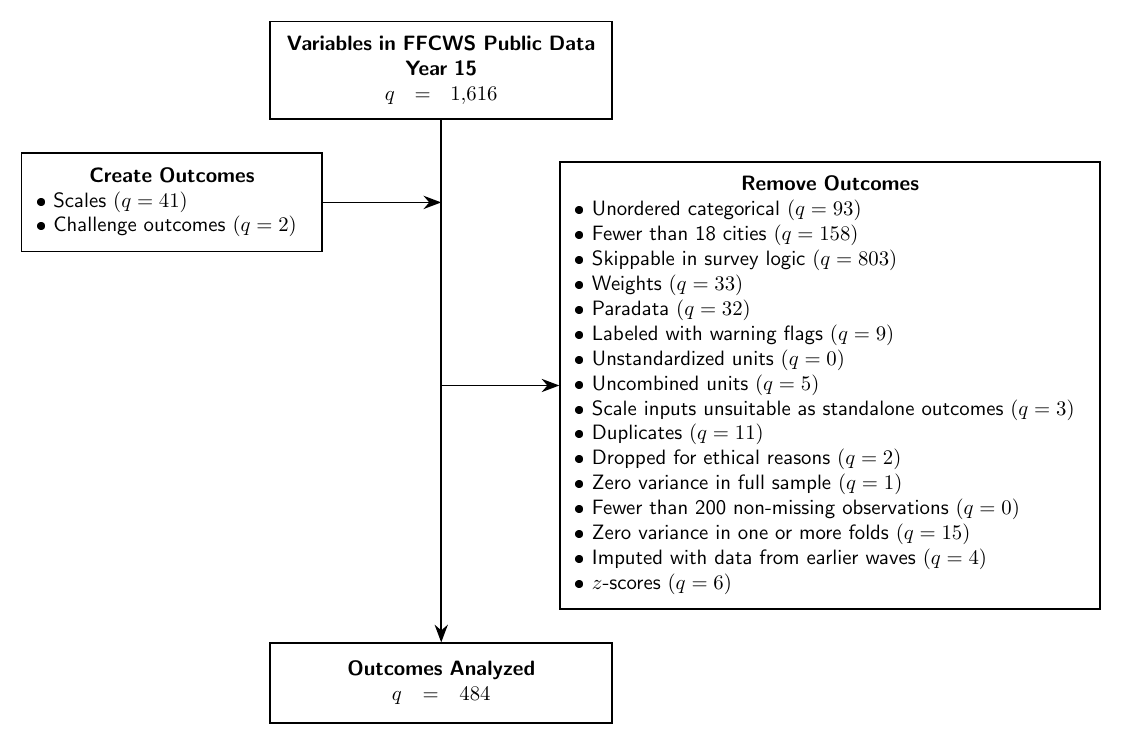}
\caption{Workflow for constructing and excluding outcomes.}
\label{fig:outcome_strobe}
\end{figure}

As described in Section~\ref{sec:outcomes} of the main text, we began with all variables in the main FFCWS data file from year 15 and then constructed additional outcomes and applied a series of exclusion criteria. These steps are summarized in Figure~\ref{fig:outcome_strobe}.

First, we constructed composite measures (scales) following FFCWS instructions \citep{FFCWSScalesDocumentation}; for more information, see Section~\ref{app:scales}. We also constructed two outcomes used in the Fragile Families Challenge that were not included among the raw year-15 variables or scales: eviction and GPA \citep{SalganikEtAl2020}.

We then excluded outcomes for several reasons. We excluded unordered categorical outcomes because $R^2_{\mathrm{CV}}$ is not defined for them. We excluded outcomes collected in fewer than 18 cities to ensure broad coverage across the FFCWS sampling sites. We excluded outcomes that could be skipped based on survey logic because they applied only to particular subpopulations, complicating comparisons of predictability across outcomes. We also excluded survey weights and paradata because they were not substantively meaningful outcomes.

Additionally, several exclusions addressed variables that would require specialized data cleaning or construction that was not feasible within our high-throughput approach. We excluded outcomes assigned certain warning codes in the FFCWS metadata \citep{KindelEtAl2019}, such as ``Variable has positive or negative outcome(s) which override a continuous answer set." We similarly excluded outcomes with heterogeneous measurement units (e.g., feet and inches), as described in Section~\ref{app:unit_standardization_and_unit_combination}. We also excluded raw variables that were designed only as components of scale measures, although most variables used to construct scales were kept in our outcome set because they could be meaningfully analyzed independently.

We further excluded duplicate outcomes that were identical or extremely similar to other outcomes,\footnote{These ``duplicate'' outcomes were either (a) identical to another outcome we were already predicting, (b) a one-to-one transformation of another outcome into different units, or (c) a ``first reading'' or ``second reading'' of height or weight, which the FFCWS combined into a constructed measure that we predicted.} as well as two outcomes excluded for ethical reasons (Section~\ref{app:ethics}). We excluded outcomes with zero variance because prediction of a constant outcome is not meaningful, and outcomes with fewer than 200 non-missing observations because we considered 200 a minimal sample size from which a machine learning algorithm might reasonably learn.

After implementing the cross-validation folds, we additionally excluded outcomes for which any inner or outer training or test fold had zero variance, because $R^2_{\mathrm{CV}}$ could not be reliably estimated for these outcomes.

During postprocessing, we excluded outcomes for which missing values had been imputed using information from earlier survey waves, because this type of imputation could artificially inflate apparent predictability. The FFCWS imputation code is not publicly available due to privacy protections, so we identified these outcomes using available documentation, consultation with the FFCWS team, and our best judgment. Finally, we excluded outcomes that are $z$-scores because they had non-missing observations with negative values, which our code was not designed to handle.

\subsubsection{Feature Construction and Exclusion}
\label{app:feature_construction_and_exclusion}

\begin{figure}[htbp]
\centering
\includegraphics[width=.9\textwidth]{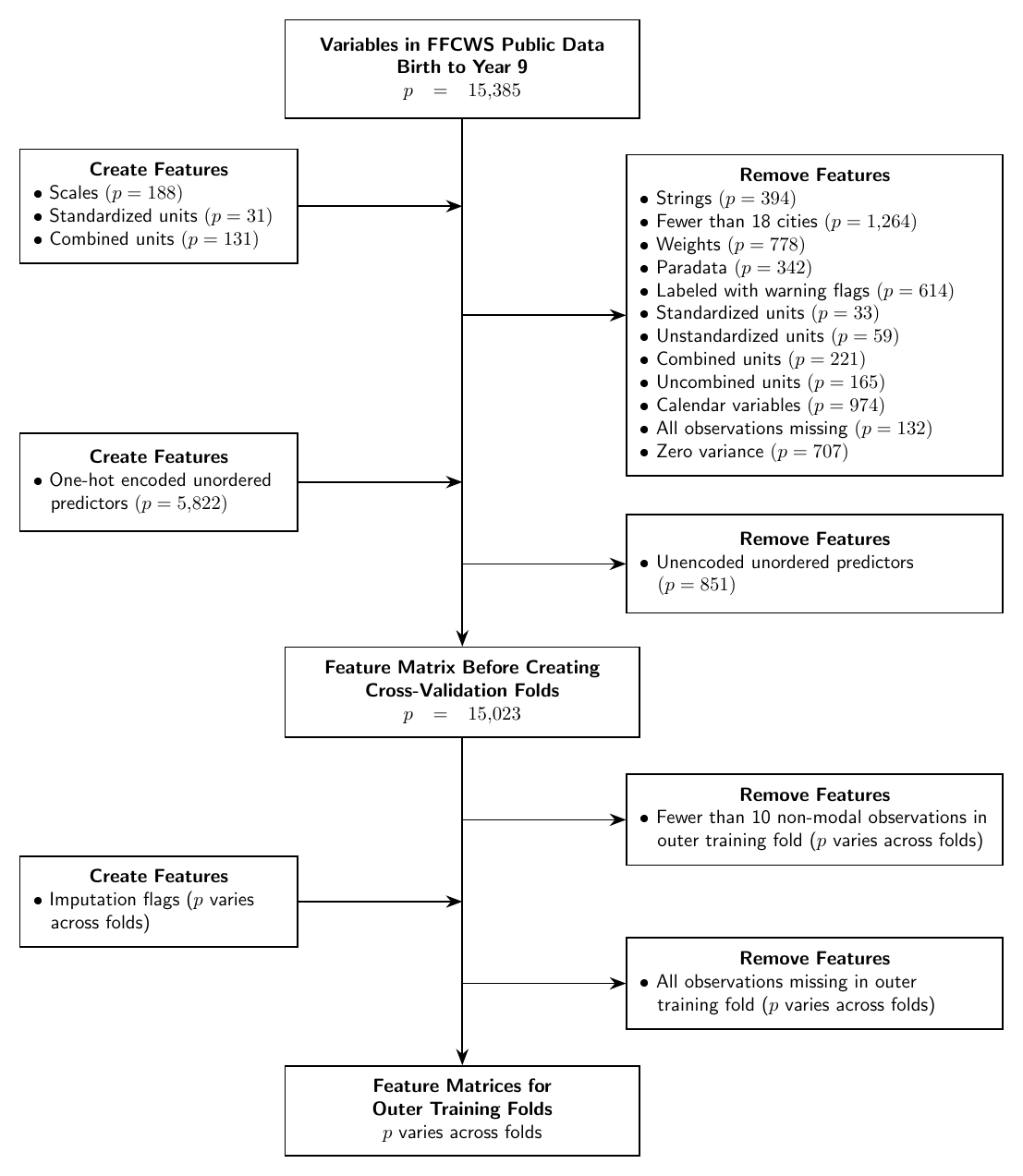}
\caption{Workflow for constructing and excluding features.}
\label{fig:feature_strobe}
\end{figure}

As described in Section~\ref{sec:features} of the main text, we constructed our feature matrices by starting with all variables in the main FFCWS data files from birth through year 9 and then applying a series of feature engineering steps and exclusion criteria designed to improve predictive performance. Excluding features can help because a high ratio of features to sample size can degrade performance via the ``curse of dimensionality'' \citep{AltmanKrzywinski2018}. Figure~\ref{fig:feature_strobe} summarizes the feature construction and exclusion process. Each outer cross-validation split utilized a slightly different feature matrix.

We first constructed composite measures (scales) for the data from birth to year 9 following the FFCWS documentation \citep{FFCWSScalesDocumentation}; for more information, see Section~\ref{app:scales}. We also constructed features that standardized or combined measurements reported in heterogeneous units (e.g., feet and inches), as described in Section~\ref{app:unit_standardization_and_unit_combination}.

We next applied a series of exclusion criteria. We excluded string variables since this data type could not be utilized by the machine learning models we used. Consistent with our outcome exclusion criteria, we also removed variables collected in fewer than 18 cities, survey weights, paradata, and variables with specified warning flags. We also excluded the raw versions of any features for which we had applied unit standardization or unit combination.\footnote{Each standardized or combined feature is associated with two raw features. However, in the flowchart, the numbers of raw features removed after unit standardization and unit combination do not correspond directly to the numbers in the ``Create Features’’ box because some of the raw features were removed by earlier exclusion criteria.} Additionally, we dropped raw features with heterogeneous units when standardizing or combining the units would have required strong assumptions (see Section~\ref{app:unit_standardization_and_unit_combination}). We also removed variables containing detailed employment- or childcare-calendar information; incorporating these variables would have required additional specialized feature engineering that was not feasible within our high-throughput approach, and we expected them to have limited predictive value in their raw form. Finally, we removed variables that were entirely missing or had zero variance.

We then one-hot encoded unordered categorical predictors and removed their original, unencoded versions.

After constructing the main feature matrix, we generated the outer cross-validation folds (see Section~\ref{app:estimand_and_estimation_procedure}) and performed the remaining preprocessing separately within each outer training fold. Specifically, within each outer training fold, we removed features for which fewer than 10 observations in that fold had values other than the modal value. We then mean-imputed missing values and created missingness indicators identifying observations that had been imputed; conducting imputation within the training folds ensured that no information from the outer test folds was used in preprocessing. We also removed features for which all observations were missing within the outer training fold.

\subsubsection{Scales}
\label{app:scales}

As noted in Sections~\ref{app:outcome_construction_and_exclusion} and~\ref{app:feature_construction_and_exclusion}, we created composite measures called ``scales'' following the FFCWS documentation~\citep{FFCWSScalesDocumentation}. We calculated all scales from the documentation except scales for which the FFCWS data already contained constructed summary scores, year 15 scales that could be skipped in the survey logic based on answers to prior questions (since all ``skippable'' outcomes were excluded, as discussed in Section~\ref{app:outcome_construction_and_exclusion}), and a ``pubertal development'' scale at year 15 for which the FFCWS instructions and our own protocols did not provide a sufficiently clear calculation procedure.

The official instructions sometimes left details of scale construction to researchers' discretion, so we developed protocols for interpreting the instructions in a consistent manner. We generally followed the scale-specific scoring instructions provided by the FFCWS documentation; when instructions were incomplete, we drew first on instructions for the same scale in other FFCWS waves and then on external documentation when available. When no complete instructions were available, we applied a consistent set of rules for aggregating variables and handling missing data. In particular, we generally treated a scale as missing when any component item was missing unless scale-specific instructions permitted otherwise, coded items so that higher values consistently represented higher levels of the construct, and used the mean of component items when the aggregation method was unspecified. Two research assistants independently interpreted the FFCWS scales documentation and wrote the scale construction code, then compared their interpretations and outputs. Specialized cases were resolved through discussion by the research team.

\subsubsection{Unit Standardization and Unit Combination}
\label{app:unit_standardization_and_unit_combination}

As noted in Sections~\ref{app:outcome_construction_and_exclusion} and~\ref{app:feature_construction_and_exclusion}, several steps in our variable construction and exclusion process addressed variables reported in heterogeneous measurement units. Aiming to improve the performance of our machine learning models, we either converted these variables to a common unit or removed them. We distinguished between ``unit standardization'' (e.g., some respondents reported time in weeks while others reported time in months), and ``unit combination'' (e.g., each respondent reported height as a combination of feet and inches). 

For constructs requiring unit standardization, a measurement is represented by two variables: a continuous variable containing the numeric value and an ordered categorical variable indicating its unit. For example, \texttt{f1j3} records how long the focal child's father has been looking for work, while \texttt{f1j3a} indicates whether this duration is reported in days, weeks, months, or years.

For constructs requiring unit combination, a measurement is represented by two continuous variables corresponding to different units. Respondents may report the measurement in both units (``A and B'') or in one unit or the other (``A or B''). For example, when asked how long he knew the focal child's mother before the pregnancy, the focal child's father could report years in \texttt{f1b1a} and additional months in \texttt{f1b1b}, making this an ``A and B'' case. In contrast, when asked about the maximum duration for which a mother could receive welfare, the focal child's father typically reported either years in \texttt{f1f10a} or months in \texttt{f1f10b}, making this an ``A or B'' case. Because the distinction between these cases was not always apparent from the variable structure alone, classification required manual review.

To identify variables requiring unit standardization or unit combination, two research assistants independently reviewed the FFCWS questionnaires, compared their findings, and conducted additional quality-control checks. For features, we converted measurements to a common unit and removed the original variables when conversion required relatively few assumptions. When conversion required stronger assumptions---for example, because the appropriate conversion rate was unclear---we removed the original variables without creating a converted feature. For outcomes, we took a more conservative approach in which we removed all outcomes with heterogeneous units, because errors in unit conversion could substantially affect our estimates of predictability.

\subsubsection{Metadata for Scales, Unit Standardization, and Unit Combination}
\label{app:metadata_creation}

We created metadata for newly constructed scales and for variables produced through unit standardization and unit combination, and appended these metadata to the existing FFCWS metadata for use in our analyses. For most metadata fields, we assigned the common value of the component variables when they agreed and manually reviewed cases in which they differed; specialized rules were used for fields such as source, variable label, variable type, topic, and subtopic. Topics and subtopics were assigned based on the component variables, with a subtopic generally included when it applied to at least half of the component items.

\FloatBarrier

\subsection{Estimand and Estimation Procedure}
\label{app:estimand_and_estimation_procedure}

Our estimand is the expected out-of-sample predictive accuracy of a model generated by our training and selection procedure \citep{LundbergEtAl2021Estimand}. It is the accuracy that would be obtained, on average, if the procedure were applied to a new FFCWS sample of the same size, and the resulting model's predictions were evaluated on additional new observations drawn according to the FFCWS sampling design. In other words, we estimated the expected accuracy of the training and selection procedure rather than the accuracy of any particular fitted model.\footnote{For more on the difference between assessing a specific model and a training procedure, see~\citet{BatesHastieTibshirani2024}.}

Our estimation approach was nested cross-validation. Conceptually, this approach is equivalent to repeatedly implementing three-way splits of the data into training, validation, and evaluation sets. The separation between the data used for selection and the data used for evaluation avoids one source of optimistically biased performance estimates \citep{CawleyTalbot2010}. We used nested cross-validation rather than a simple three-way split to reduce the risk that our results are specific to a single split of the data. In the process detailed below, the inner training sets correspond to the concept of a training set, the inner test sets correspond to the concept of a validation set, and the outer test sets correspond to the concept of an evaluation set. 

Our process was as follows:
\begin{enumerate}
    \item \textit{Create outer folds.} We split the non-missing observations for a given outcome into five groups called outer folds. We combined four of these folds together into an outer training set, and held back the fifth fold as an outer test set. 
    \item \textit{Create inner folds.} We split each outer-fold training set (i.e., each combination of four outer folds) into five inner folds. We combined four inner folds together as an inner training set, and held back the fifth fold as an inner test set. 
    \item \textit{Use the inner folds to tune hyperparameters and select a pipeline.} Using one of our inner training sets, we trained thousands of models to predict the given outcome, using different pipelines and hyperparameter draws. Given our total sample size in the draft submitted for journal review ($n = 2{,}121$) and our use of five outer folds and five inner folds, candidate pipeline-hyperparameter combinations were trained on up to 1,357 families ($\frac{4}{5} \times \frac{4}{5} \times 2{,}121$) during this step. We then tested the resulting models on the associated inner test set. Within each of the five outer training sets, we repeated this process for all five combinations of inner training set and inner test set. The pipeline and associated hyperparameter values with the highest $\widehat{R^2}_{\mathrm{CV}}$ across the five inner test folds within an outer training set was the “winning” pipeline-hyperparameter combination for that outer training set (see Equation~\ref{eq:r2_cv} in the main text). The winning pipeline-hyperparameter combination could differ across the five outer training sets.
    \item \textit{Train and evaluate the selected pipeline on the outer folds.} On each outer training set, we trained the pipeline-hyperparameter combination that had been selected as the winner in the associated inner folds. The sample size for the training set in this step was up to 1,697 families ($\frac{4}{5} \times 2{,}121$). We repeated this process for all five outer training sets. We then used $R^2_{\mathrm{CV}}$ to summarize performance across all five outer test sets, and reported the resulting performance estimate in the paper.
\end{enumerate}

The sample sizes noted above will increase when we test our infrastructure on a sample of $n = 4{,}242$ after the journal review process, as discussed in Section~\ref{sec:eval_strategy}, ``Evaluation Strategy,'' in the main text. Since missingness patterns differ across outcomes, the sample size for each individual outcome may be lower.

\subsection{Uncertainty Quantification}
\label{app:uncertainty_quantification}

Our main performance metric is $R^2_{\mathrm{CV}}$ as defined in the main text. To quantify uncertainty, we find it helpful to re-express the estimate $\widehat{R^2}_{\mathrm{CV}}$ as
\begin{equation}
\widehat{R^2}_{\mathrm{CV}} = 1 - \hat{B}
\end{equation}
where 
\begin{equation}
\hat{B} = \frac{\sum_{i} e_{i, \text{pipeline}}}{\sum_{i} e_{i, \text{null}}}
\label{eq:hat_b}
\end{equation}
and $e_{i, \text{pipeline}}$ is the error given the chosen loss function (recall that in our case ``error'' was defined to be squared error) and $e_{i, \text{null}}$ is defined analogously for the null model (recall that in our case, the null model was defined to be the mean of the training data).

Since $\widehat{R^2}_{\mathrm{CV}} = 1 - \hat{B}$, the variance of $\widehat{R^2}_{\mathrm{CV}}$ is the variance of $\hat{B}$. $\hat{B}$ is a ratio, and therefore we follow standard practice for estimating the variance of a ratio. Following~\citet[p.~179]{SarndalSwenssonWretman1992}, we re-express the ratio as a first-order Taylor series (neglecting the finite population correction):
\begin{equation}
\widehat{se}(\widehat{R^2}_{\mathrm{CV}}) = \sqrt{\frac{1}{(\bar{e}_{\text{null}})^2} \frac{1}{n} \left( S^2_{e_{\text{pipeline}}} + \hat{B}^2 S^2_{e_{\text{null}}} - 2 \hat{B} S_{e_{\text{pipeline}}, e_{\text{null}}} \right)}
\end{equation}
where
\begin{equation}
S_{e_{\text{pipeline}}, e_{\text{null}}} = \frac{1}{n-1} \sum_{i} (e_{i, \text{pipeline}} - \bar{e}_{\text{pipeline}}) (e_{i, \text{null}} - \bar{e}_{\text{null}}) 
\end{equation}
\begin{equation}
S^2_{e_{\text{pipeline}}} = \frac{1}{n-1} \sum_{i} (e_{i, \text{pipeline}} - \bar{e}_{\text{pipeline}})^2
\end{equation}
and $S^2_{e_{\text{null}}}$ is defined analogously to $S^2_{e_{\text{pipeline}}}$.

We interpret this standard error as reflecting the sample-to-sample variability over different datasets of the same size drawn from the same data generating process~\citep{BatesHastieTibshirani2024}. In other words, our standard error is estimating the sample-to-sample variability of repeated, parallel versions of the FFCWS.

The 95 percent confidence interval for the original quantity of interest $\widehat{R^2}_{\mathrm{CV}}$ is then calculated using a normal approximation: 
\begin{equation}
CI = \widehat{R^2}_{\mathrm{CV}} \;\pm\; 1.96 \cdot \widehat{se}(\widehat{R^2}_{\mathrm{CV}})
\end{equation}

The method that we use above assumes that the errors ($e_i$) are independent. This assumption is widely used, but it should be considered an approximation that is more or less appropriate depending on the setting~\citep{Bengio_Grandvalet_2004}. Recently,~\citet{BatesHastieTibshirani2024} proposed a new method for estimating uncertainty after cross-validation that relaxes the need for this assumption but at the cost of requiring more computation. We use the standard method rather than the newer method for two reasons. First, based on the results in~\citet{BatesHastieTibshirani2024} (especially Figure 10), we believe we are in a setting where the standard approach should provide a reasonable approximation. Second, the newer method is substantially more computationally intensive because it adds another layer of nesting to the cross-validation. The required increase in computation time, cost, and complexity is impractical in a high-throughput setting. The core claims in our paper do not hinge on the exact coverage probabilities of these confidence intervals. 

\subsection{Pipeline Components and Hyperparameter Tuning}
\label{app:pipeline}

As described in the main text, we tested hundreds of machine learning pipelines for each outcome. We define a ``pipeline'' as a combination of options for: (1) scaling features, (2) selecting features, and (3) modeling the relationship between the selected features and the outcome. We also conducted hyperparameter tuning, thus testing thousands of trained models per outcome. Below we describe our approach to each of these components. 

\subsubsection{Scaling}
\label{app:scaling}

We considered six scaler options, selected based on availability in \texttt{scikit-learn}: standard scaler, robust scaler, quantile transformer with uniform distribution, quantile transformer with normal distribution, min-max scaler, and max-abs scaler. To determine which of these scalers were worth the computation cost of testing for all outcomes, we tested all six in a smaller job with approximately 50 outcomes. We found that for most outcomes, the six scalers generally performed similarly to one another in terms of the winning $\widehat{R^2}_{\mathrm{CV}}$ scores. However, for some outcomes, the winning $\widehat{R^2}_{\mathrm{CV}}$ scores differed across scalers. In particular, the quantile transformers performed better than the other scalers for some outcomes, but worse than the other scalers for other outcomes. We therefore chose two scalers to include in our final job: (1) standard scaler, and (2) quantile transformer with uniform distribution.

\subsubsection{Feature Selection}
\label{app:feature_selection}

We tested two methods of feature selection: (1) F-tests for linear association, and (2) Spearman rank correlations. For each of these two methods, we tested a variety of values for the number of selected features ($k$), ranging from $k = 5$ to $k = 100$. Based on exploratory analyses with approximately 50 outcomes, we opted to include $k = 5$, $k = 10$, $k = 25$, and $k = 50$ in our set of pipeline options. Selecting more than 50 features did not substantially improve predictive performance in these exploratory analyses. We also included an option with $k = 1$ so that a simple one-feature benchmark model could be selected within the same pipeline framework (see Section~\ref{sec:skill_is_low} of the main text). In total, this yielded 10 feature selection options (2 methods $\times$ 5 values of $k$). To avoid information leakage, feature selection was repeated independently for every training set in every iteration of the inner and outer cross-validation loops.

\subsubsection{Modeling}
\label{app:modeling}

We included the following 17 model classes in our pipeline options:

\begin{itemize}
    \item \textbf{Linear models:} linear regression, logistic regression, ridge regression, lasso regression, stochastic gradient descent regression, and Bayesian ridge regression.
    \item \textbf{Support vector models:} epsilon support vector regression, linear support vector regression, and nu support vector regression.
    \item \textbf{Tree-based models:} decision tree regression, extra trees regression, random forest regression, AdaBoost regression, and XGBoost regression.
    \item \textbf{Instance-based models:} $k$-nearest neighbors regression.
    \item \textbf{Neural network models:} multilayer perceptron regression.
    \item \textbf{Baseline model:} dummy regression.
\end{itemize}

We included dummy regression, which predicts the mean of the training data for all cases, to allow the pipeline selection procedure to choose a model with $R^2_{\mathrm{CV}} = 0$ if no alternative model provided predictive value. In practice, however, some other models typically achieved slightly positive $\widehat{R^2}_{\mathrm{CV}}$ in the inner cross-validation folds and were therefore selected, even when their performance in the outer evaluation folds was $\widehat{R^2}_{\mathrm{CV}} \leq 0$ due to sampling variation between folds. The dummy regressor is distinct from the ``simple benchmark'' described in Section~\ref{sec:skill_is_low} of the main text.

\subsubsection{Hyperparameter Tuning}
\label{app:hyperparameter_tuning}

We tested the default hyperparameter values plus up to 63 random draws from a grid of candidate hyperparameter values for each model class \citep{BergstraBengio2012}. For each outcome, we chose a winning pipeline and winning hyperparameter values for that pipeline based on performance on the inner folds of nested cross-validation. The winning pipeline and hyperparameters for each outcome could differ across the five outer folds (see Section~\ref{app:estimand_and_estimation_procedure}).

For each model class, we hard-coded one of the hyperparameter draws to match the default settings in our statistical learning packages, which often preconfigure hyperparameters to produce reasonable off-the-shelf performance. We then used expert judgment and guidance from \citet{Hastie2009ESL} to determine which hyperparameters to tune and select an initial set of values for the tuning grid, from which we drew random combinations of hyperparameter values as mentioned above. We then expanded the range of values in the grid based on the performance of those initial values. Specifically, we ran a job with approximately 50 outcomes using our initial training sample ($n = 2{,}121$), and checked whether the highest grid value of each hyperparameter increased the top inner fold $\widehat{R^2}_{\mathrm{CV}}$ by at least 0.01 for the associated model class for at least 10 percent of the outer cross-validation folds across those outcomes. If yes, we expanded the grid to include higher values for that hyperparameter. We repeated this process until the maximum value in the grid did \textit{not} increase the top inner fold $\widehat{R^2}_{\mathrm{CV}}$ by at least 0.01 for the associated model class for at least 10 percent of the outer folds across those outcomes, or until the maximum possible hyperparameter value was reached. We conducted the same procedure to determine when to add lower values for hyperparameters in our grid. Changing the values in the grid for one hyperparameter could affect the performance of the values for another hyperparameter in the same model, so we re-evaluated the sets of values for all hyperparameters in a model after changing the values for one or more hyperparameters in that model. The \href{https://github.com/emilycantrell/predicting-life-outcomes-ffcws/blob/main/models/paramgrid.py}{final grid} is available on our GitHub repository. It is possible that the ideal values for the hyperparameter grid could change when we test our infrastructure on a sample of $n = 4{,}242$ after the journal review process, but we will maintain our current hyperparameter grid values to avoid making data-dependent refinements to the search space after evaluating the final sample.


\subsection{Simple Benchmark Models}
\label{app:simple_benchmark_models}

Our simple benchmark models in Section~\ref{sec:skill_is_low} of the main text were constructed using a restricted version of the same pipeline components considered in our full combinatorial infrastructure. Specifically, features were standardized using a standard scaler, a single feature was selected based on its univariate association with the outcome via an F-test, and the model class was linear regression for continuous and ordered categorical outcomes or logistic regression for binary outcomes. The logistic regression models used an \( \ell_2 \) penalty, following \texttt{scikit-learn} defaults. These combinations of scaler, feature selection method, and model class were also included among the candidate pipelines in our full search space.
  
\subsection{Computational Reproducibility}

We wrote our code with the aim that other researchers can run the code and produce the same results. To support this goal we took a number of steps such as: (1) using virtual machines from Microsoft Azure, where other researchers can access a similar hardware environment, (2) using a dependency manager (\texttt{Miniconda}) to specify the required software environment, and (3) setting seeds for all instances involving pseudorandomness to further increase reproducibility~\citep{Liu2019}. 

To assess computational reproducibility, we ran our code twice on the same 46-outcome jobfiles; a reproducibility test on the full set of 484 outcomes was cost prohibitive. As expected, we did not detect any differences between the results of the two test jobs. The winning pipelines produced identical predictions for each outcome in each outer fold. We also confirmed that we pseudorandomly drew the same combinations of hyperparameter values from the hyperparameter grid in the tuning process. While simpler than the final job used in the paper, this test job used the same feature matrix, scalers, feature selectors, model classes, cross-validation fold structure, and hyperparameter grid tuning process as the final job. More information about the code can be found on our \href{https://github.com/emilycantrell/predicting-life-outcomes-ffcws}{GitHub repository}.

\clearpage

\setcounter{figure}{0}
\setcounter{table}{0}

\section{REFORMS Checklist}\label{sec:REFORMS}

To document our methodology and scientific claims in accordance with recommended practices, we completed the ``Recommendations for Machine-learning-based Science'' (REFORMS) checklist from \citet{Kapoor2024}.

\subsection{Module 1: Study goals}

\textbf{1a. Population or distribution about which the scientific claim is made.}

\noindent There are two ways to define the ``population of interest" for this study: the population of people and the population of outcomes. For the population of people, we focus on the FFCWS sample; we do not use sampling weights to generalize to the broader population targeted by the FFCWS sampling design \citep{ReichmanEtAl2001}. For the population of outcomes, we define the 484 outcomes we study as the full population of interest, because these outcomes are not statistically representative of a larger population of outcomes (see Section~\ref{sec:outcomes} of the main text). Therefore, we do not make statistical claims about generalizability to other outcomes. Instead, Section~\ref{sec:discussion} of the main text provides a qualitative discussion of the contexts in which we expect our findings to generalize and the limitations of that generalizability.

\noindent \textbf{1b. Motivation for choosing this population or distribution (1a.).}

\noindent Regarding both the population of people and the population of outcomes, the population of interest follows from our choice of dataset. We explain why we chose the FFCWS dataset in Section~\ref{sec:data} of the main text.

\noindent \textbf{1c. Motivation for the use of machine learning methods in the study.} 

\noindent The introduction to the main text explains our motivation for studying the predictability of life outcomes and our motivation for using a high-throughput approach with a wide range of machine learning methods.

\noindent 

\subsection{Module 2: Computational reproducibility}

\noindent \textbf{2a. Dataset used for training and evaluating the model along with link or DOI to uniquely identify the dataset.} 

\noindent See Section~\ref{sec:ffcws_data} of the main text, ``Overview of the FFCWS Data." \href{https://github.com/emilycantrell/predicting-life-outcomes-ffcws\#data-access}{Instructions} for accessing the version of the FFCWS data that we utilized in this manuscript are available on our GitHub repository.

\noindent \textbf{2b. Code used to train and evaluate the model and produce the results reported in the paper along with link or DOI to uniquely identify the version of the code used.}

\noindent The code is available on our \href{https://github.com/emilycantrell/predicting-life-outcomes-ffcws}{GitHub repository}.

\noindent \textbf{2c. Description of the computing infrastructure used.}

\noindent Our models were trained on a cloud computing environment (Microsoft Azure) using CPUs. Information about our computing infrastructure is available in \href{https://github.com/emilycantrell/predicting-life-outcomes-ffcws/blob/main/azure/worker.bicep}{\texttt{azure/worker.bicep}} on our GitHub repository. Additionally, a schematic describing the parallelization in our computing infrastructure is available in a \href{https://github.com/emilycantrell/predicting-life-outcomes-ffcws\#architecture}{README} on our GitHub repository. The Python version and package versions are pinned in \href{https://github.com/emilycantrell/predicting-life-outcomes-ffcws/blob/main/environment_pinned_linux.yml}{\texttt{environment\_pinned\_linux.yml}}. As stated in the main text, generating the complete set of results in the draft submitted to the journal for review took 959 compute hours.

\noindent \textbf{2d. README file which contains instructions for generating the results using the provided dataset and code.}

\noindent We provide instructions for \href{https://github.com/emilycantrell/predicting-life-outcomes-ffcws/blob/main/docs/1-Local-Setup.md}{running jobs on a local machine} and \href{https://github.com/emilycantrell/predicting-life-outcomes-ffcws/blob/main/docs/2-Azure-Setup.md}{running jobs on Azure} on our GitHub repository.

\noindent \textbf{2e. Reproduction script to produce all results reported in the paper.}

\noindent Reproducing the full analysis requires applying for access to the FFCWS data and provisioning Azure computing infrastructure. Since these steps cannot be fully automated, we provide the instructions linked above in checklist item (2d) rather than a reproduction script.

\subsection{Module 3: Data quality}\label{module3}

\noindent \textbf{3a. Source(s) of data, separately for the training and evaluation datasets (if applicable), along with the time when the dataset(s) are collected, the source and process of ground-truth annotations, and other data documentation.}

\noindent For information about the FFCWS sampling design, see \citet{ReichmanEtAl2001}. Additional information is available on the \href{https://ffcws.princeton.edu/data-and-documentation/getting-started}{FFCWS website}.

\noindent \textbf{3b. Distribution or set from which the dataset is sampled (i.e., the sampling frame).}

\noindent The sampling strategy is detailed in \citet{ReichmanEtAl2001}.

\noindent \textbf{3c. Justification for why the dataset is useful for the modeling task at hand.}

\noindent We explain why we chose the FFCWS dataset in Section~\ref{sec:data} of the main text.

\noindent \textbf{3d. The definition of the outcome variable of the model along with descriptive statistics, if applicable.}

\noindent In this paper, we predict hundreds of outcome variables. The full list of outcomes is available in Appendix~\ref{app:performance_for_all_outcomes}, and additional metadata for all outcomes is available in a \href{https://github.com/emilycantrell/predicting-life-outcomes-ffcws/blob/main/analysis/figures_and_analyses_for_paper/precomputed_results/results_and_metadata_for_all_outcomes_manuscript_submission.csv}{CSV file} on our GitHub repository. Codebooks are available on the \href{https://ffcws.princeton.edu/data-and-documentation/getting-started}{FFCWS website} for readers seeking more information about how each outcome is defined. 

\noindent \textbf{3e. Number of samples in the dataset.}

\noindent The aforementioned \href{https://github.com/emilycantrell/predicting-life-outcomes-ffcws/blob/main/analysis/figures_and_analyses_for_paper/precomputed_results/results_and_metadata_for_all_outcomes_manuscript_submission.csv}{CSV file} on our GitHub repository lists the sample size for each outcome before cross-validation splits are generated. This refers to the sample size within the sample of $n = 2{,}121$ that we used in the draft submitted for journal review. The sample size varies across outcomes due to missingness.

\noindent \textbf{3f. Percentage of missing data, split by class for a categorical outcome variable.}

\noindent We do not provide summary statistics on missing data. Missingness patterns in our current study are similar to the patterns described in Section S1.1 of the appendix of \citet{SalganikEtAl2020}, although we use a slightly different version of the FFCWS data. \citet{SalganikEtAl2020} write, ``Of the approximately 55 million possible entries in the background dataset ($4{,}242 \times 12{,}942$), about 73\% of possible entries did not have a value." They note that the largest source of missingness is survey skip patterns (i.e., many questions were only asked of certain subpopulations of respondents).

\noindent \textbf{3g. Justification for why the distribution or set from which the dataset is drawn (3b.) is representative of the one about which the scientific claim is being made (1a.).}

\noindent Regarding the population of people, our scientific claims concern predictive performance within the FFCWS sample used in our analysis rather than population-representative predictive performance. Regarding the population of outcomes, the 484 outcomes we study are defined as our whole population of interest, as discussed in Section~\ref{sec:outcomes} of the main text.

\subsection{Module 4: Data preprocessing}\label{module4}

\noindent \textbf{4a. Identification of whether any samples are excluded with a rationale for why they are excluded.}

\noindent For each outcome, we exclude observations for which the outcome value is missing, since model training and evaluation require an observed outcome value.

\noindent \textbf{4b. How impossible or corrupt samples are dealt with.}

\noindent We did not identify any impossible or corrupt samples.

\noindent \textbf{4c. All transformations of the dataset from its raw form (3a.) to the form used in the model, for instance, treatment of missing data and normalization.}

\noindent We describe our data preparation steps in Section~\ref{sec:data_preparation} of the main text. In addition to the data preparation steps described there, we also rescaled the features as described in Section~\ref{sec:machine_learning_pipelines} of the main text and Section~\ref{app:scaling} of the appendix.

\subsection{Module 5: Modeling}\label{module5}

\noindent \textbf{5a. Detailed descriptions of all models trained, including: all features used in the model (including any feature selection); types of models implemented (e.g., Random Forests, Neural Networks); loss function used.}

\noindent Section~\ref{app:feature_construction_and_exclusion} of the appendix shows how the raw FFCWS variables were transformed into the feature matrices that were utilized in our machine learning pipelines. We provide an overview of our feature selection and modeling methods in Section~\ref{sec:machine_learning_pipelines} of the main text, and we provide additional details in Section~\ref{app:pipeline} of the appendix.

\noindent \textbf{5b. Justification for the choice of model types implemented.}

\noindent As stated in Section~\ref{sec:machine_learning_pipelines} of the main text, we restricted our approach to techniques that were commonly available in mature open-source libraries such as \texttt{scikit-learn} \citep{PedregosaEtAl2011}.

\noindent \textbf{5c. Method for evaluating the model(s) reported in the paper, including details of train-test splits or cross-validation folds.}

\noindent See Section~\ref{sec:eval_strategy} of the main text and Section~\ref{app:estimand_and_estimation_procedure} of the appendix.

\noindent \textbf{5d. Method for selecting the model(s) reported in the paper.}

\noindent See Section~\ref{sec:eval_strategy} of the main text and Section~\ref{app:estimand_and_estimation_procedure} of the appendix.

\noindent \textbf{5e. For the model(s) reported in the paper, specify details about the hyperparameter tuning.}

\noindent See Section~\ref{sec:machine_learning_pipelines} of the main text and Section~\ref{app:hyperparameter_tuning} of the appendix.

\noindent \textbf{5f. Justification that model comparisons are against appropriate baselines.}

\noindent We justify our benchmark procedure in Section~\ref{sec:skill_is_low} of the main text.

\subsection{Module 6: Data leakage}

\noindent \textbf{6a. Justification that pre-processing (\hyperref[module4]{Module 4}) and modeling (\hyperref[module5]{Module 5}) steps only use information from the training dataset (and not the test dataset).}

\noindent No information from the outer test folds was used for preprocessing, model selection, or model fitting. Within each outer training fold, imputation was performed before the inner cross-validation splits, while scaling, feature selection, and model fitting were performed separately within each inner training fold.

\noindent \textbf{6b. Methods to address dependencies or duplicates between the training and test datasets (e.g., different samples from the same patients are kept in the same dataset partition).}

\noindent Not applicable. Each focal family contributes at most one observation; we therefore do not have repeated observations from the same family split across training and test folds.

\noindent \textbf{6c. Justification that each feature or input used in the model is legitimate for the task at hand and does not lead to leakage.}

\noindent Our goal was to predict outcomes measured at survey year 15 using data from earlier years. Accordingly, the feature matrix only includes data from years 0--9. Additionally, we omitted four outcomes for which the FFCWS data contained imputed values based on data from prior years, as described in Section~\ref{app:outcome_construction_and_exclusion} of the appendix, since predicting those outcomes with the same features that were used for imputation could artificially inflate performance.

\subsection{Module 7: Metrics and uncertainty}

\noindent \textbf{7a. All metrics used to assess and compare model performance (e.g., accuracy, AUROC, etc.). Justify that the metric used to select the final model is suitable for the task.}

\noindent We utilized $R^2_{\mathrm{CV}}$ for model selection and evaluation. See Section~\ref{sec:metric} of the main text for our motivations for choosing this metric.

\noindent \textbf{7b. Uncertainty estimates (e.g., confidence intervals, standard deviations), and details of how these are calculated.}

\noindent See Section~\ref{app:uncertainty_quantification} of the appendix for information about how we calculated confidence intervals.

\noindent \textbf{7c. Justification for the choice of statistical tests (if used) and a check for the assumptions of the statistical test.}

\noindent Not applicable.

\subsection{Module 8: Generalizability and limitations}

\noindent \textbf{8a. Evidence of external validity.}

\noindent See Section~\ref{sec:discussion} of the main text for a discussion of generalizability across machine learning methods, data, and prediction tasks, as well as a comparison to prediction in U.S. social services settings.

\noindent \textbf{8b. Contexts in which the authors do not expect the study’s findings to hold.}

\noindent The discussion of generalizability in Section~\ref{sec:discussion} of the main text also considers the boundaries of our findings' applicability to other contexts.

\end{document}